\documentclass[usenatbib,onecolumn]{mn2e}

\usepackage{amsmath}
\usepackage{graphicx}
\usepackage{color}\definecolor{AliceBlue}{rgb}{0.94,0.97,1.00}
\definecolor{AntiqueWhite1}{rgb}{1.00,0.94,0.86}
\definecolor{AntiqueWhite2}{rgb}{0.93,0.87,0.80}
\definecolor{AntiqueWhite3}{rgb}{0.80,0.75,0.69}
\definecolor{AntiqueWhite4}{rgb}{0.55,0.51,0.47}
\definecolor{AntiqueWhite}{rgb}{0.98,0.92,0.84}
\definecolor{BlanchedAlmond}{rgb}{1.00,0.92,0.80}
\definecolor{BlueViolet}{rgb}{0.54,0.17,0.89}
\definecolor{CadetBlue1}{rgb}{0.60,0.96,1.00}
\definecolor{CadetBlue2}{rgb}{0.56,0.90,0.93}
\definecolor{CadetBlue3}{rgb}{0.48,0.77,0.80}
\definecolor{CadetBlue4}{rgb}{0.33,0.53,0.55}
\definecolor{CadetBlue}{rgb}{0.37,0.62,0.63}
\definecolor{CornflowerBlue}{rgb}{0.39,0.58,0.93}
\definecolor{DarkBlue}{rgb}{0.00,0.00,0.55}
\definecolor{DarkCyan}{rgb}{0.00,0.55,0.55}
\definecolor{DarkGoldenrod1}{rgb}{1.00,0.73,0.06}
\definecolor{DarkGoldenrod2}{rgb}{0.93,0.68,0.05}
\definecolor{DarkGoldenrod3}{rgb}{0.80,0.58,0.05}
\definecolor{DarkGoldenrod4}{rgb}{0.55,0.40,0.03}
\definecolor{DarkGoldenrod}{rgb}{0.72,0.53,0.04}
\definecolor{DarkGray}{rgb}{0.66,0.66,0.66}
\definecolor{DarkGreen}{rgb}{0.00,0.39,0.00}
\definecolor{DarkGrey}{rgb}{0.66,0.66,0.66}
\definecolor{DarkKhaki}{rgb}{0.74,0.72,0.42}
\definecolor{DarkMagenta}{rgb}{0.55,0.00,0.55}
\definecolor{DarkOliveGreen1}{rgb}{0.79,1.00,0.44}
\definecolor{DarkOliveGreen2}{rgb}{0.74,0.93,0.41}
\definecolor{DarkOliveGreen3}{rgb}{0.64,0.80,0.35}
\definecolor{DarkOliveGreen4}{rgb}{0.43,0.55,0.24}
\definecolor{DarkOliveGreen}{rgb}{0.33,0.42,0.18}
\definecolor{DarkOrange1}{rgb}{1.00,0.50,0.00}
\definecolor{DarkOrange2}{rgb}{0.93,0.46,0.00}
\definecolor{DarkOrange3}{rgb}{0.80,0.40,0.00}
\definecolor{DarkOrange4}{rgb}{0.55,0.27,0.00}
\definecolor{DarkOrange}{rgb}{1.00,0.55,0.00}
\definecolor{DarkOrchid1}{rgb}{0.75,0.24,1.00}
\definecolor{DarkOrchid2}{rgb}{0.70,0.23,0.93}
\definecolor{DarkOrchid3}{rgb}{0.60,0.20,0.80}
\definecolor{DarkOrchid4}{rgb}{0.41,0.13,0.55}
\definecolor{DarkOrchid}{rgb}{0.60,0.20,0.80}
\definecolor{DarkRed}{rgb}{0.55,0.00,0.00}
\definecolor{DarkSalmon}{rgb}{0.91,0.59,0.48}
\definecolor{DarkSeaGreen1}{rgb}{0.76,1.00,0.76}
\definecolor{DarkSeaGreen2}{rgb}{0.71,0.93,0.71}
\definecolor{DarkSeaGreen3}{rgb}{0.61,0.80,0.61}
\definecolor{DarkSeaGreen4}{rgb}{0.41,0.55,0.41}
\definecolor{DarkSeaGreen}{rgb}{0.56,0.74,0.56}
\definecolor{DarkSlateBlue}{rgb}{0.28,0.24,0.55}
\definecolor{DarkSlateGray1}{rgb}{0.59,1.00,1.00}
\definecolor{DarkSlateGray2}{rgb}{0.55,0.93,0.93}
\definecolor{DarkSlateGray3}{rgb}{0.47,0.80,0.80}
\definecolor{DarkSlateGray4}{rgb}{0.32,0.55,0.55}
\definecolor{DarkSlateGray}{rgb}{0.18,0.31,0.31}
\definecolor{DarkSlateGrey}{rgb}{0.18,0.31,0.31}
\definecolor{DarkTurquoise}{rgb}{0.00,0.81,0.82}
\definecolor{DarkViolet}{rgb}{0.58,0.00,0.83}
\definecolor{DeepPink1}{rgb}{1.00,0.08,0.58}
\definecolor{DeepPink2}{rgb}{0.93,0.07,0.54}
\definecolor{DeepPink3}{rgb}{0.80,0.06,0.46}
\definecolor{DeepPink4}{rgb}{0.55,0.04,0.31}
\definecolor{DeepPink}{rgb}{1.00,0.08,0.58}
\definecolor{DeepSkyBlue1}{rgb}{0.00,0.75,1.00}
\definecolor{DeepSkyBlue2}{rgb}{0.00,0.70,0.93}
\definecolor{DeepSkyBlue3}{rgb}{0.00,0.60,0.80}
\definecolor{DeepSkyBlue4}{rgb}{0.00,0.41,0.55}
\definecolor{DeepSkyBlue}{rgb}{0.00,0.75,1.00}
\definecolor{DimGray}{rgb}{0.41,0.41,0.41}
\definecolor{DimGrey}{rgb}{0.41,0.41,0.41}
\definecolor{DodgerBlue1}{rgb}{0.12,0.56,1.00}
\definecolor{DodgerBlue2}{rgb}{0.11,0.53,0.93}
\definecolor{DodgerBlue3}{rgb}{0.09,0.45,0.80}
\definecolor{DodgerBlue4}{rgb}{0.06,0.31,0.55}
\definecolor{DodgerBlue}{rgb}{0.12,0.56,1.00}
\definecolor{FloralWhite}{rgb}{1.00,0.98,0.94}
\definecolor{ForestGreen}{rgb}{0.13,0.55,0.13}
\definecolor{GhostWhite}{rgb}{0.97,0.97,1.00}
\definecolor{GreenYellow}{rgb}{0.68,1.00,0.18}
\definecolor{HotPink1}{rgb}{1.00,0.43,0.71}
\definecolor{HotPink2}{rgb}{0.93,0.42,0.65}
\definecolor{HotPink3}{rgb}{0.80,0.38,0.56}
\definecolor{HotPink4}{rgb}{0.55,0.23,0.38}
\definecolor{HotPink}{rgb}{1.00,0.41,0.71}
\definecolor{IndianRed1}{rgb}{1.00,0.42,0.42}
\definecolor{IndianRed2}{rgb}{0.93,0.39,0.39}
\definecolor{IndianRed3}{rgb}{0.80,0.33,0.33}
\definecolor{IndianRed4}{rgb}{0.55,0.23,0.23}
\definecolor{IndianRed}{rgb}{0.80,0.36,0.36}
\definecolor{LavenderBlush1}{rgb}{1.00,0.94,0.96}
\definecolor{LavenderBlush2}{rgb}{0.93,0.88,0.90}
\definecolor{LavenderBlush3}{rgb}{0.80,0.76,0.77}
\definecolor{LavenderBlush4}{rgb}{0.55,0.51,0.53}
\definecolor{LavenderBlush}{rgb}{1.00,0.94,0.96}
\definecolor{LawnGreen}{rgb}{0.49,0.99,0.00}
\definecolor{LemonChiffon1}{rgb}{1.00,0.98,0.80}
\definecolor{LemonChiffon2}{rgb}{0.93,0.91,0.75}
\definecolor{LemonChiffon3}{rgb}{0.80,0.79,0.65}
\definecolor{LemonChiffon4}{rgb}{0.55,0.54,0.44}
\definecolor{LemonChiffon}{rgb}{1.00,0.98,0.80}
\definecolor{LightBlue1}{rgb}{0.75,0.94,1.00}
\definecolor{LightBlue2}{rgb}{0.70,0.87,0.93}
\definecolor{LightBlue3}{rgb}{0.60,0.75,0.80}
\definecolor{LightBlue4}{rgb}{0.41,0.51,0.55}
\definecolor{LightBlue}{rgb}{0.68,0.85,0.90}
\definecolor{LightCoral}{rgb}{0.94,0.50,0.50}
\definecolor{LightCyan1}{rgb}{0.88,1.00,1.00}
\definecolor{LightCyan2}{rgb}{0.82,0.93,0.93}
\definecolor{LightCyan3}{rgb}{0.71,0.80,0.80}
\definecolor{LightCyan4}{rgb}{0.48,0.55,0.55}
\definecolor{LightCyan}{rgb}{0.88,1.00,1.00}
\definecolor{LightGoldenrod1}{rgb}{1.00,0.93,0.55}
\definecolor{LightGoldenrod2}{rgb}{0.93,0.86,0.51}
\definecolor{LightGoldenrod3}{rgb}{0.80,0.75,0.44}
\definecolor{LightGoldenrod4}{rgb}{0.55,0.51,0.30}
\definecolor{LightGoldenrodYellow}{rgb}{0.98,0.98,0.82}
\definecolor{LightGoldenrod}{rgb}{0.93,0.87,0.51}
\definecolor{LightGray}{rgb}{0.83,0.83,0.83}
\definecolor{LightGreen}{rgb}{0.56,0.93,0.56}
\definecolor{LightGrey}{rgb}{0.83,0.83,0.83}
\definecolor{LightPink1}{rgb}{1.00,0.68,0.73}
\definecolor{LightPink2}{rgb}{0.93,0.64,0.68}
\definecolor{LightPink3}{rgb}{0.80,0.55,0.58}
\definecolor{LightPink4}{rgb}{0.55,0.37,0.40}
\definecolor{LightPink}{rgb}{1.00,0.71,0.76}
\definecolor{LightSalmon1}{rgb}{1.00,0.63,0.48}
\definecolor{LightSalmon2}{rgb}{0.93,0.58,0.45}
\definecolor{LightSalmon3}{rgb}{0.80,0.51,0.38}
\definecolor{LightSalmon4}{rgb}{0.55,0.34,0.26}
\definecolor{LightSalmon}{rgb}{1.00,0.63,0.48}
\definecolor{LightSeaGreen}{rgb}{0.13,0.70,0.67}
\definecolor{LightSkyBlue1}{rgb}{0.69,0.89,1.00}
\definecolor{LightSkyBlue2}{rgb}{0.64,0.83,0.93}
\definecolor{LightSkyBlue3}{rgb}{0.55,0.71,0.80}
\definecolor{LightSkyBlue4}{rgb}{0.38,0.48,0.55}
\definecolor{LightSkyBlue}{rgb}{0.53,0.81,0.98}
\definecolor{LightSlateBlue}{rgb}{0.52,0.44,1.00}
\definecolor{LightSlateGray}{rgb}{0.47,0.53,0.60}
\definecolor{LightSlateGrey}{rgb}{0.47,0.53,0.60}
\definecolor{LightSteelBlue1}{rgb}{0.79,0.88,1.00}
\definecolor{LightSteelBlue2}{rgb}{0.74,0.82,0.93}
\definecolor{LightSteelBlue3}{rgb}{0.64,0.71,0.80}
\definecolor{LightSteelBlue4}{rgb}{0.43,0.48,0.55}
\definecolor{LightSteelBlue}{rgb}{0.69,0.77,0.87}
\definecolor{LightYellow1}{rgb}{1.00,1.00,0.88}
\definecolor{LightYellow2}{rgb}{0.93,0.93,0.82}
\definecolor{LightYellow3}{rgb}{0.80,0.80,0.71}
\definecolor{LightYellow4}{rgb}{0.55,0.55,0.48}
\definecolor{LightYellow}{rgb}{1.00,1.00,0.88}
\definecolor{LimeGreen}{rgb}{0.20,0.80,0.20}
\definecolor{MediumAquamarine}{rgb}{0.40,0.80,0.67}
\definecolor{MediumBlue}{rgb}{0.00,0.00,0.80}
\definecolor{MediumOrchid1}{rgb}{0.88,0.40,1.00}
\definecolor{MediumOrchid2}{rgb}{0.82,0.37,0.93}
\definecolor{MediumOrchid3}{rgb}{0.71,0.32,0.80}
\definecolor{MediumOrchid4}{rgb}{0.48,0.22,0.55}
\definecolor{MediumOrchid}{rgb}{0.73,0.33,0.83}
\definecolor{MediumPurple1}{rgb}{0.67,0.51,1.00}
\definecolor{MediumPurple2}{rgb}{0.62,0.47,0.93}
\definecolor{MediumPurple3}{rgb}{0.54,0.41,0.80}
\definecolor{MediumPurple4}{rgb}{0.36,0.28,0.55}
\definecolor{MediumPurple}{rgb}{0.58,0.44,0.86}
\definecolor{MediumSeaGreen}{rgb}{0.24,0.70,0.44}
\definecolor{MediumSlateBlue}{rgb}{0.48,0.41,0.93}
\definecolor{MediumSpringGreen}{rgb}{0.00,0.98,0.60}
\definecolor{MediumTurquoise}{rgb}{0.28,0.82,0.80}
\definecolor{MediumVioletRed}{rgb}{0.78,0.08,0.52}
\definecolor{MidnightBlue}{rgb}{0.10,0.10,0.44}
\definecolor{MintCream}{rgb}{0.96,1.00,0.98}
\definecolor{MistyRose1}{rgb}{1.00,0.89,0.88}
\definecolor{MistyRose2}{rgb}{0.93,0.84,0.82}
\definecolor{MistyRose3}{rgb}{0.80,0.72,0.71}
\definecolor{MistyRose4}{rgb}{0.55,0.49,0.48}
\definecolor{MistyRose}{rgb}{1.00,0.89,0.88}
\definecolor{NavajoWhite1}{rgb}{1.00,0.87,0.68}
\definecolor{NavajoWhite2}{rgb}{0.93,0.81,0.63}
\definecolor{NavajoWhite3}{rgb}{0.80,0.70,0.55}
\definecolor{NavajoWhite4}{rgb}{0.55,0.47,0.37}
\definecolor{NavajoWhite}{rgb}{1.00,0.87,0.68}
\definecolor{NavyBlue}{rgb}{0.00,0.00,0.50}
\definecolor{OldLace}{rgb}{0.99,0.96,0.90}
\definecolor{OliveDrab1}{rgb}{0.75,1.00,0.24}
\definecolor{OliveDrab2}{rgb}{0.70,0.93,0.23}
\definecolor{OliveDrab3}{rgb}{0.60,0.80,0.20}
\definecolor{OliveDrab4}{rgb}{0.41,0.55,0.13}
\definecolor{OliveDrab}{rgb}{0.42,0.56,0.14}
\definecolor{OrangeRed1}{rgb}{1.00,0.27,0.00}
\definecolor{OrangeRed2}{rgb}{0.93,0.25,0.00}
\definecolor{OrangeRed3}{rgb}{0.80,0.22,0.00}
\definecolor{OrangeRed4}{rgb}{0.55,0.15,0.00}
\definecolor{OrangeRed}{rgb}{1.00,0.27,0.00}
\definecolor{PaleGoldenrod}{rgb}{0.93,0.91,0.67}
\definecolor{PaleGreen1}{rgb}{0.60,1.00,0.60}
\definecolor{PaleGreen2}{rgb}{0.56,0.93,0.56}
\definecolor{PaleGreen3}{rgb}{0.49,0.80,0.49}
\definecolor{PaleGreen4}{rgb}{0.33,0.55,0.33}
\definecolor{PaleGreen}{rgb}{0.60,0.98,0.60}
\definecolor{PaleTurquoise1}{rgb}{0.73,1.00,1.00}
\definecolor{PaleTurquoise2}{rgb}{0.68,0.93,0.93}
\definecolor{PaleTurquoise3}{rgb}{0.59,0.80,0.80}
\definecolor{PaleTurquoise4}{rgb}{0.40,0.55,0.55}
\definecolor{PaleTurquoise}{rgb}{0.69,0.93,0.93}
\definecolor{PaleVioletRed1}{rgb}{1.00,0.51,0.67}
\definecolor{PaleVioletRed2}{rgb}{0.93,0.47,0.62}
\definecolor{PaleVioletRed3}{rgb}{0.80,0.41,0.54}
\definecolor{PaleVioletRed4}{rgb}{0.55,0.28,0.36}
\definecolor{PaleVioletRed}{rgb}{0.86,0.44,0.58}
\definecolor{PapayaWhip}{rgb}{1.00,0.94,0.84}
\definecolor{PeachPuff1}{rgb}{1.00,0.85,0.73}
\definecolor{PeachPuff2}{rgb}{0.93,0.80,0.68}
\definecolor{PeachPuff3}{rgb}{0.80,0.69,0.58}
\definecolor{PeachPuff4}{rgb}{0.55,0.47,0.40}
\definecolor{PeachPuff}{rgb}{1.00,0.85,0.73}
\definecolor{PowderBlue}{rgb}{0.69,0.88,0.90}
\definecolor{RosyBrown1}{rgb}{1.00,0.76,0.76}
\definecolor{RosyBrown2}{rgb}{0.93,0.71,0.71}
\definecolor{RosyBrown3}{rgb}{0.80,0.61,0.61}
\definecolor{RosyBrown4}{rgb}{0.55,0.41,0.41}
\definecolor{RosyBrown}{rgb}{0.74,0.56,0.56}
\definecolor{RoyalBlue1}{rgb}{0.28,0.46,1.00}
\definecolor{RoyalBlue2}{rgb}{0.26,0.43,0.93}
\definecolor{RoyalBlue3}{rgb}{0.23,0.37,0.80}
\definecolor{RoyalBlue4}{rgb}{0.15,0.25,0.55}
\definecolor{RoyalBlue}{rgb}{0.25,0.41,0.88}
\definecolor{SaddleBrown}{rgb}{0.55,0.27,0.07}
\definecolor{SandyBrown}{rgb}{0.96,0.64,0.38}
\definecolor{SeaGreen1}{rgb}{0.33,1.00,0.62}
\definecolor{SeaGreen2}{rgb}{0.31,0.93,0.58}
\definecolor{SeaGreen3}{rgb}{0.26,0.80,0.50}
\definecolor{SeaGreen4}{rgb}{0.18,0.55,0.34}
\definecolor{SeaGreen}{rgb}{0.18,0.55,0.34}
\definecolor{SkyBlue1}{rgb}{0.53,0.81,1.00}
\definecolor{SkyBlue2}{rgb}{0.49,0.75,0.93}
\definecolor{SkyBlue3}{rgb}{0.42,0.65,0.80}
\definecolor{SkyBlue4}{rgb}{0.29,0.44,0.55}
\definecolor{SkyBlue}{rgb}{0.53,0.81,0.92}
\definecolor{SlateBlue1}{rgb}{0.51,0.44,1.00}
\definecolor{SlateBlue2}{rgb}{0.48,0.40,0.93}
\definecolor{SlateBlue3}{rgb}{0.41,0.35,0.80}
\definecolor{SlateBlue4}{rgb}{0.28,0.24,0.55}
\definecolor{SlateBlue}{rgb}{0.42,0.35,0.80}
\definecolor{SlateGray1}{rgb}{0.78,0.89,1.00}
\definecolor{SlateGray2}{rgb}{0.73,0.83,0.93}
\definecolor{SlateGray3}{rgb}{0.62,0.71,0.80}
\definecolor{SlateGray4}{rgb}{0.42,0.48,0.55}
\definecolor{SlateGray}{rgb}{0.44,0.50,0.56}
\definecolor{SlateGrey}{rgb}{0.44,0.50,0.56}
\definecolor{SpringGreen1}{rgb}{0.00,1.00,0.50}
\definecolor{SpringGreen2}{rgb}{0.00,0.93,0.46}
\definecolor{SpringGreen3}{rgb}{0.00,0.80,0.40}
\definecolor{SpringGreen4}{rgb}{0.00,0.55,0.27}
\definecolor{SpringGreen}{rgb}{0.00,1.00,0.50}
\definecolor{SteelBlue1}{rgb}{0.39,0.72,1.00}
\definecolor{SteelBlue2}{rgb}{0.36,0.67,0.93}
\definecolor{SteelBlue3}{rgb}{0.31,0.58,0.80}
\definecolor{SteelBlue4}{rgb}{0.21,0.39,0.55}
\definecolor{SteelBlue}{rgb}{0.27,0.51,0.71}
\definecolor{VioletRed1}{rgb}{1.00,0.24,0.59}
\definecolor{VioletRed2}{rgb}{0.93,0.23,0.55}
\definecolor{VioletRed3}{rgb}{0.80,0.20,0.47}
\definecolor{VioletRed4}{rgb}{0.55,0.13,0.32}
\definecolor{VioletRed}{rgb}{0.82,0.13,0.56}
\definecolor{WhiteSmoke}{rgb}{0.96,0.96,0.96}
\definecolor{YellowGreen}{rgb}{0.60,0.80,0.20}
\definecolor{aliceblue}{rgb}{0.94,0.97,1.00}
\definecolor{antiquewhite}{rgb}{0.98,0.92,0.84}
\definecolor{aquamarine1}{rgb}{0.50,1.00,0.83}
\definecolor{aquamarine2}{rgb}{0.46,0.93,0.78}
\definecolor{aquamarine3}{rgb}{0.40,0.80,0.67}
\definecolor{aquamarine4}{rgb}{0.27,0.55,0.45}
\definecolor{aquamarine}{rgb}{0.50,1.00,0.83}
\definecolor{azure1}{rgb}{0.94,1.00,1.00}
\definecolor{azure2}{rgb}{0.88,0.93,0.93}
\definecolor{azure3}{rgb}{0.76,0.80,0.80}
\definecolor{azure4}{rgb}{0.51,0.55,0.55}
\definecolor{azure}{rgb}{0.94,1.00,1.00}
\definecolor{beige}{rgb}{0.96,0.96,0.86}
\definecolor{bisque1}{rgb}{1.00,0.89,0.77}
\definecolor{bisque2}{rgb}{0.93,0.84,0.72}
\definecolor{bisque3}{rgb}{0.80,0.72,0.62}
\definecolor{bisque4}{rgb}{0.55,0.49,0.42}
\definecolor{bisque}{rgb}{1.00,0.89,0.77}
\definecolor{black}{rgb}{0.00,0.00,0.00}
\definecolor{blanchedalmond}{rgb}{1.00,0.92,0.80}
\definecolor{blue1}{rgb}{0.00,0.00,1.00}
\definecolor{blue2}{rgb}{0.00,0.00,0.93}
\definecolor{blue3}{rgb}{0.00,0.00,0.80}
\definecolor{blue4}{rgb}{0.00,0.00,0.55}
\definecolor{blueviolet}{rgb}{0.54,0.17,0.89}
\definecolor{blue}{rgb}{0.00,0.00,1.00}
\definecolor{brown1}{rgb}{1.00,0.25,0.25}
\definecolor{brown2}{rgb}{0.93,0.23,0.23}
\definecolor{brown3}{rgb}{0.80,0.20,0.20}
\definecolor{brown4}{rgb}{0.55,0.14,0.14}
\definecolor{brown}{rgb}{0.65,0.16,0.16}
\definecolor{burlywood1}{rgb}{1.00,0.83,0.61}
\definecolor{burlywood2}{rgb}{0.93,0.77,0.57}
\definecolor{burlywood3}{rgb}{0.80,0.67,0.49}
\definecolor{burlywood4}{rgb}{0.55,0.45,0.33}
\definecolor{burlywood}{rgb}{0.87,0.72,0.53}
\definecolor{cadetblue}{rgb}{0.37,0.62,0.63}
\definecolor{chartreuse1}{rgb}{0.50,1.00,0.00}
\definecolor{chartreuse2}{rgb}{0.46,0.93,0.00}
\definecolor{chartreuse3}{rgb}{0.40,0.80,0.00}
\definecolor{chartreuse4}{rgb}{0.27,0.55,0.00}
\definecolor{chartreuse}{rgb}{0.50,1.00,0.00}
\definecolor{chocolate1}{rgb}{1.00,0.50,0.14}
\definecolor{chocolate2}{rgb}{0.93,0.46,0.13}
\definecolor{chocolate3}{rgb}{0.80,0.40,0.11}
\definecolor{chocolate4}{rgb}{0.55,0.27,0.07}
\definecolor{chocolate}{rgb}{0.82,0.41,0.12}
\definecolor{coral1}{rgb}{1.00,0.45,0.34}
\definecolor{coral2}{rgb}{0.93,0.42,0.31}
\definecolor{coral3}{rgb}{0.80,0.36,0.27}
\definecolor{coral4}{rgb}{0.55,0.24,0.18}
\definecolor{coral}{rgb}{1.00,0.50,0.31}
\definecolor{cornflowerblue}{rgb}{0.39,0.58,0.93}
\definecolor{cornsilk1}{rgb}{1.00,0.97,0.86}
\definecolor{cornsilk2}{rgb}{0.93,0.91,0.80}
\definecolor{cornsilk3}{rgb}{0.80,0.78,0.69}
\definecolor{cornsilk4}{rgb}{0.55,0.53,0.47}
\definecolor{cornsilk}{rgb}{1.00,0.97,0.86}
\definecolor{cyan1}{rgb}{0.00,1.00,1.00}
\definecolor{cyan2}{rgb}{0.00,0.93,0.93}
\definecolor{cyan3}{rgb}{0.00,0.80,0.80}
\definecolor{cyan4}{rgb}{0.00,0.55,0.55}
\definecolor{cyan}{rgb}{0.00,1.00,1.00}
\definecolor{darkblue}{rgb}{0.00,0.00,0.55}
\definecolor{darkcyan}{rgb}{0.00,0.55,0.55}
\definecolor{darkgoldenrod}{rgb}{0.72,0.53,0.04}
\definecolor{darkgray}{rgb}{0.66,0.66,0.66}
\definecolor{darkgreen}{rgb}{0.00,0.39,0.00}
\definecolor{darkgrey}{rgb}{0.66,0.66,0.66}
\definecolor{darkkhaki}{rgb}{0.74,0.72,0.42}
\definecolor{darkmagenta}{rgb}{0.55,0.00,0.55}
\definecolor{darkolive}{rgb}{0.33,0.42,0.18}
\definecolor{darkorange}{rgb}{1.00,0.55,0.00}
\definecolor{darkorchid}{rgb}{0.60,0.20,0.80}
\definecolor{darkred}{rgb}{0.55,0.00,0.00}
\definecolor{darksalmon}{rgb}{0.91,0.59,0.48}
\definecolor{darksea}{rgb}{0.56,0.74,0.56}
\definecolor{darkslate}{rgb}{0.18,0.31,0.31}
\definecolor{darkslate}{rgb}{0.18,0.31,0.31}
\definecolor{darkslate}{rgb}{0.28,0.24,0.55}
\definecolor{darkturquoise}{rgb}{0.00,0.81,0.82}
\definecolor{darkviolet}{rgb}{0.58,0.00,0.83}
\definecolor{deeppink}{rgb}{1.00,0.08,0.58}
\definecolor{deepsky}{rgb}{0.00,0.75,1.00}
\definecolor{dimgray}{rgb}{0.41,0.41,0.41}
\definecolor{dimgrey}{rgb}{0.41,0.41,0.41}
\definecolor{dodgerblue}{rgb}{0.12,0.56,1.00}
\definecolor{firebrick1}{rgb}{1.00,0.19,0.19}
\definecolor{firebrick2}{rgb}{0.93,0.17,0.17}
\definecolor{firebrick3}{rgb}{0.80,0.15,0.15}
\definecolor{firebrick4}{rgb}{0.55,0.10,0.10}
\definecolor{firebrick}{rgb}{0.70,0.13,0.13}
\definecolor{floralwhite}{rgb}{1.00,0.98,0.94}
\definecolor{forestgreen}{rgb}{0.13,0.55,0.13}
\definecolor{gainsboro}{rgb}{0.86,0.86,0.86}
\definecolor{ghostwhite}{rgb}{0.97,0.97,1.00}
\definecolor{gold1}{rgb}{1.00,0.84,0.00}
\definecolor{gold2}{rgb}{0.93,0.79,0.00}
\definecolor{gold3}{rgb}{0.80,0.68,0.00}
\definecolor{gold4}{rgb}{0.55,0.46,0.00}
\definecolor{goldenrod1}{rgb}{1.00,0.76,0.15}
\definecolor{goldenrod2}{rgb}{0.93,0.71,0.13}
\definecolor{goldenrod3}{rgb}{0.80,0.61,0.11}
\definecolor{goldenrod4}{rgb}{0.55,0.41,0.08}
\definecolor{goldenrod}{rgb}{0.85,0.65,0.13}
\definecolor{gold}{rgb}{1.00,0.84,0.00}
\definecolor{gray0}{rgb}{0.00,0.00,0.00}
\definecolor{gray100}{rgb}{1.00,1.00,1.00}
\definecolor{gray10}{rgb}{0.10,0.10,0.10}
\definecolor{gray11}{rgb}{0.11,0.11,0.11}
\definecolor{gray12}{rgb}{0.12,0.12,0.12}
\definecolor{gray13}{rgb}{0.13,0.13,0.13}
\definecolor{gray14}{rgb}{0.14,0.14,0.14}
\definecolor{gray15}{rgb}{0.15,0.15,0.15}
\definecolor{gray16}{rgb}{0.16,0.16,0.16}
\definecolor{gray17}{rgb}{0.17,0.17,0.17}
\definecolor{gray18}{rgb}{0.18,0.18,0.18}
\definecolor{gray19}{rgb}{0.19,0.19,0.19}
\definecolor{gray1}{rgb}{0.01,0.01,0.01}
\definecolor{gray20}{rgb}{0.20,0.20,0.20}
\definecolor{gray21}{rgb}{0.21,0.21,0.21}
\definecolor{gray22}{rgb}{0.22,0.22,0.22}
\definecolor{gray23}{rgb}{0.23,0.23,0.23}
\definecolor{gray24}{rgb}{0.24,0.24,0.24}
\definecolor{gray25}{rgb}{0.25,0.25,0.25}
\definecolor{gray26}{rgb}{0.26,0.26,0.26}
\definecolor{gray27}{rgb}{0.27,0.27,0.27}
\definecolor{gray28}{rgb}{0.28,0.28,0.28}
\definecolor{gray29}{rgb}{0.29,0.29,0.29}
\definecolor{gray2}{rgb}{0.02,0.02,0.02}
\definecolor{gray30}{rgb}{0.30,0.30,0.30}
\definecolor{gray31}{rgb}{0.31,0.31,0.31}
\definecolor{gray32}{rgb}{0.32,0.32,0.32}
\definecolor{gray33}{rgb}{0.33,0.33,0.33}
\definecolor{gray34}{rgb}{0.34,0.34,0.34}
\definecolor{gray35}{rgb}{0.35,0.35,0.35}
\definecolor{gray36}{rgb}{0.36,0.36,0.36}
\definecolor{gray37}{rgb}{0.37,0.37,0.37}
\definecolor{gray38}{rgb}{0.38,0.38,0.38}
\definecolor{gray39}{rgb}{0.39,0.39,0.39}
\definecolor{gray3}{rgb}{0.03,0.03,0.03}
\definecolor{gray40}{rgb}{0.40,0.40,0.40}
\definecolor{gray41}{rgb}{0.41,0.41,0.41}
\definecolor{gray42}{rgb}{0.42,0.42,0.42}
\definecolor{gray43}{rgb}{0.43,0.43,0.43}
\definecolor{gray44}{rgb}{0.44,0.44,0.44}
\definecolor{gray45}{rgb}{0.45,0.45,0.45}
\definecolor{gray46}{rgb}{0.46,0.46,0.46}
\definecolor{gray47}{rgb}{0.47,0.47,0.47}
\definecolor{gray48}{rgb}{0.48,0.48,0.48}
\definecolor{gray49}{rgb}{0.49,0.49,0.49}
\definecolor{gray4}{rgb}{0.04,0.04,0.04}
\definecolor{gray50}{rgb}{0.50,0.50,0.50}
\definecolor{gray51}{rgb}{0.51,0.51,0.51}
\definecolor{gray52}{rgb}{0.52,0.52,0.52}
\definecolor{gray53}{rgb}{0.53,0.53,0.53}
\definecolor{gray54}{rgb}{0.54,0.54,0.54}
\definecolor{gray55}{rgb}{0.55,0.55,0.55}
\definecolor{gray56}{rgb}{0.56,0.56,0.56}
\definecolor{gray57}{rgb}{0.57,0.57,0.57}
\definecolor{gray58}{rgb}{0.58,0.58,0.58}
\definecolor{gray59}{rgb}{0.59,0.59,0.59}
\definecolor{gray5}{rgb}{0.05,0.05,0.05}
\definecolor{gray60}{rgb}{0.60,0.60,0.60}
\definecolor{gray61}{rgb}{0.61,0.61,0.61}
\definecolor{gray62}{rgb}{0.62,0.62,0.62}
\definecolor{gray63}{rgb}{0.63,0.63,0.63}
\definecolor{gray64}{rgb}{0.64,0.64,0.64}
\definecolor{gray65}{rgb}{0.65,0.65,0.65}
\definecolor{gray66}{rgb}{0.66,0.66,0.66}
\definecolor{gray67}{rgb}{0.67,0.67,0.67}
\definecolor{gray68}{rgb}{0.68,0.68,0.68}
\definecolor{gray69}{rgb}{0.69,0.69,0.69}
\definecolor{gray6}{rgb}{0.06,0.06,0.06}
\definecolor{gray70}{rgb}{0.70,0.70,0.70}
\definecolor{gray71}{rgb}{0.71,0.71,0.71}
\definecolor{gray72}{rgb}{0.72,0.72,0.72}
\definecolor{gray73}{rgb}{0.73,0.73,0.73}
\definecolor{gray74}{rgb}{0.74,0.74,0.74}
\definecolor{gray75}{rgb}{0.75,0.75,0.75}
\definecolor{gray76}{rgb}{0.76,0.76,0.76}
\definecolor{gray77}{rgb}{0.77,0.77,0.77}
\definecolor{gray78}{rgb}{0.78,0.78,0.78}
\definecolor{gray79}{rgb}{0.79,0.79,0.79}
\definecolor{gray7}{rgb}{0.07,0.07,0.07}
\definecolor{gray80}{rgb}{0.80,0.80,0.80}
\definecolor{gray81}{rgb}{0.81,0.81,0.81}
\definecolor{gray82}{rgb}{0.82,0.82,0.82}
\definecolor{gray83}{rgb}{0.83,0.83,0.83}
\definecolor{gray84}{rgb}{0.84,0.84,0.84}
\definecolor{gray85}{rgb}{0.85,0.85,0.85}
\definecolor{gray86}{rgb}{0.86,0.86,0.86}
\definecolor{gray87}{rgb}{0.87,0.87,0.87}
\definecolor{gray88}{rgb}{0.88,0.88,0.88}
\definecolor{gray89}{rgb}{0.89,0.89,0.89}
\definecolor{gray8}{rgb}{0.08,0.08,0.08}
\definecolor{gray90}{rgb}{0.90,0.90,0.90}
\definecolor{gray91}{rgb}{0.91,0.91,0.91}
\definecolor{gray92}{rgb}{0.92,0.92,0.92}
\definecolor{gray93}{rgb}{0.93,0.93,0.93}
\definecolor{gray94}{rgb}{0.94,0.94,0.94}
\definecolor{gray95}{rgb}{0.95,0.95,0.95}
\definecolor{gray96}{rgb}{0.96,0.96,0.96}
\definecolor{gray97}{rgb}{0.97,0.97,0.97}
\definecolor{gray98}{rgb}{0.98,0.98,0.98}
\definecolor{gray99}{rgb}{0.99,0.99,0.99}
\definecolor{gray9}{rgb}{0.09,0.09,0.09}
\definecolor{gray}{rgb}{0.75,0.75,0.75}
\definecolor{green1}{rgb}{0.00,1.00,0.00}
\definecolor{green2}{rgb}{0.00,0.93,0.00}
\definecolor{green3}{rgb}{0.00,0.80,0.00}
\definecolor{green4}{rgb}{0.00,0.55,0.00}
\definecolor{greenyellow}{rgb}{0.68,1.00,0.18}
\definecolor{green}{rgb}{0.00,1.00,0.00}
\definecolor{grey0}{rgb}{0.00,0.00,0.00}
\definecolor{grey100}{rgb}{1.00,1.00,1.00}
\definecolor{grey10}{rgb}{0.10,0.10,0.10}
\definecolor{grey11}{rgb}{0.11,0.11,0.11}
\definecolor{grey12}{rgb}{0.12,0.12,0.12}
\definecolor{grey13}{rgb}{0.13,0.13,0.13}
\definecolor{grey14}{rgb}{0.14,0.14,0.14}
\definecolor{grey15}{rgb}{0.15,0.15,0.15}
\definecolor{grey16}{rgb}{0.16,0.16,0.16}
\definecolor{grey17}{rgb}{0.17,0.17,0.17}
\definecolor{grey18}{rgb}{0.18,0.18,0.18}
\definecolor{grey19}{rgb}{0.19,0.19,0.19}
\definecolor{grey1}{rgb}{0.01,0.01,0.01}
\definecolor{grey20}{rgb}{0.20,0.20,0.20}
\definecolor{grey21}{rgb}{0.21,0.21,0.21}
\definecolor{grey22}{rgb}{0.22,0.22,0.22}
\definecolor{grey23}{rgb}{0.23,0.23,0.23}
\definecolor{grey24}{rgb}{0.24,0.24,0.24}
\definecolor{grey25}{rgb}{0.25,0.25,0.25}
\definecolor{grey26}{rgb}{0.26,0.26,0.26}
\definecolor{grey27}{rgb}{0.27,0.27,0.27}
\definecolor{grey28}{rgb}{0.28,0.28,0.28}
\definecolor{grey29}{rgb}{0.29,0.29,0.29}
\definecolor{grey2}{rgb}{0.02,0.02,0.02}
\definecolor{grey30}{rgb}{0.30,0.30,0.30}
\definecolor{grey31}{rgb}{0.31,0.31,0.31}
\definecolor{grey32}{rgb}{0.32,0.32,0.32}
\definecolor{grey33}{rgb}{0.33,0.33,0.33}
\definecolor{grey34}{rgb}{0.34,0.34,0.34}
\definecolor{grey35}{rgb}{0.35,0.35,0.35}
\definecolor{grey36}{rgb}{0.36,0.36,0.36}
\definecolor{grey37}{rgb}{0.37,0.37,0.37}
\definecolor{grey38}{rgb}{0.38,0.38,0.38}
\definecolor{grey39}{rgb}{0.39,0.39,0.39}
\definecolor{grey3}{rgb}{0.03,0.03,0.03}
\definecolor{grey40}{rgb}{0.40,0.40,0.40}
\definecolor{grey41}{rgb}{0.41,0.41,0.41}
\definecolor{grey42}{rgb}{0.42,0.42,0.42}
\definecolor{grey43}{rgb}{0.43,0.43,0.43}
\definecolor{grey44}{rgb}{0.44,0.44,0.44}
\definecolor{grey45}{rgb}{0.45,0.45,0.45}
\definecolor{grey46}{rgb}{0.46,0.46,0.46}
\definecolor{grey47}{rgb}{0.47,0.47,0.47}
\definecolor{grey48}{rgb}{0.48,0.48,0.48}
\definecolor{grey49}{rgb}{0.49,0.49,0.49}
\definecolor{grey4}{rgb}{0.04,0.04,0.04}
\definecolor{grey50}{rgb}{0.50,0.50,0.50}
\definecolor{grey51}{rgb}{0.51,0.51,0.51}
\definecolor{grey52}{rgb}{0.52,0.52,0.52}
\definecolor{grey53}{rgb}{0.53,0.53,0.53}
\definecolor{grey54}{rgb}{0.54,0.54,0.54}
\definecolor{grey55}{rgb}{0.55,0.55,0.55}
\definecolor{grey56}{rgb}{0.56,0.56,0.56}
\definecolor{grey57}{rgb}{0.57,0.57,0.57}
\definecolor{grey58}{rgb}{0.58,0.58,0.58}
\definecolor{grey59}{rgb}{0.59,0.59,0.59}
\definecolor{grey5}{rgb}{0.05,0.05,0.05}
\definecolor{grey60}{rgb}{0.60,0.60,0.60}
\definecolor{grey61}{rgb}{0.61,0.61,0.61}
\definecolor{grey62}{rgb}{0.62,0.62,0.62}
\definecolor{grey63}{rgb}{0.63,0.63,0.63}
\definecolor{grey64}{rgb}{0.64,0.64,0.64}
\definecolor{grey65}{rgb}{0.65,0.65,0.65}
\definecolor{grey66}{rgb}{0.66,0.66,0.66}
\definecolor{grey67}{rgb}{0.67,0.67,0.67}
\definecolor{grey68}{rgb}{0.68,0.68,0.68}
\definecolor{grey69}{rgb}{0.69,0.69,0.69}
\definecolor{grey6}{rgb}{0.06,0.06,0.06}
\definecolor{grey70}{rgb}{0.70,0.70,0.70}
\definecolor{grey71}{rgb}{0.71,0.71,0.71}
\definecolor{grey72}{rgb}{0.72,0.72,0.72}
\definecolor{grey73}{rgb}{0.73,0.73,0.73}
\definecolor{grey74}{rgb}{0.74,0.74,0.74}
\definecolor{grey75}{rgb}{0.75,0.75,0.75}
\definecolor{grey76}{rgb}{0.76,0.76,0.76}
\definecolor{grey77}{rgb}{0.77,0.77,0.77}
\definecolor{grey78}{rgb}{0.78,0.78,0.78}
\definecolor{grey79}{rgb}{0.79,0.79,0.79}
\definecolor{grey7}{rgb}{0.07,0.07,0.07}
\definecolor{grey80}{rgb}{0.80,0.80,0.80}
\definecolor{grey81}{rgb}{0.81,0.81,0.81}
\definecolor{grey82}{rgb}{0.82,0.82,0.82}
\definecolor{grey83}{rgb}{0.83,0.83,0.83}
\definecolor{grey84}{rgb}{0.84,0.84,0.84}
\definecolor{grey85}{rgb}{0.85,0.85,0.85}
\definecolor{grey86}{rgb}{0.86,0.86,0.86}
\definecolor{grey87}{rgb}{0.87,0.87,0.87}
\definecolor{grey88}{rgb}{0.88,0.88,0.88}
\definecolor{grey89}{rgb}{0.89,0.89,0.89}
\definecolor{grey8}{rgb}{0.08,0.08,0.08}
\definecolor{grey90}{rgb}{0.90,0.90,0.90}
\definecolor{grey91}{rgb}{0.91,0.91,0.91}
\definecolor{grey92}{rgb}{0.92,0.92,0.92}
\definecolor{grey93}{rgb}{0.93,0.93,0.93}
\definecolor{grey94}{rgb}{0.94,0.94,0.94}
\definecolor{grey95}{rgb}{0.95,0.95,0.95}
\definecolor{grey96}{rgb}{0.96,0.96,0.96}
\definecolor{grey97}{rgb}{0.97,0.97,0.97}
\definecolor{grey98}{rgb}{0.98,0.98,0.98}
\definecolor{grey99}{rgb}{0.99,0.99,0.99}
\definecolor{grey9}{rgb}{0.09,0.09,0.09}
\definecolor{grey}{rgb}{0.75,0.75,0.75}
\definecolor{honeydew1}{rgb}{0.94,1.00,0.94}
\definecolor{honeydew2}{rgb}{0.88,0.93,0.88}
\definecolor{honeydew3}{rgb}{0.76,0.80,0.76}
\definecolor{honeydew4}{rgb}{0.51,0.55,0.51}
\definecolor{honeydew}{rgb}{0.94,1.00,0.94}
\definecolor{hotpink}{rgb}{1.00,0.41,0.71}
\definecolor{indianred}{rgb}{0.80,0.36,0.36}
\definecolor{ivory1}{rgb}{1.00,1.00,0.94}
\definecolor{ivory2}{rgb}{0.93,0.93,0.88}
\definecolor{ivory3}{rgb}{0.80,0.80,0.76}
\definecolor{ivory4}{rgb}{0.55,0.55,0.51}
\definecolor{ivory}{rgb}{1.00,1.00,0.94}
\definecolor{khaki1}{rgb}{1.00,0.96,0.56}
\definecolor{khaki2}{rgb}{0.93,0.90,0.52}
\definecolor{khaki3}{rgb}{0.80,0.78,0.45}
\definecolor{khaki4}{rgb}{0.55,0.53,0.31}
\definecolor{khaki}{rgb}{0.94,0.90,0.55}
\definecolor{lavenderblush}{rgb}{1.00,0.94,0.96}
\definecolor{lavender}{rgb}{0.90,0.90,0.98}
\definecolor{lawngreen}{rgb}{0.49,0.99,0.00}
\definecolor{lemonchiffon}{rgb}{1.00,0.98,0.80}
\definecolor{lightblue}{rgb}{0.68,0.85,0.90}
\definecolor{lightcoral}{rgb}{0.94,0.50,0.50}
\definecolor{lightcyan}{rgb}{0.88,1.00,1.00}
\definecolor{lightgoldenrod}{rgb}{0.93,0.87,0.51}
\definecolor{lightgoldenrod}{rgb}{0.98,0.98,0.82}
\definecolor{lightgray}{rgb}{0.83,0.83,0.83}
\definecolor{lightgreen}{rgb}{0.56,0.93,0.56}
\definecolor{lightgrey}{rgb}{0.83,0.83,0.83}
\definecolor{lightpink}{rgb}{1.00,0.71,0.76}
\definecolor{lightsalmon}{rgb}{1.00,0.63,0.48}
\definecolor{lightsea}{rgb}{0.13,0.70,0.67}
\definecolor{lightsky}{rgb}{0.53,0.81,0.98}
\definecolor{lightslate}{rgb}{0.47,0.53,0.60}
\definecolor{lightslate}{rgb}{0.47,0.53,0.60}
\definecolor{lightslate}{rgb}{0.52,0.44,1.00}
\definecolor{lightsteel}{rgb}{0.69,0.77,0.87}
\definecolor{lightyellow}{rgb}{1.00,1.00,0.88}
\definecolor{limegreen}{rgb}{0.20,0.80,0.20}
\definecolor{linen}{rgb}{0.98,0.94,0.90}
\definecolor{magenta1}{rgb}{1.00,0.00,1.00}
\definecolor{magenta2}{rgb}{0.93,0.00,0.93}
\definecolor{magenta3}{rgb}{0.80,0.00,0.80}
\definecolor{magenta4}{rgb}{0.55,0.00,0.55}
\definecolor{magenta}{rgb}{1.00,0.00,1.00}
\definecolor{maroon1}{rgb}{1.00,0.20,0.70}
\definecolor{maroon2}{rgb}{0.93,0.19,0.65}
\definecolor{maroon3}{rgb}{0.80,0.16,0.56}
\definecolor{maroon4}{rgb}{0.55,0.11,0.38}
\definecolor{maroon}{rgb}{0.69,0.19,0.38}
\definecolor{mediumaquamarine}{rgb}{0.40,0.80,0.67}
\definecolor{mediumblue}{rgb}{0.00,0.00,0.80}
\definecolor{mediumorchid}{rgb}{0.73,0.33,0.83}
\definecolor{mediumpurple}{rgb}{0.58,0.44,0.86}
\definecolor{mediumsea}{rgb}{0.24,0.70,0.44}
\definecolor{mediumslate}{rgb}{0.48,0.41,0.93}
\definecolor{mediumspring}{rgb}{0.00,0.98,0.60}
\definecolor{mediumturquoise}{rgb}{0.28,0.82,0.80}
\definecolor{mediumviolet}{rgb}{0.78,0.08,0.52}
\definecolor{midnightblue}{rgb}{0.10,0.10,0.44}
\definecolor{mintcream}{rgb}{0.96,1.00,0.98}
\definecolor{mistyrose}{rgb}{1.00,0.89,0.88}
\definecolor{moccasin}{rgb}{1.00,0.89,0.71}
\definecolor{navajowhite}{rgb}{1.00,0.87,0.68}
\definecolor{navyblue}{rgb}{0.00,0.00,0.50}
\definecolor{navy}{rgb}{0.00,0.00,0.50}
\definecolor{oldlace}{rgb}{0.99,0.96,0.90}
\definecolor{olivedrab}{rgb}{0.42,0.56,0.14}
\definecolor{orange1}{rgb}{1.00,0.65,0.00}
\definecolor{orange2}{rgb}{0.93,0.60,0.00}
\definecolor{orange3}{rgb}{0.80,0.52,0.00}
\definecolor{orange4}{rgb}{0.55,0.35,0.00}
\definecolor{orangered}{rgb}{1.00,0.27,0.00}
\definecolor{orange}{rgb}{1.00,0.65,0.00}
\definecolor{orchid1}{rgb}{1.00,0.51,0.98}
\definecolor{orchid2}{rgb}{0.93,0.48,0.91}
\definecolor{orchid3}{rgb}{0.80,0.41,0.79}
\definecolor{orchid4}{rgb}{0.55,0.28,0.54}
\definecolor{orchid}{rgb}{0.85,0.44,0.84}
\definecolor{palegoldenrod}{rgb}{0.93,0.91,0.67}
\definecolor{palegreen}{rgb}{0.60,0.98,0.60}
\definecolor{paleturquoise}{rgb}{0.69,0.93,0.93}
\definecolor{paleviolet}{rgb}{0.86,0.44,0.58}
\definecolor{papayawhip}{rgb}{1.00,0.94,0.84}
\definecolor{peachpuff}{rgb}{1.00,0.85,0.73}
\definecolor{peru}{rgb}{0.80,0.52,0.25}
\definecolor{pink1}{rgb}{1.00,0.71,0.77}
\definecolor{pink2}{rgb}{0.93,0.66,0.72}
\definecolor{pink3}{rgb}{0.80,0.57,0.62}
\definecolor{pink4}{rgb}{0.55,0.39,0.42}
\definecolor{pink}{rgb}{1.00,0.75,0.80}
\definecolor{plum1}{rgb}{1.00,0.73,1.00}
\definecolor{plum2}{rgb}{0.93,0.68,0.93}
\definecolor{plum3}{rgb}{0.80,0.59,0.80}
\definecolor{plum4}{rgb}{0.55,0.40,0.55}
\definecolor{plum}{rgb}{0.87,0.63,0.87}
\definecolor{powderblue}{rgb}{0.69,0.88,0.90}
\definecolor{purple1}{rgb}{0.61,0.19,1.00}
\definecolor{purple2}{rgb}{0.57,0.17,0.93}
\definecolor{purple3}{rgb}{0.49,0.15,0.80}
\definecolor{purple4}{rgb}{0.33,0.10,0.55}
\definecolor{purple}{rgb}{0.63,0.13,0.94}
\definecolor{red1}{rgb}{1.00,0.00,0.00}
\definecolor{red2}{rgb}{0.93,0.00,0.00}
\definecolor{red3}{rgb}{0.80,0.00,0.00}
\definecolor{red4}{rgb}{0.55,0.00,0.00}
\definecolor{red}{rgb}{1.00,0.00,0.00}
\definecolor{rosybrown}{rgb}{0.74,0.56,0.56}
\definecolor{royalblue}{rgb}{0.25,0.41,0.88}
\definecolor{saddlebrown}{rgb}{0.55,0.27,0.07}
\definecolor{salmon1}{rgb}{1.00,0.55,0.41}
\definecolor{salmon2}{rgb}{0.93,0.51,0.38}
\definecolor{salmon3}{rgb}{0.80,0.44,0.33}
\definecolor{salmon4}{rgb}{0.55,0.30,0.22}
\definecolor{salmon}{rgb}{0.98,0.50,0.45}
\definecolor{sandybrown}{rgb}{0.96,0.64,0.38}
\definecolor{seagreen}{rgb}{0.18,0.55,0.34}
\definecolor{seashell1}{rgb}{1.00,0.96,0.93}
\definecolor{seashell2}{rgb}{0.93,0.90,0.87}
\definecolor{seashell3}{rgb}{0.80,0.77,0.75}
\definecolor{seashell4}{rgb}{0.55,0.53,0.51}
\definecolor{seashell}{rgb}{1.00,0.96,0.93}
\definecolor{sienna1}{rgb}{1.00,0.51,0.28}
\definecolor{sienna2}{rgb}{0.93,0.47,0.26}
\definecolor{sienna3}{rgb}{0.80,0.41,0.22}
\definecolor{sienna4}{rgb}{0.55,0.28,0.15}
\definecolor{sienna}{rgb}{0.63,0.32,0.18}
\definecolor{skyblue}{rgb}{0.53,0.81,0.92}
\definecolor{slateblue}{rgb}{0.42,0.35,0.80}
\definecolor{slategray}{rgb}{0.44,0.50,0.56}
\definecolor{slategrey}{rgb}{0.44,0.50,0.56}
\definecolor{snow1}{rgb}{1.00,0.98,0.98}
\definecolor{snow2}{rgb}{0.93,0.91,0.91}
\definecolor{snow3}{rgb}{0.80,0.79,0.79}
\definecolor{snow4}{rgb}{0.55,0.54,0.54}
\definecolor{snow}{rgb}{1.00,0.98,0.98}
\definecolor{springgreen}{rgb}{0.00,1.00,0.50}
\definecolor{steelblue}{rgb}{0.27,0.51,0.71}
\definecolor{tan1}{rgb}{1.00,0.65,0.31}
\definecolor{tan2}{rgb}{0.93,0.60,0.29}
\definecolor{tan3}{rgb}{0.80,0.52,0.25}
\definecolor{tan4}{rgb}{0.55,0.35,0.17}
\definecolor{tan}{rgb}{0.82,0.71,0.55}
\definecolor{thistle1}{rgb}{1.00,0.88,1.00}
\definecolor{thistle2}{rgb}{0.93,0.82,0.93}
\definecolor{thistle3}{rgb}{0.80,0.71,0.80}
\definecolor{thistle4}{rgb}{0.55,0.48,0.55}
\definecolor{thistle}{rgb}{0.85,0.75,0.85}
\definecolor{tomato1}{rgb}{1.00,0.39,0.28}
\definecolor{tomato2}{rgb}{0.93,0.36,0.26}
\definecolor{tomato3}{rgb}{0.80,0.31,0.22}
\definecolor{tomato4}{rgb}{0.55,0.21,0.15}
\definecolor{tomato}{rgb}{1.00,0.39,0.28}
\definecolor{turquoise1}{rgb}{0.00,0.96,1.00}
\definecolor{turquoise2}{rgb}{0.00,0.90,0.93}
\definecolor{turquoise3}{rgb}{0.00,0.77,0.80}
\definecolor{turquoise4}{rgb}{0.00,0.53,0.55}
\definecolor{turquoise}{rgb}{0.25,0.88,0.82}
\definecolor{violetred}{rgb}{0.82,0.13,0.56}
\definecolor{violet}{rgb}{0.93,0.51,0.93}
\definecolor{wheat1}{rgb}{1.00,0.91,0.73}
\definecolor{wheat2}{rgb}{0.93,0.85,0.68}
\definecolor{wheat3}{rgb}{0.80,0.73,0.59}
\definecolor{wheat4}{rgb}{0.55,0.49,0.40}
\definecolor{wheat}{rgb}{0.96,0.87,0.70}
\definecolor{whitesmoke}{rgb}{0.96,0.96,0.96}
\definecolor{white}{rgb}{1.00,1.00,1.00}
\definecolor{yellow1}{rgb}{1.00,1.00,0.00}
\definecolor{yellow2}{rgb}{0.93,0.93,0.00}
\definecolor{yellow3}{rgb}{0.80,0.80,0.00}
\definecolor{yellow4}{rgb}{0.55,0.55,0.00}
\definecolor{yellowgreen}{rgb}{0.60,0.80,0.20}
\definecolor{yellow}{rgb}{1.00,1.00,0.00}
\usepackage{multicol}

\title{Mass-Galaxy offsets in Abell 3827, 2218 and 1689: intrinsic properties or line-of-sight substructures?}

\author[I. Mohammed et al.] {Irshad Mohammed\thanks{irshad@physik.uzh.ch},$^1$
Jori Liesenborgs,$^2$ Prasenjit Saha$^1$ and Liliya L. R. Williams$^3$\\
$^1${Institute for Theoretical Physics, University of Z$\ddot{u}$rich, 8057 Z$\ddot{u}$rich Switzerland}\\
$^2${Expertisecentrum voor Digitale Media, Universiteit Hasselt, Wetenschapspark 2, B-3590, Diepenbeek, Belgium}\\
$^3${School of Physics \& Astronomy, University of Minnesota, 116 Church Street SE, Minneapolis, MN 55455, USA}
}

\begin{document}

\maketitle

\begin{abstract}
We have made mass maps of three strong-lensing clusters, Abell 3827,
Abell 2218 and Abell 1689, in order to test for mass-light offsets.
The technique used is GRALE, which enables lens reconstruction with
minimal assumptions, and specifically with no information about the cluster
light being given.  In the first two of these clusters, we find local
mass peaks in the central regions that are displaced from the nearby
galaxies by a few to several kpc.  These offsets {\em could\/} be due
to line of sight structure unrelated to the clusters, but that is very
unlikely, given the typical levels of chance line-of-sight
coincidences in $\Lambda CDM$ simulations --- for Abell 3827 and Abell
2218 the offsets appear to be intrinsic.  In the case of Abell 1689,
we see no significant offsets in the central region, but we do detect
a possible line of sight structure: it appears only when sources at
$z\ga 3$ are used for reconstructing the mass.  We discuss possible
origins of the mass-galaxy offsets in Abell 3827 and Abell 2218: these
include pure gravitational effects like dynamical friction, but also
non-standard mechanisms like self-interacting dark-matter.
\end{abstract}


\begin{multicols}{2}

%%%%%%%%%%%%%%%%%%%%%%%%%%%%%%%%%%%%%%%%%%%%%%%%%%%%%%%%%%%%%%%%%%%%%%%%%%
\section{Introduction}

Our current understanding of the universe and its dynamics indicates
that its major components are dark: cold dark-matter (CDM) and the
so-called ``dark-energy''.  Unlike baryons, dark-matter
interacts only gravitationally and provides the deep potential wells
which are followed by the baryons. The baryons form clumps at these
potential wells and cool down to form stars.  The standard
$\Lambda$CDM model explains a range of observed processes pretty well,
from the angular power spectrum of the cosmic microwave background \citep{2013arXiv1303.5075P} to
the baryonic acoustic oscillations \citep{2013MNRAS.433.1202S} in the large scale structure and
the number counts of clusters.  However, the intrinsic properties and
behaviour of dark-matter and dark-energy remain an open problem in
cosmology.

In the picture of hierarchical structure formation in $\Lambda$CDM
model, galaxy-clusters are the most recently formed structures that
are gravitationally bound. They are cosmic laboratories to test
the laws of gravity, structure formations and the interaction of
different species of particles. A galaxy cluster contains lots of
galaxies --- tens to thousands, hot intra-cluster plasma visible in
X-rays, a variety of of relativistic particles and finally dark-matter
which dominates its mass budget.  Measuring the mass of the galaxy-cluster 
is an essential aspect of using the cluster to study many other things.
There are several physical processes that enable one to measure
the mass: the kinematics of cluster galaxies
\citep{2013ApJ...772...47S}, the hydrodynamics of hot gas emitting
X-rays \citep{2009ApJ...692.1033V}, and gravitational lensing.  Lensing
is particular interesting, because it relies only on gravity and does
not itself require any luminous objects in the cluster being studied.
One of the questions that lensing can address is how well the luminous matter
traces the distribution of total mass. Deviations, or lack thereof, from the
mass-follows-light hypothesis will provide important information about the 
physical processes going in within clusters. The first lensing-based detection 
of deviations from mass-follows-light goes back to the late 1990's 
\citep{1998AJ....116.1541A} but the observation that generated a wide interest 
in these deviations was that of the Bullet Cluster \citep{2006ApJ...648L.109C}, 
which showed unambiguously that dark matter is quite collisionless compared to 
the gas phase baryonic matter \citep{2008ApJ...679.1173R}. While the properties 
of dark matter are probably not the only reason for deviations from mass-follows-light
in galaxy clusters, dark matter self-interaction cross-section and how to optimally
extract it from observations is an exciting avenue of research 
\citep{2013arXiv1310.1731H,2013MNRAS.433.1517H}.

This work uses strong gravitational lensing to look for deviations from 
mass-follows-light, i.e. it explores the correspondence on the sky between the
dark-matter peaks with the galaxies in the central parts of three
galaxy clusters, Abell 3827, 2218 and 1689. These clusters are very
different from each other in morphology and redshift. 
%A manifestation about nature/properties of dark-matter is also proposed.
As we discuss in Section~\ref{disc}, some deviations we find may be due to the 
non-standard properties of dark matter, but others could be the result of 
superimposed substructure, or hydrodynamics within the cluster. 

We use GRALE \citep{2006MNRAS.367.1209L,2007MNRAS.380.1729L}, a
strong-gravitational lensing tool to reconstruct the mass map of the
clusters.  There is no overall parametric form for the mass
distribution, but rather an adaptive grid.  Other than the redshift,
no information about the cluster is required as input, not even its
location or morphology.  This makes GRALE well-suited to reconstruction of
mass maps before comparison with light.

%%%%%%%%%%%%%%%%%%%%%%%%%%%%%%%%%%%%%%%%%%%%%%%%%%%%%%%%%%%%%%%%%%%%%%%%%%

\section{The lens-reconstruction technique}

GRALE has been applied to other strong-lensing clusters
\citep{2008MNRAS.389..415L,2009MNRAS.397..341L}
and compared with other techniques
\citep{2010MNRAS.408.1916Z,2011MNRAS.413.1753Z}, so here we just give
a general description and then some tests.

\subsection{Grale}

The data given to GRALE consist of the identified multiple-image
systems and their redshifts, along with possible regions where
additional images are guessed to be likely.  No information about the
light from the lens is given.  The mass maps in GRALE are free-form,
being made up of a superposition of many components.  In the present
work, each component is taken as a Plummer lens, that is, the usual
Plummer sphere
\begin{equation}
\rho = \frac{3M}{4\pi} \frac{a^2}{(r^2+a^2)^{5/2}}
\end{equation}
projected to two dimensions.  Other choices of lens component, such
as square tiles, are also possible.

Any mass distribution in GRALE is assigned a fitness with respect to the
given data.  The fitness has two components, as follows.

\begin{enumerate}
\item For a given mass map, the input images are ray-traced back to
  the source, using the lens equation.  The more nearly these
  back-projected images coincide for any multiple-image system, the
  fitter the mass map.  If the fitness measure were simply the
  source-plane distance between the back-projected images, that would
  favour extreme magnification (tiny sources); accordingly, the fitnes
  measure is scaled to the source size.
\item There could be further places in the image plane that, when
  ray-traced back to the source, coincide with the sources
  corresponding to the observed images.  These correspond to extra
  images, and would be favoured by the above fitness measure.  There
  may indeed be undiscovered extra images in certain regions, but in
  most of the image plane, extra images can be ruled out with high
  confidence.  The area of no images present is referred to in GRALE
  as the null space.  For each image system, the user specifies a null
  space, which is simply the image plane with the images themselves
  cut out, and (optionally) further cutouts where incipient images
  could potentially be present.  Images in the null space lead to a
  fitness penalty for the mass map.
\end{enumerate}

It is possible to have other components to the fitness, such as time
delays for quasar source \citep{2009MNRAS.397..341L}, but the present
work uses these two.  The null space, item (ii) above, is a unique
aspect of GRALE.  There are other techniques that allow the mass
distribution to be very general in form, as with GRALE, but they make
additional assumptions in order to suppress extra images, such as constraining
local density gradients \citep{2006ApJ...652L...5S} or applying smooth
interpolation schemes \cite{2008ApJ...681..814C}.  Only GRALE
incorporates the absence of images as useful data.

The computational part of GRALE is optimizing the fitness function for
the given data, using a genetic algorithm.  The basic idea, inspired
by Darwinian evolution, is to generate a population of trial
solutions. A fitness measure is assigned to each trial solution and
then these solutions are combined, cloned and mutated to get the next
generation of populations supported by a better fitness function.
Genetic algorithms have long been used in astrophysics for hard
optimization problems \citep[for a somewhat old but readable review,
  see][]{1995ApJS..101..309C}.  They tend to be computationally
expensive, but are often effective on otherwise intractable problems.
GRALE uses a multi-objective genetic algorithm, meaning that the
different components of the fitness function are compared
individually, not just combined into a single function.  Only the
fitness ranking matters in genetic algorithms, not the actual values
of the fitness.  In terms of likelihoods and posterior probabilities,
models with better fitness are considered more probable, that is, the
fitness components are monotonic in the posterior probability, but
there is no known or assumed functional relation between likelihood
and fitness.

The locations and masses of the Plummer components are chosen by the
genetic algorithm.  The algorithm also adapts the number of Plummers,
but an allowed range is specified by the user.  That is, the user
specifies the level of substructure.  For the GRALE fitness measure,
lower is better, and it decreases as we increase the resolution of the
map.  This is quite intuitive as more Plummer spheres naturally result
in a better fit.  So the overall criterion should be somehow a
function of the GRALE fitness measure and the number of Plummers.  We
are not aware of any theoretical argument that yields the appropriate
criterion, but after some experimentation we found one that works
reasonably well in test cases.  This is an `unfitness' or
\begin{equation} \label{eq:badness}
\hbox{badness} = \ln\left(\hbox{GRALE fitness} \times
                     \sqrt{\hbox{number of components}}\,\right) .
\end{equation}
If we think of the GRALE fitness measure as a mismatch distance, and
the number of Plummers as the inverse resolution length, the badness
criterion appears natural.

To choose the number of Plummer components, we adopted the following
procedure.  First, we have GRALE reconstruct the lens with a
comparatively low number of Plummers.  Then we let GRALE improve the
fit with progressively more Plummers, allowing more substructure to be
introduced.  After that, we let GRALE continue to adapt the fit with
progressively fewer Plummers.  The mass distribution with the minimum
badness \eqref{eq:badness} is taken as the result.

\def\apjs{APJS}
%%%%%%%%%%%%%%%%%%%%%%%%%%%%%%%%%%%%%%%%%%%%%%%%%%%%%%%%%%%%%%%%%%%%%%%%%%

We now report on two simulated lenses, which we generated and then
reconstructed with GRALE, in order to check the pipeline and calibrate
the error estimates.  

\subsection{A simple lens}

\begin{figure}
  \includegraphics[width=0.24\hsize]{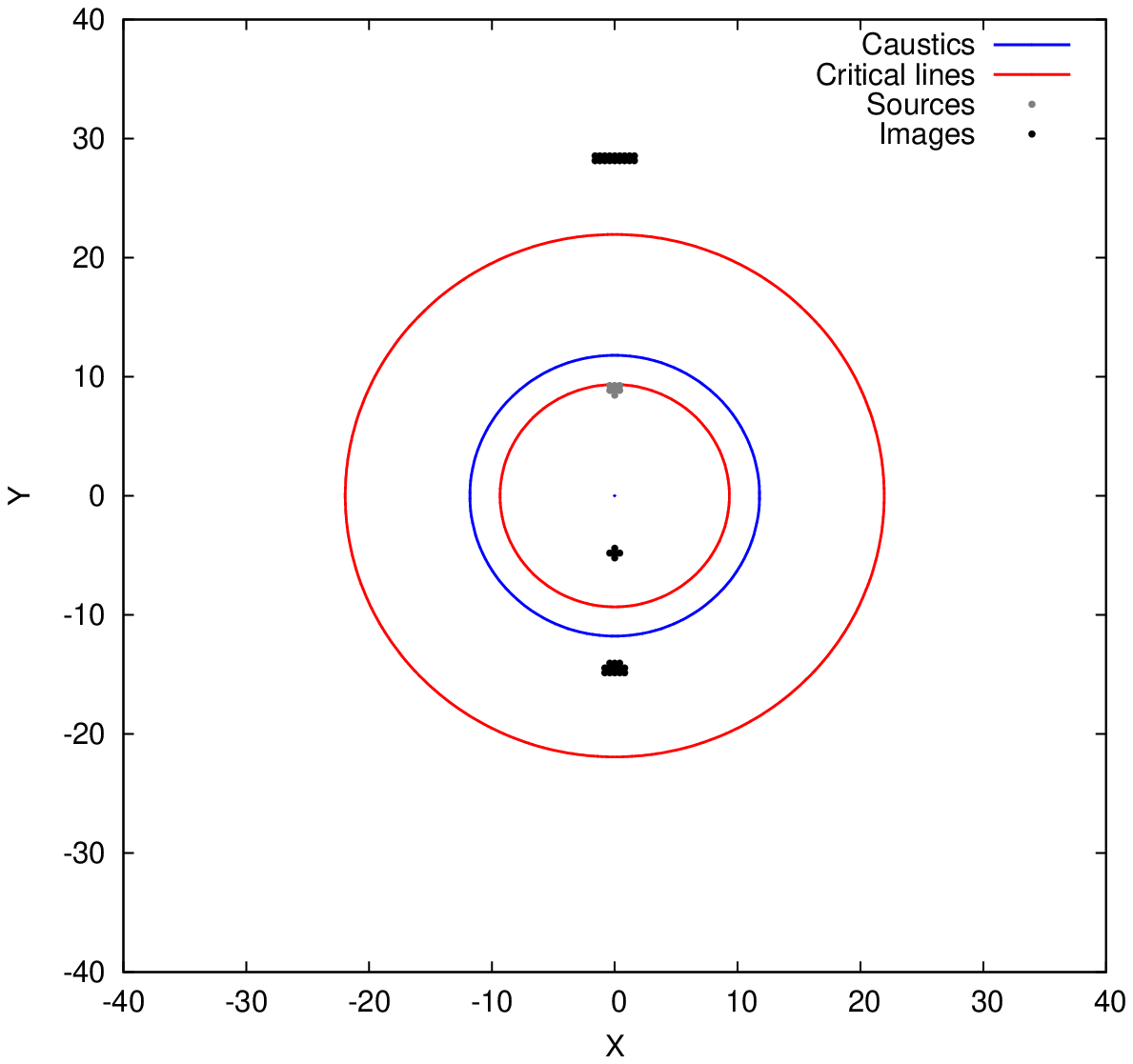}
  \hfil
  \includegraphics[width=0.3\hsize]{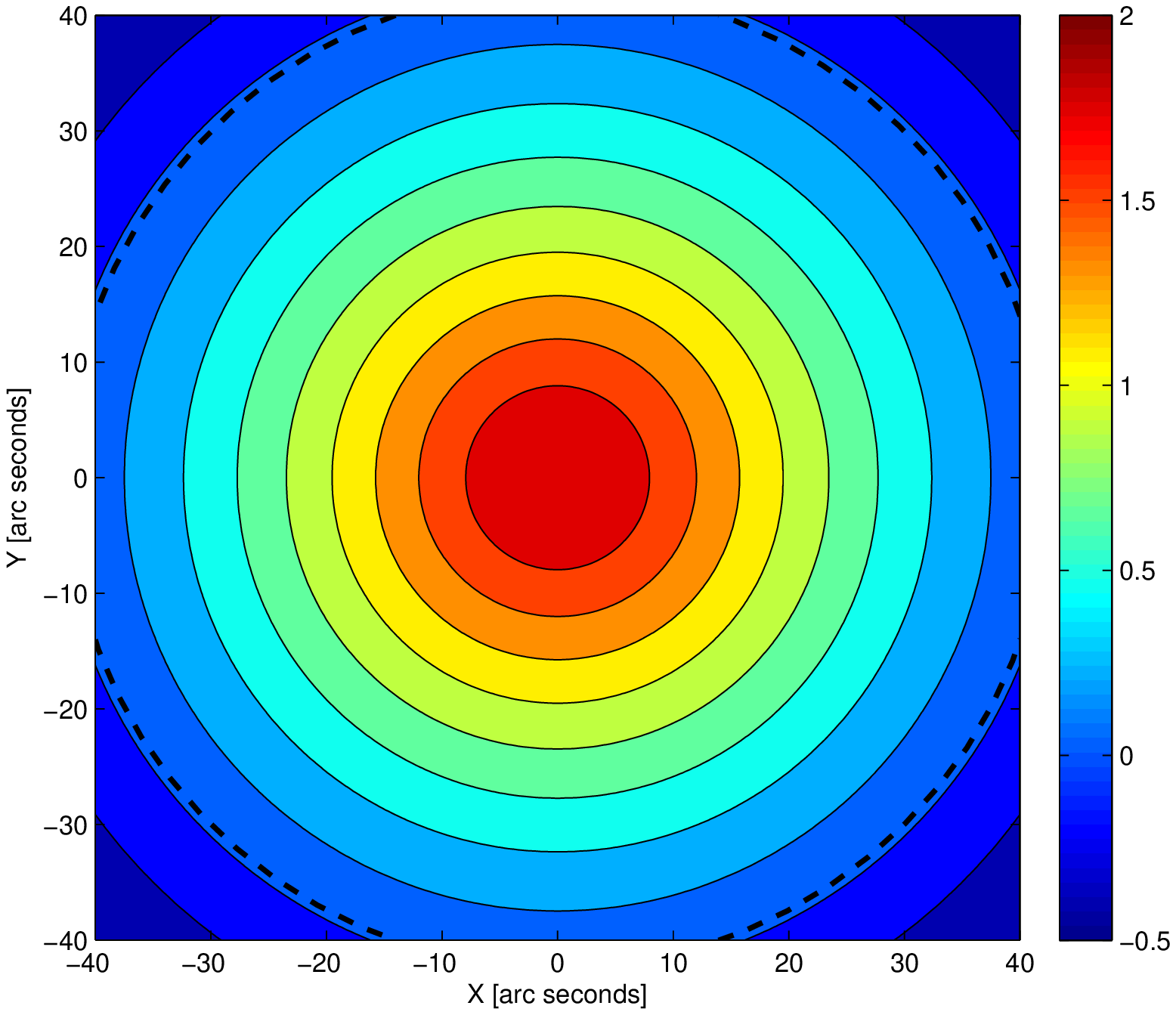}
  \hfil
  \includegraphics[width=0.24\hsize]{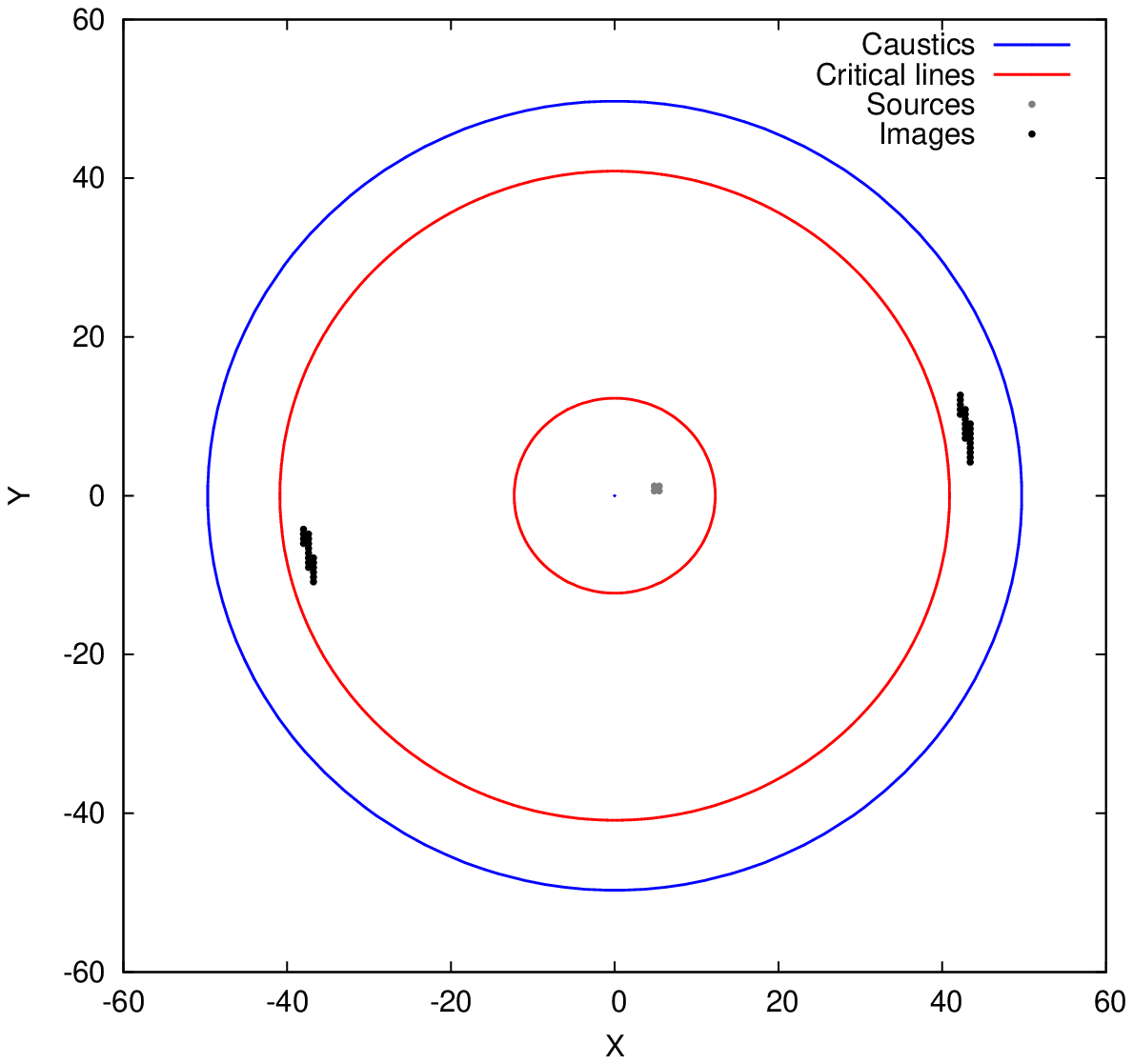}\\
  \includegraphics[width=0.24\hsize]{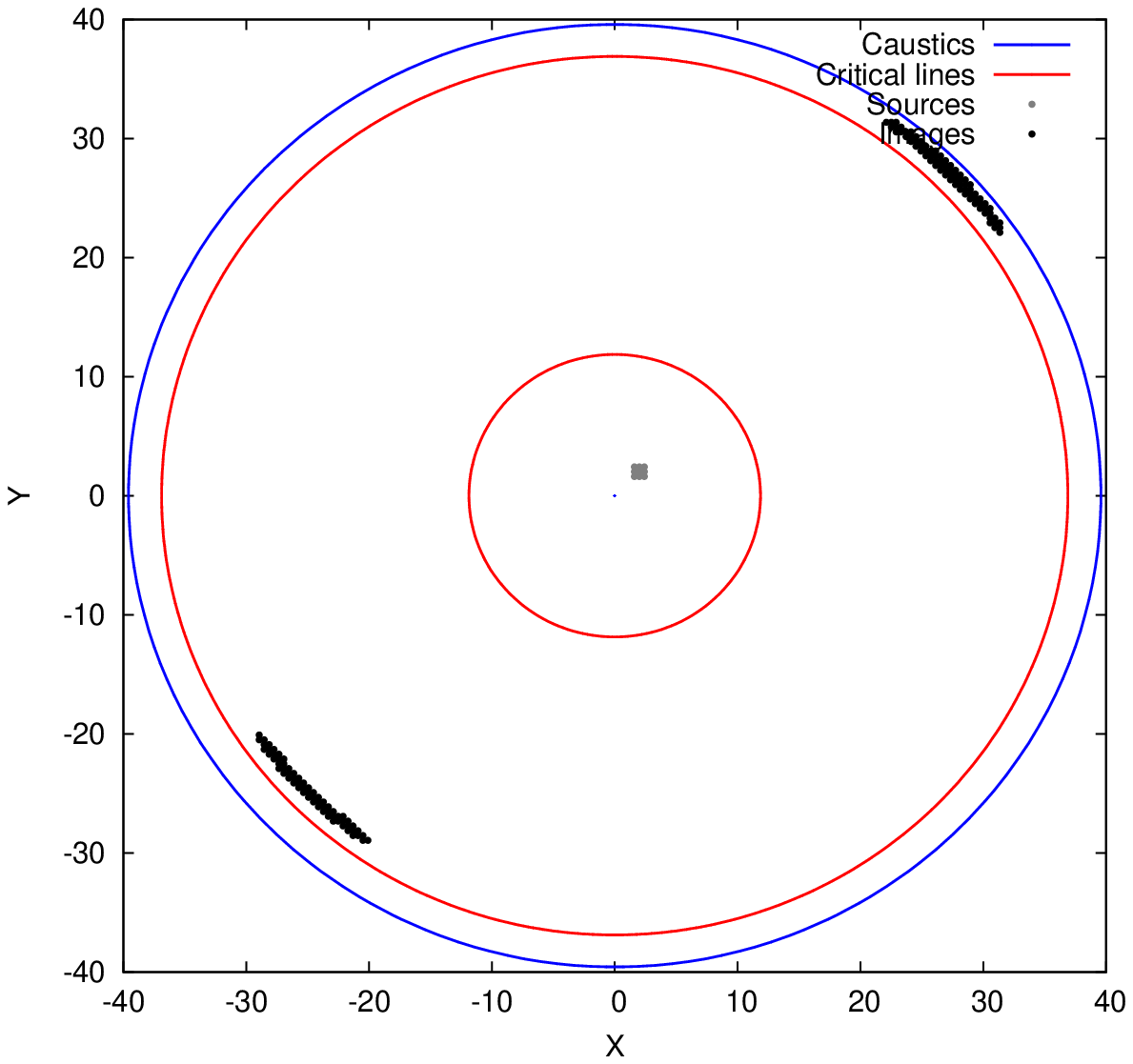}
  \includegraphics[width=0.24\hsize]{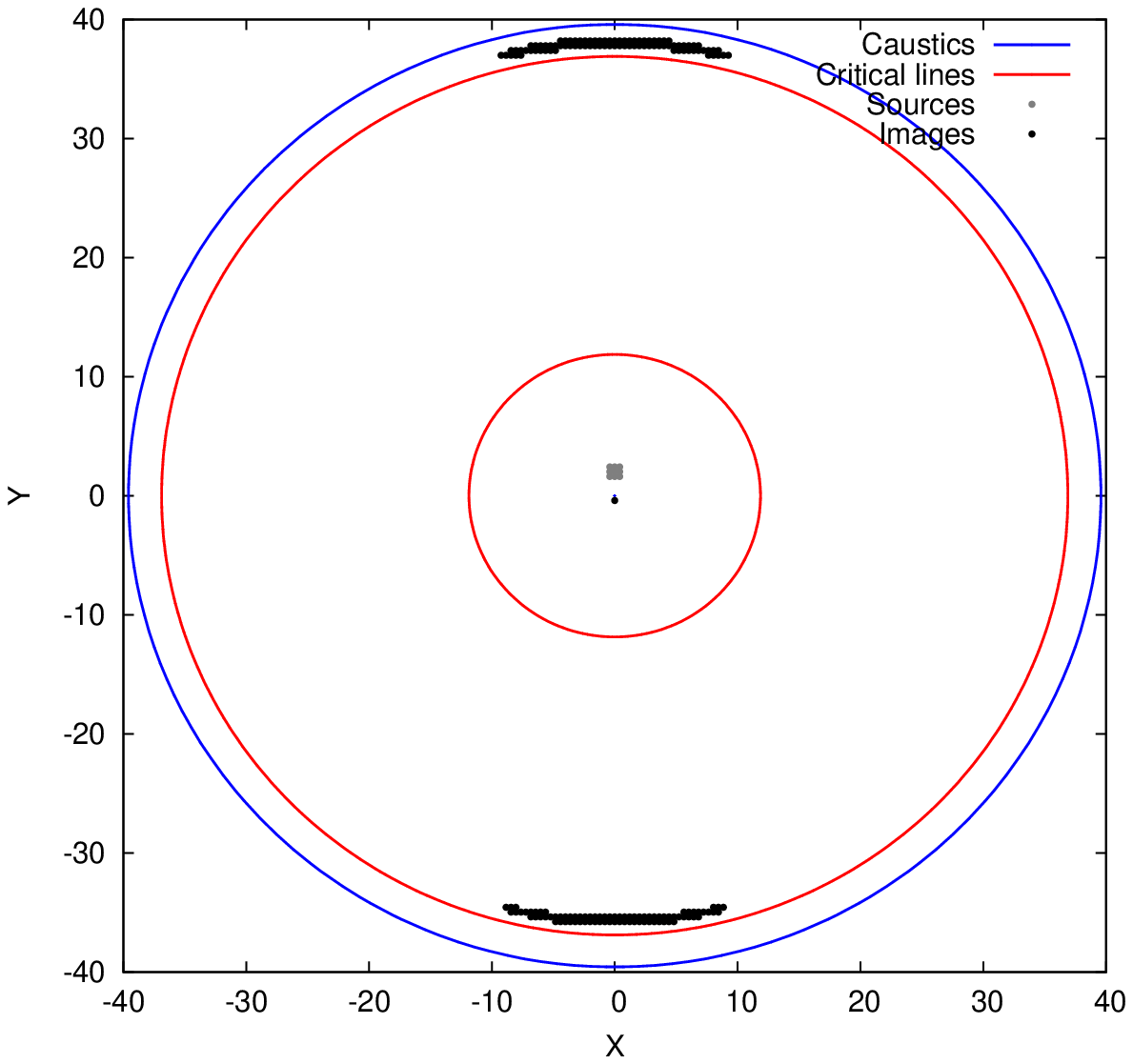}
  \includegraphics[width=0.24\hsize]{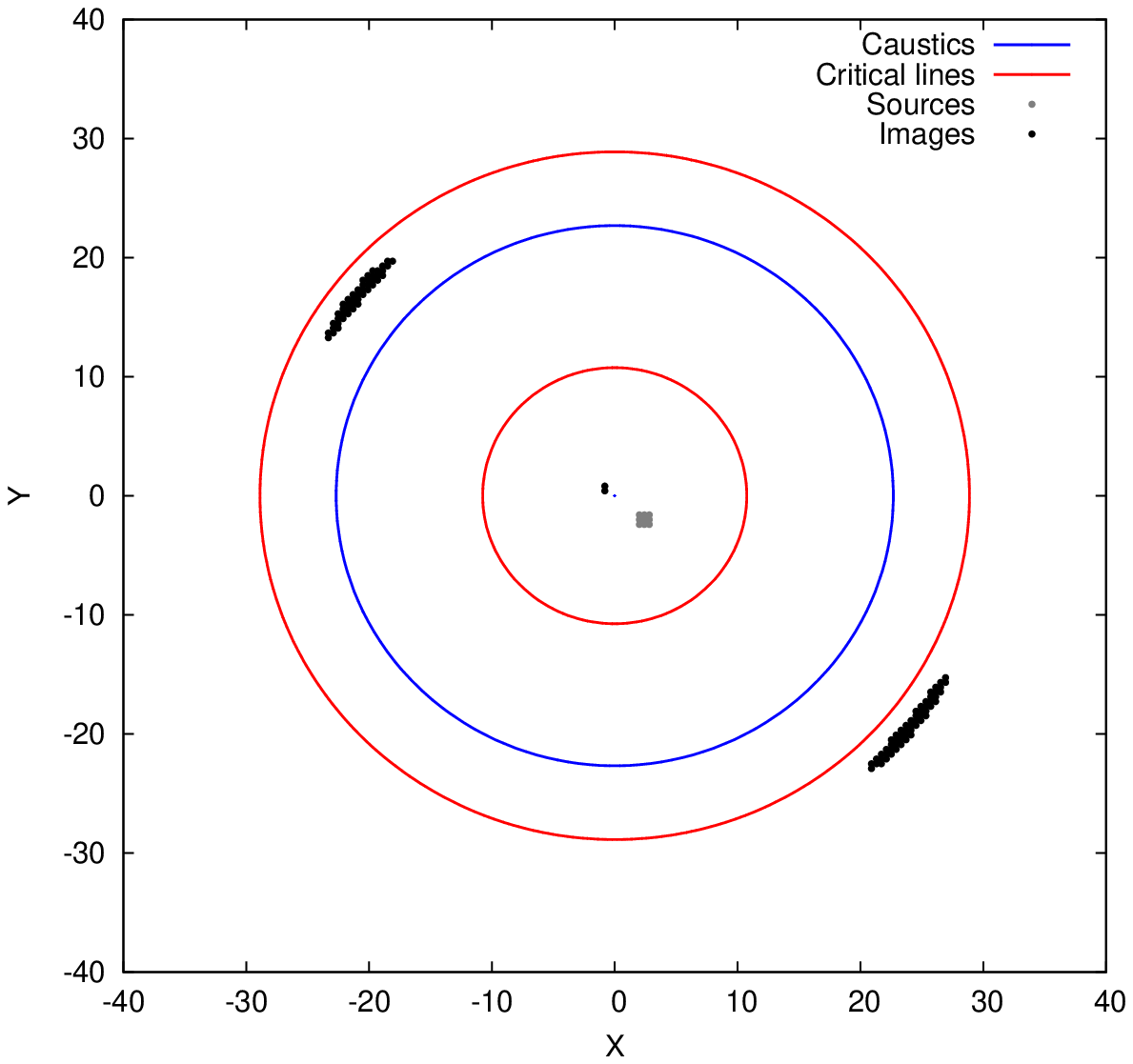}
  \includegraphics[width=0.24\hsize]{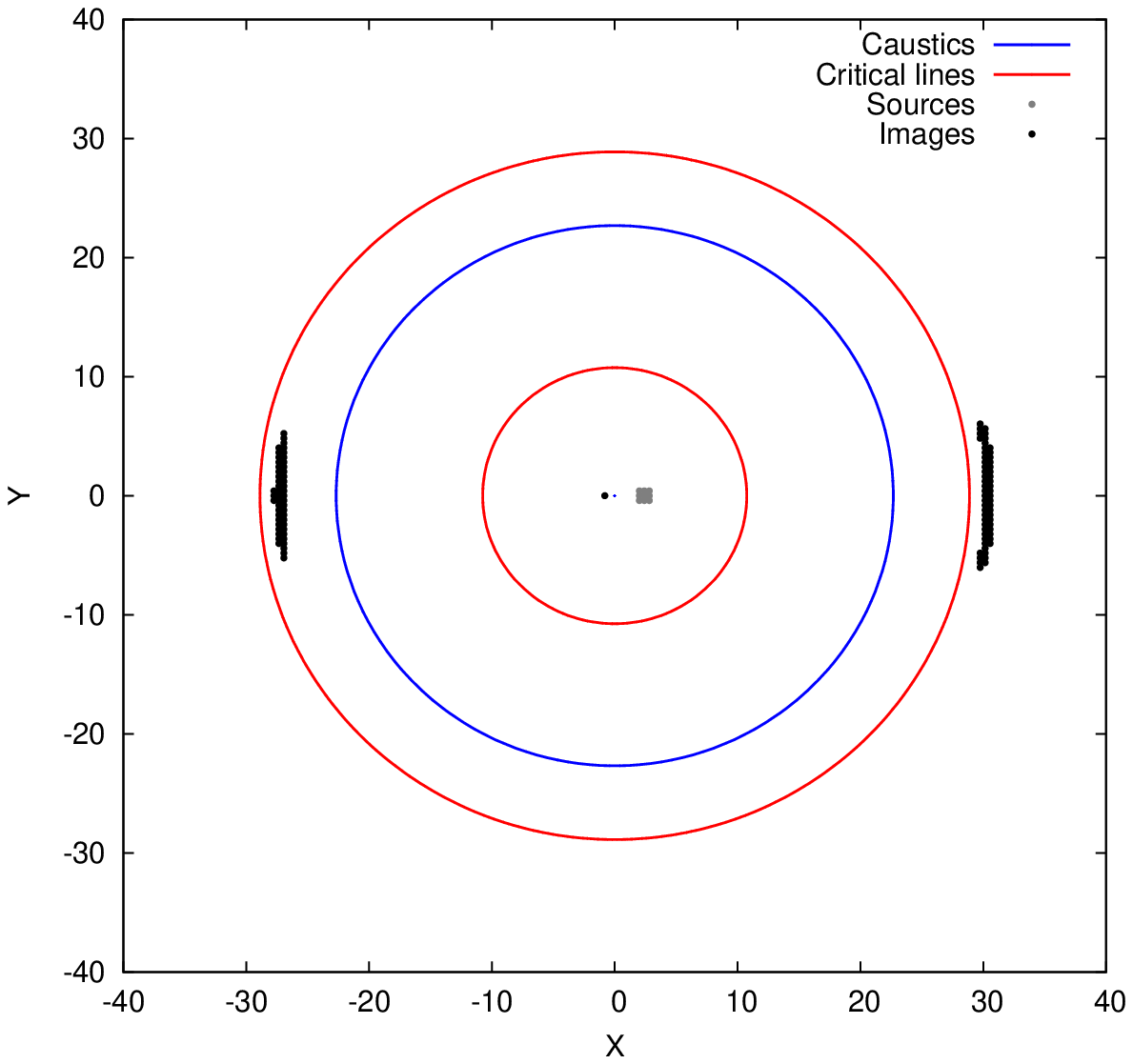}
\caption{\label{fig:plummerin} A circularly symmetric synthetic lens
  (centre top panel) and six image systems from sources at different
  redshifts. Sources are in grey, caustics are in blue, critical
  curves are in red. The contour lines in the synthetic lens are those
  of constant surface mass density; the color scale is in units of
  log~(kg~m$^{-2}$). The same scale is used in all figures in this
  paper. For reference, $\Sigma_{\rm crit}$ for $z_l=0.1$ and $z_s=0.2$
  in a standard $\Lambda$CDM cosmology is $18.7$ kg~m$^{-2}$.}
\end{figure}

\begin{figure}
  \includegraphics[width=0.3\hsize]{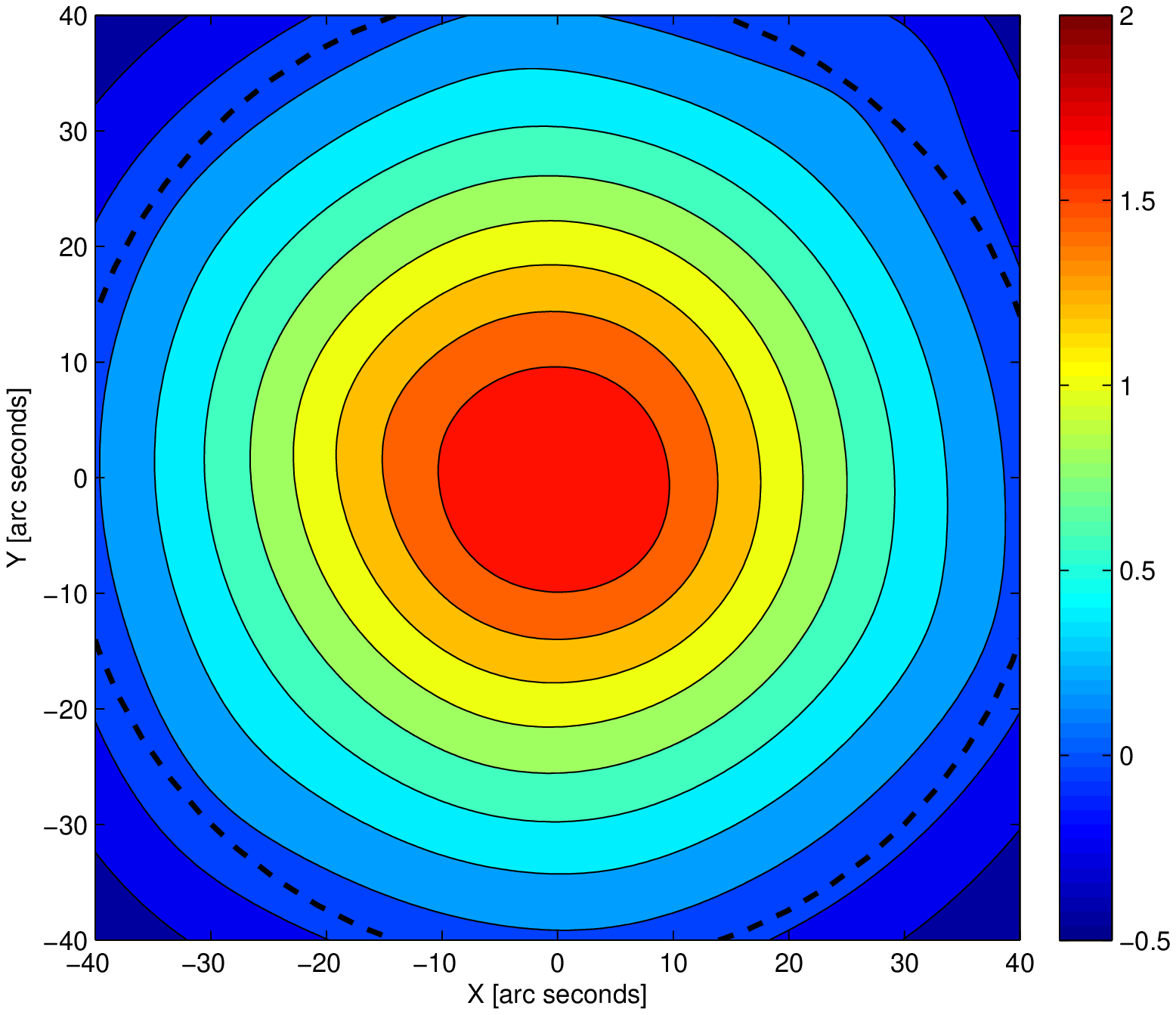}
  \includegraphics[width=0.3\hsize]{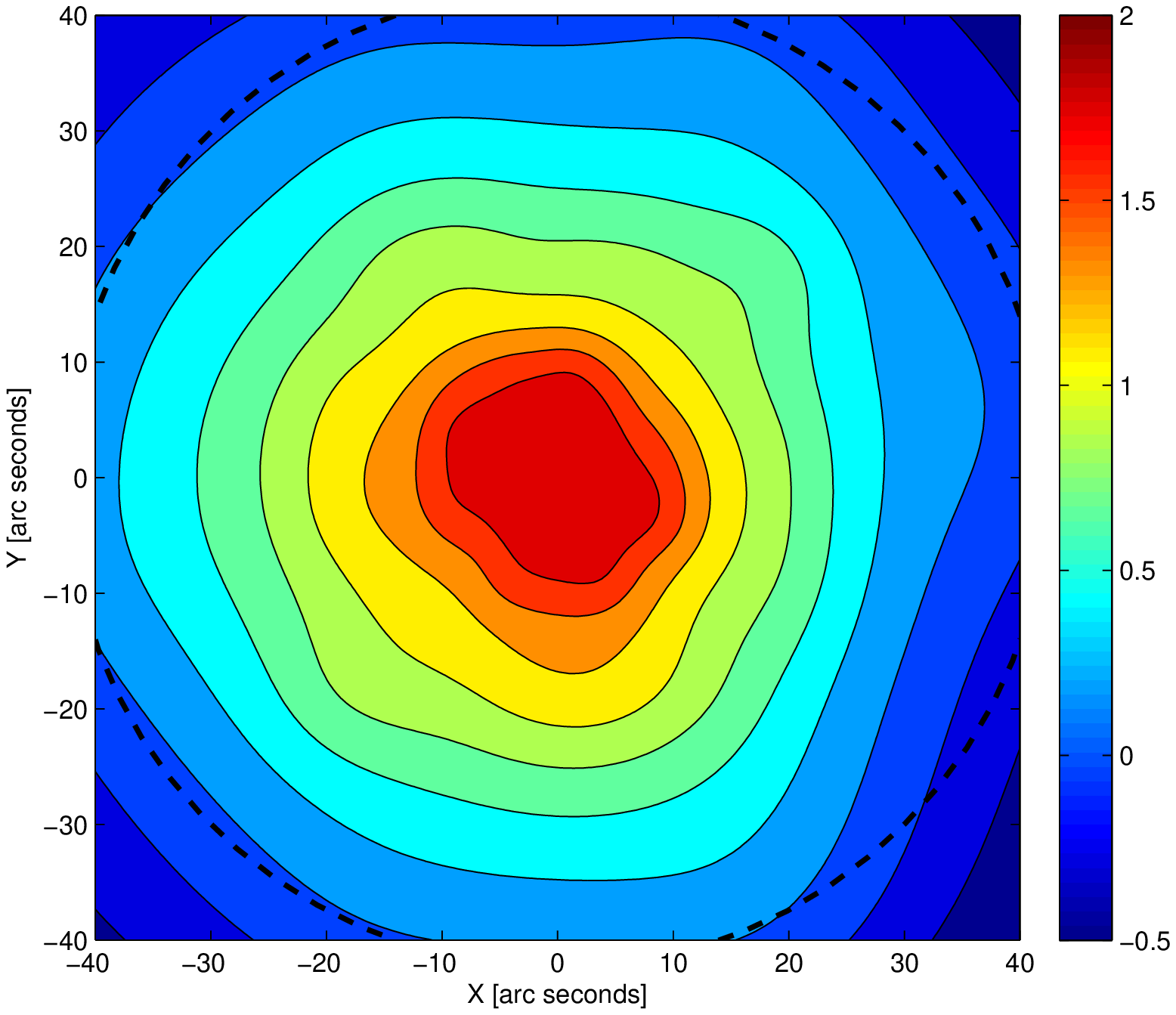} 
  \includegraphics[width=0.3\hsize]{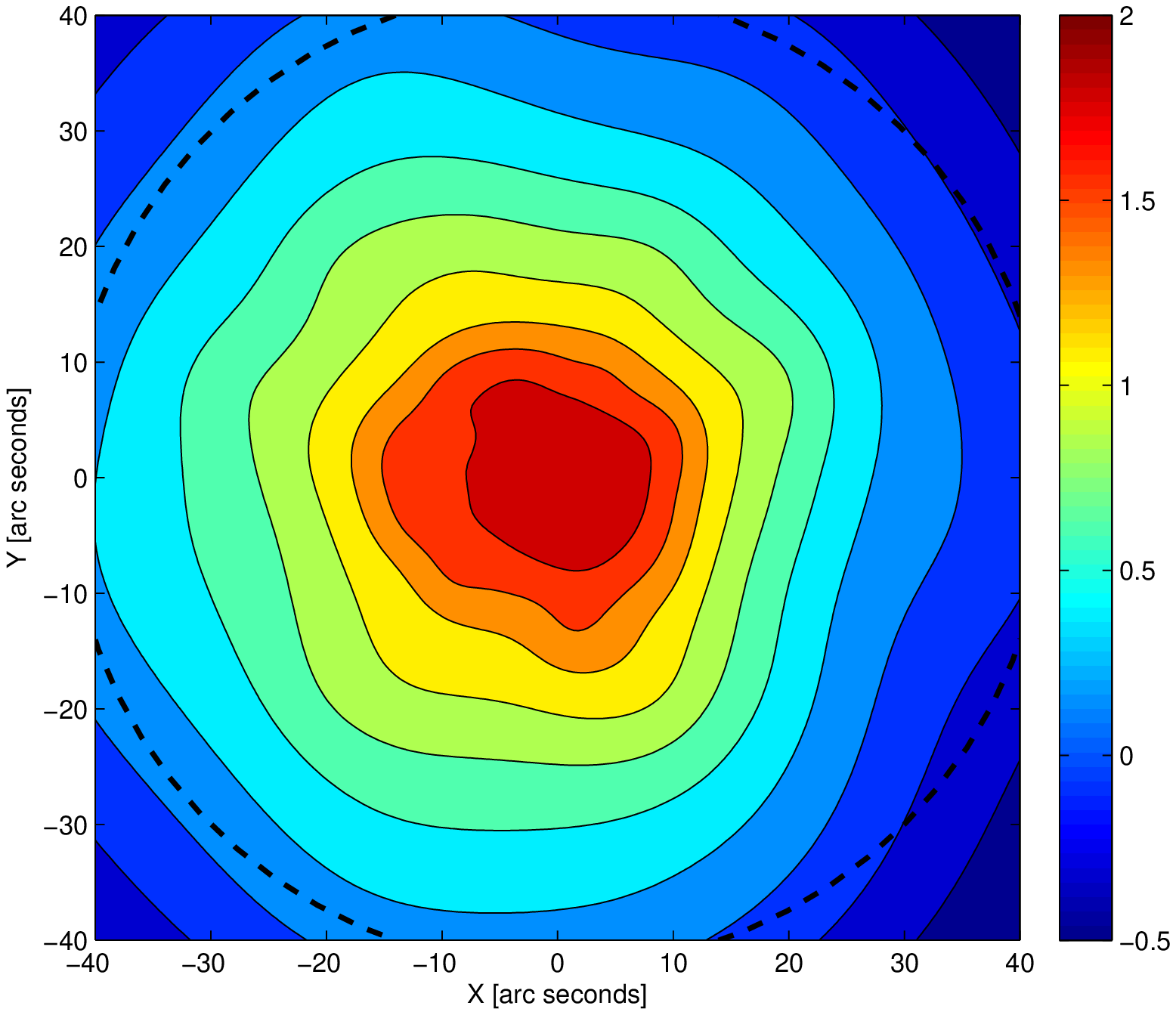}\\
  \includegraphics[width=0.3\hsize]{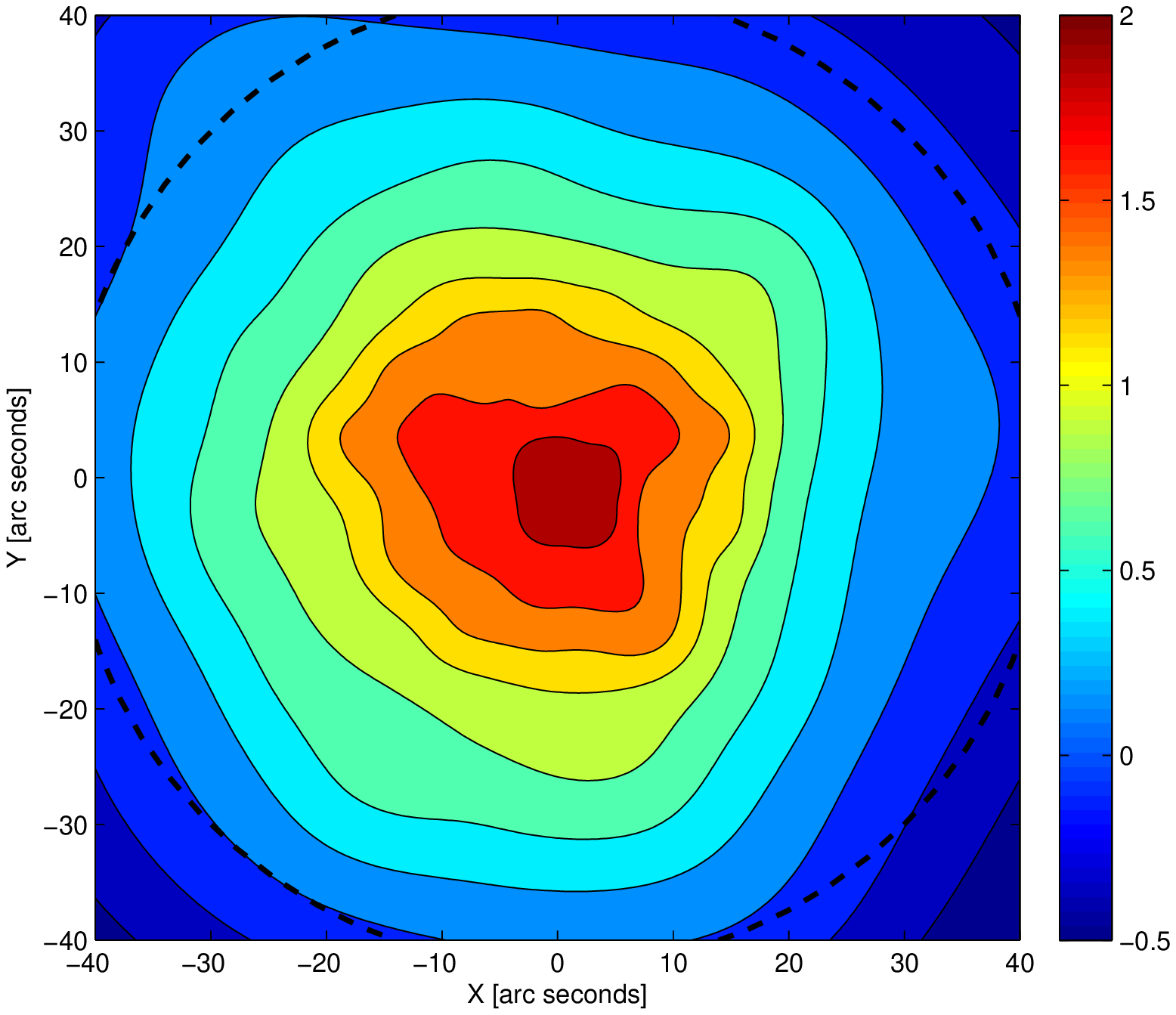} 
  \includegraphics[width=0.3\hsize]{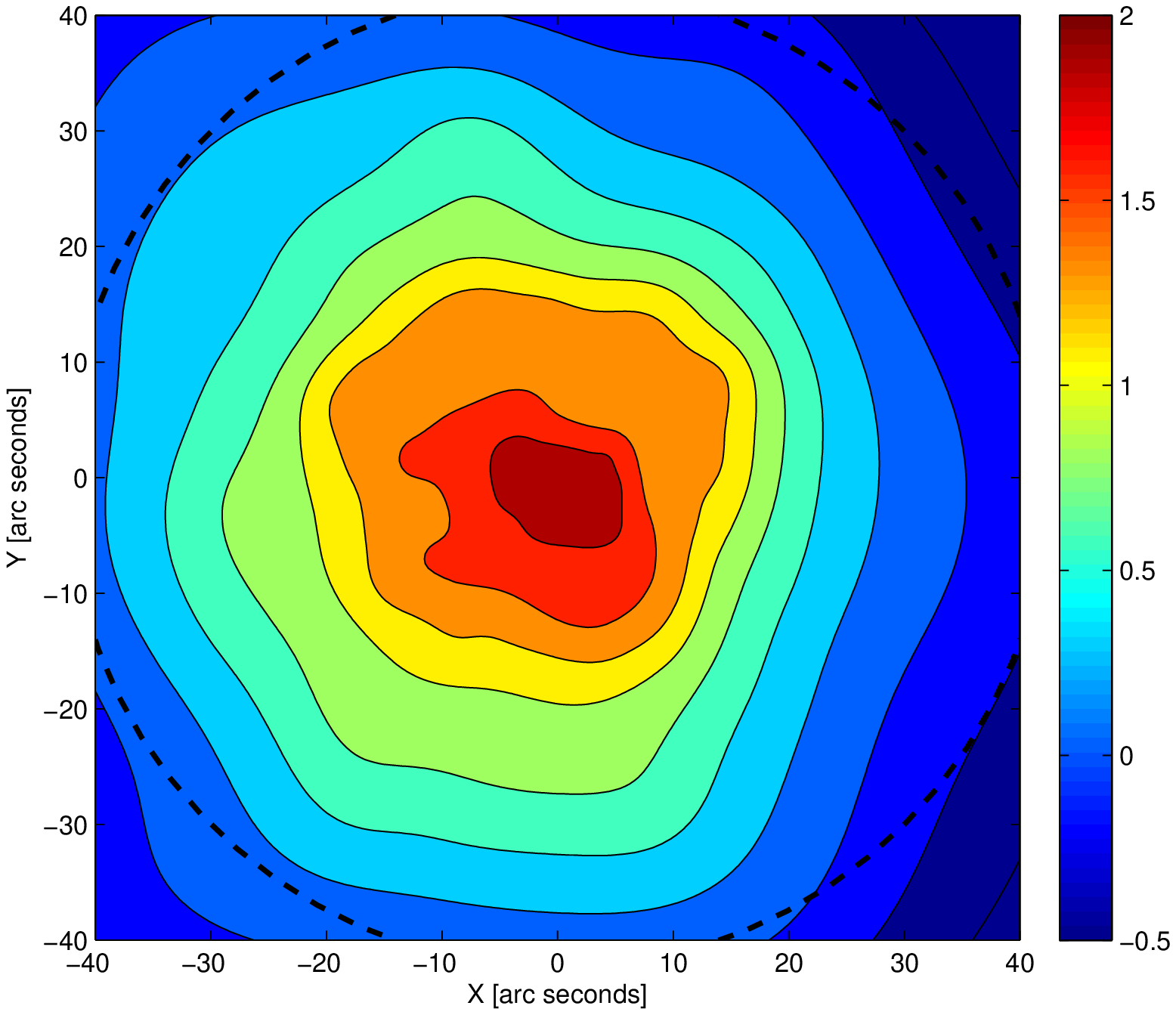}
  \includegraphics[width=0.3\hsize]{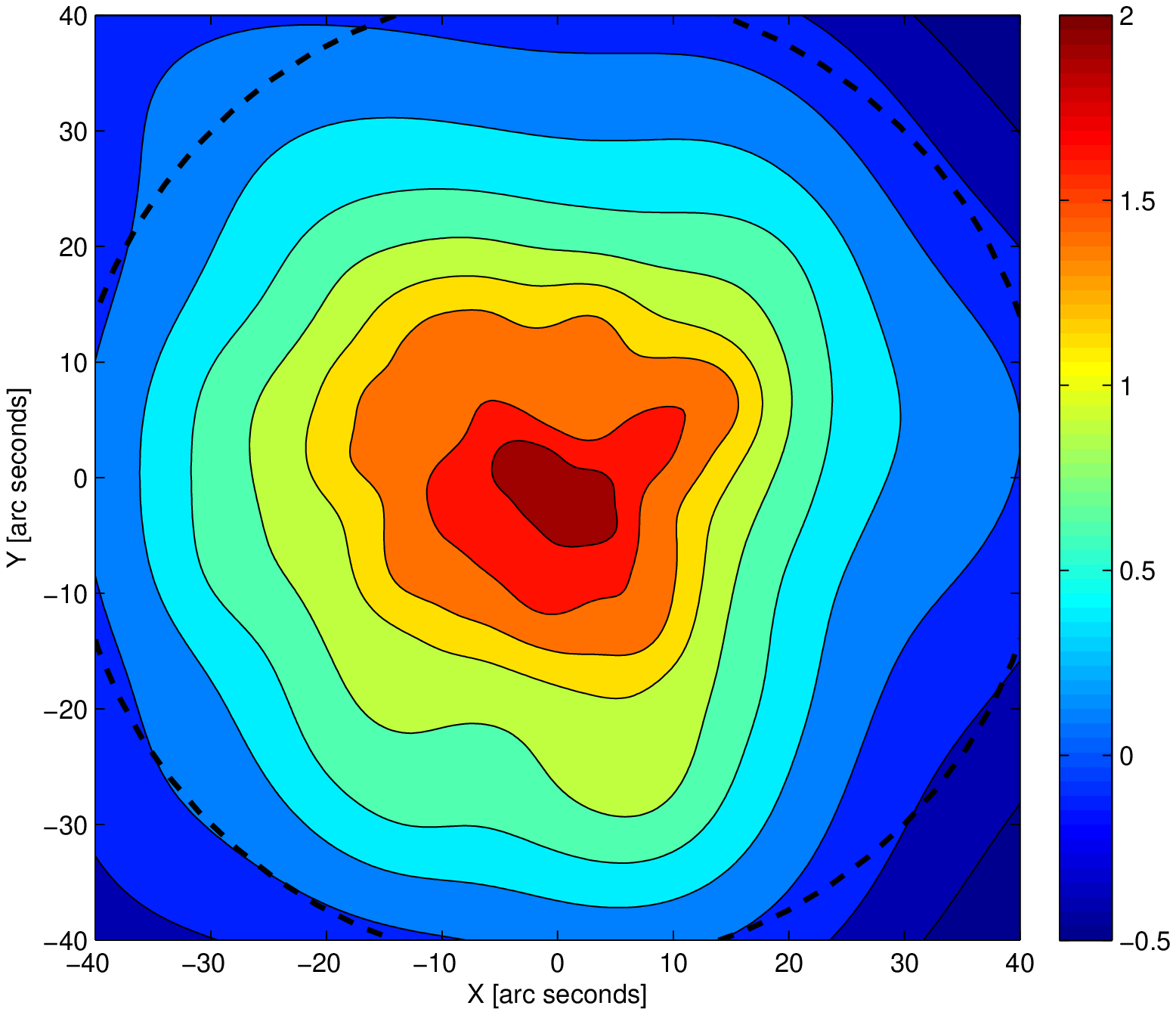}\\
  \includegraphics[width=0.3\hsize]{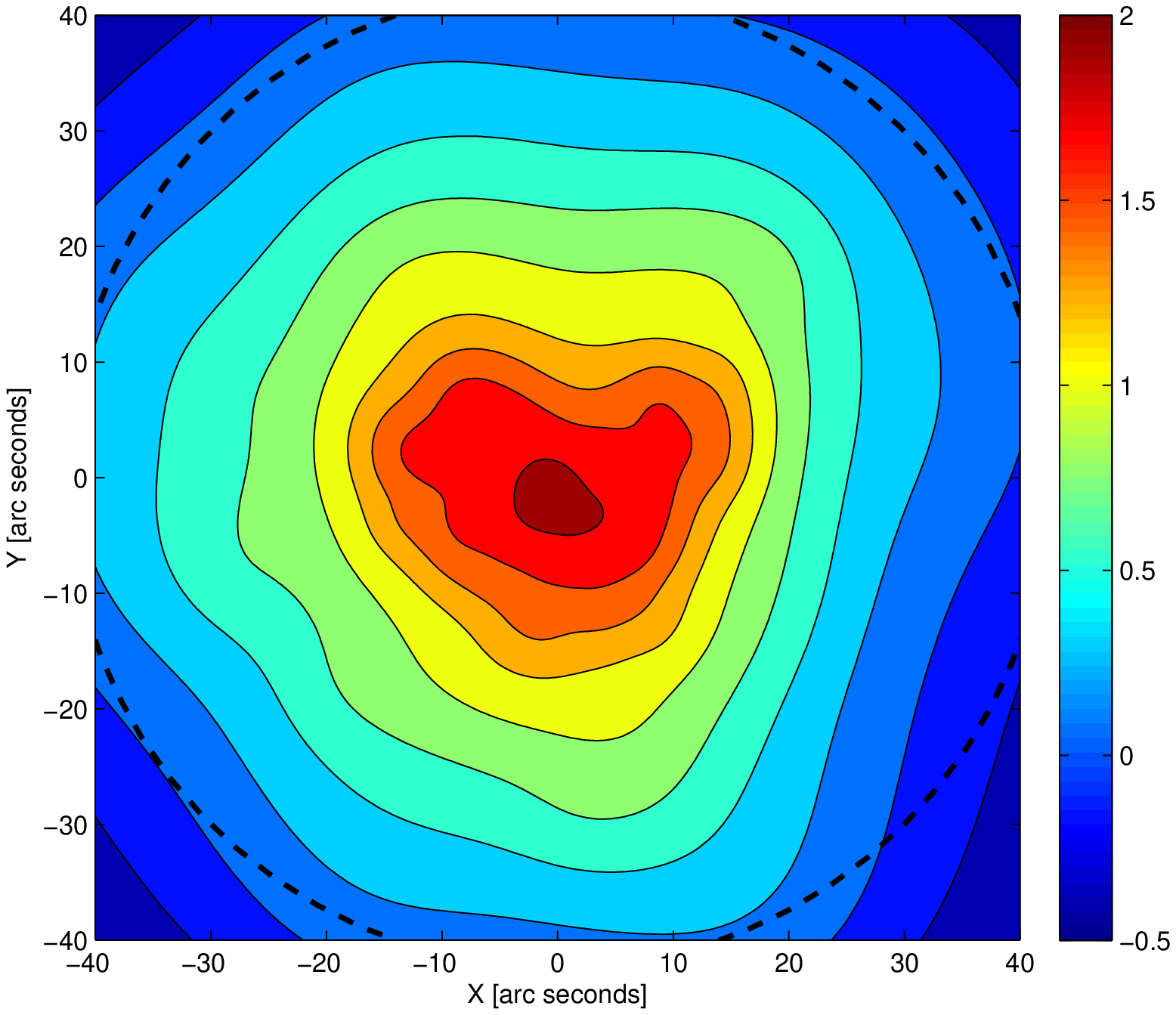}
  \includegraphics[width=0.3\hsize]{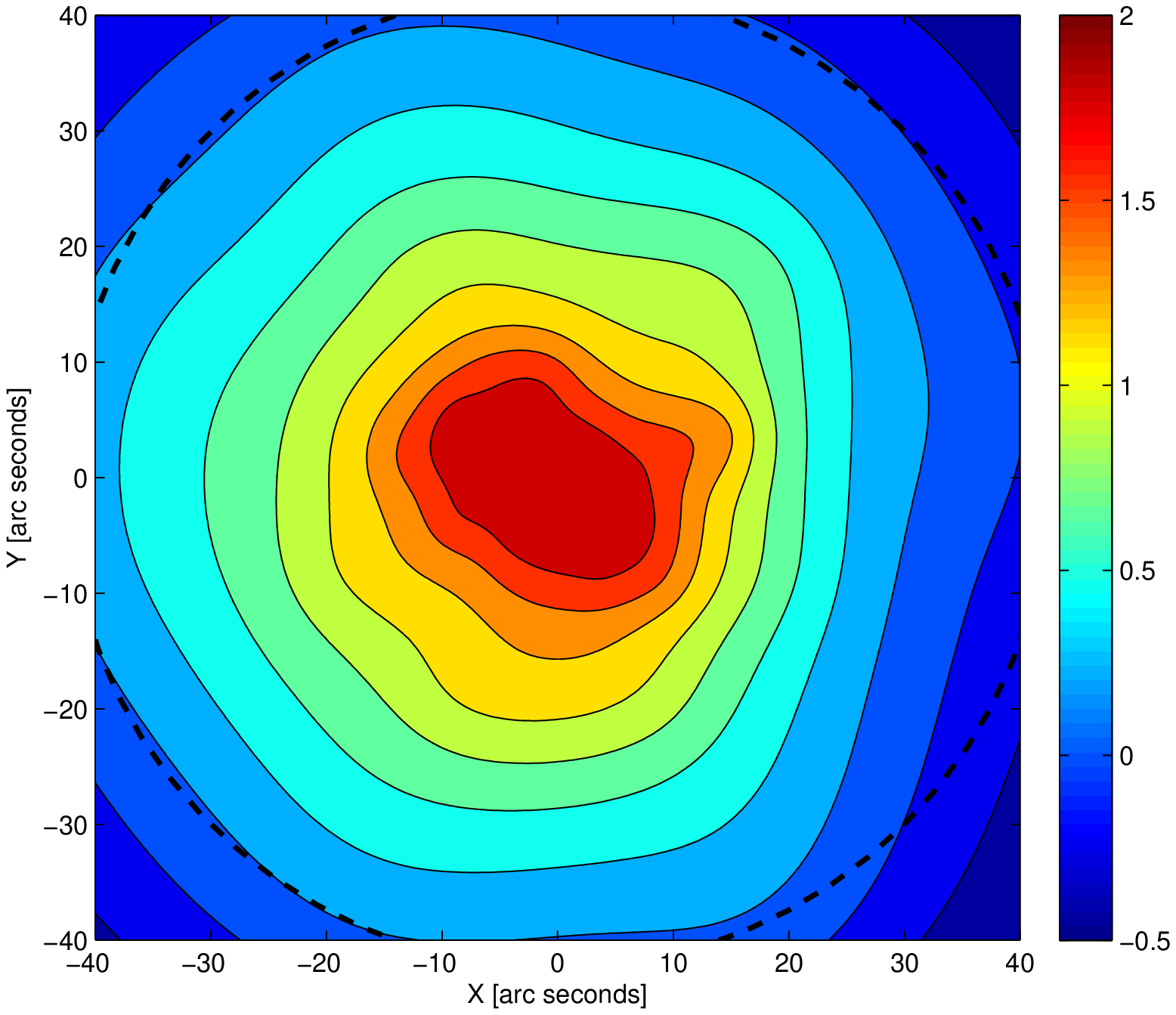} 
  \includegraphics[width=0.3\hsize]{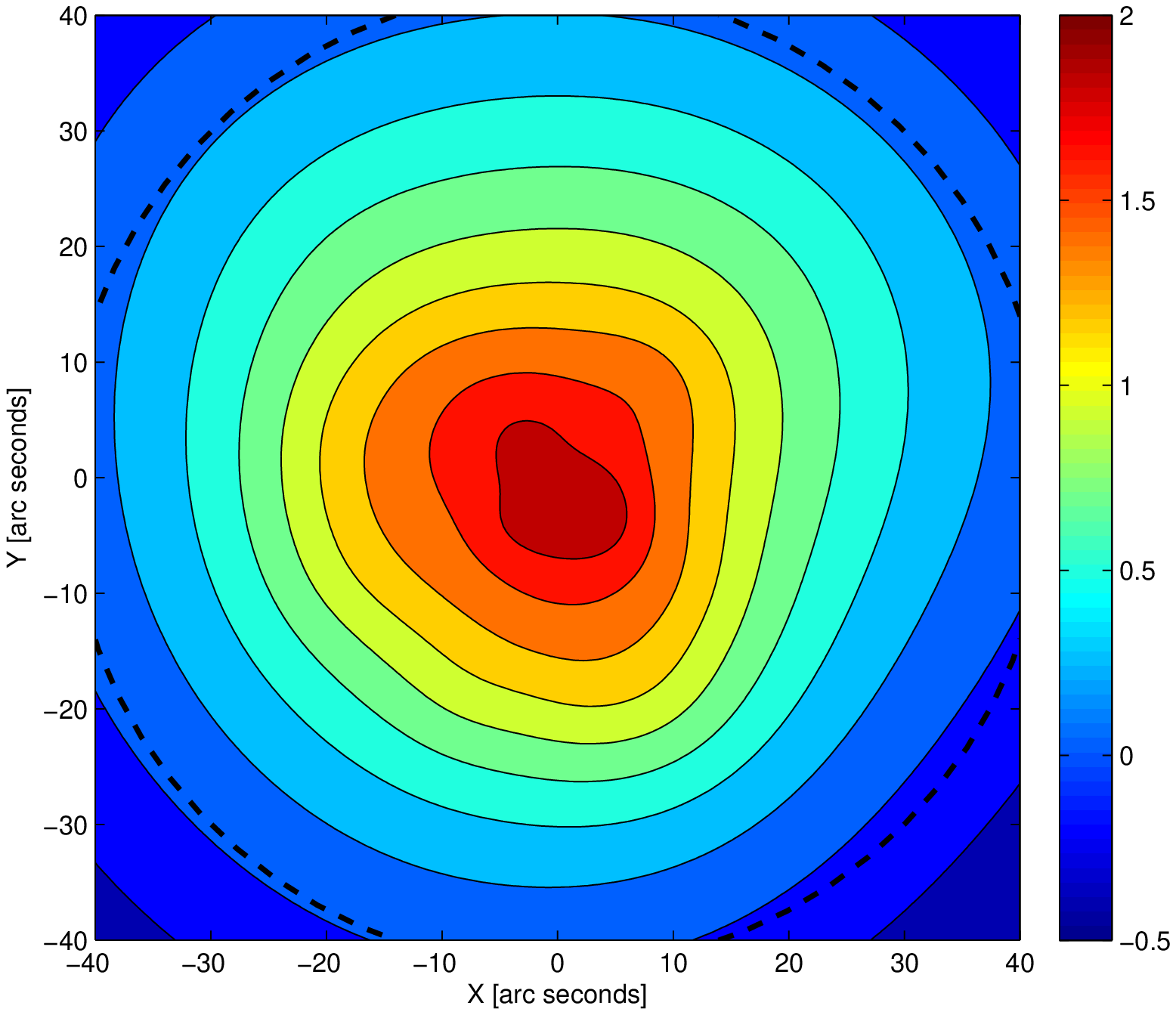}\\
  \includegraphics[width=0.4\hsize]{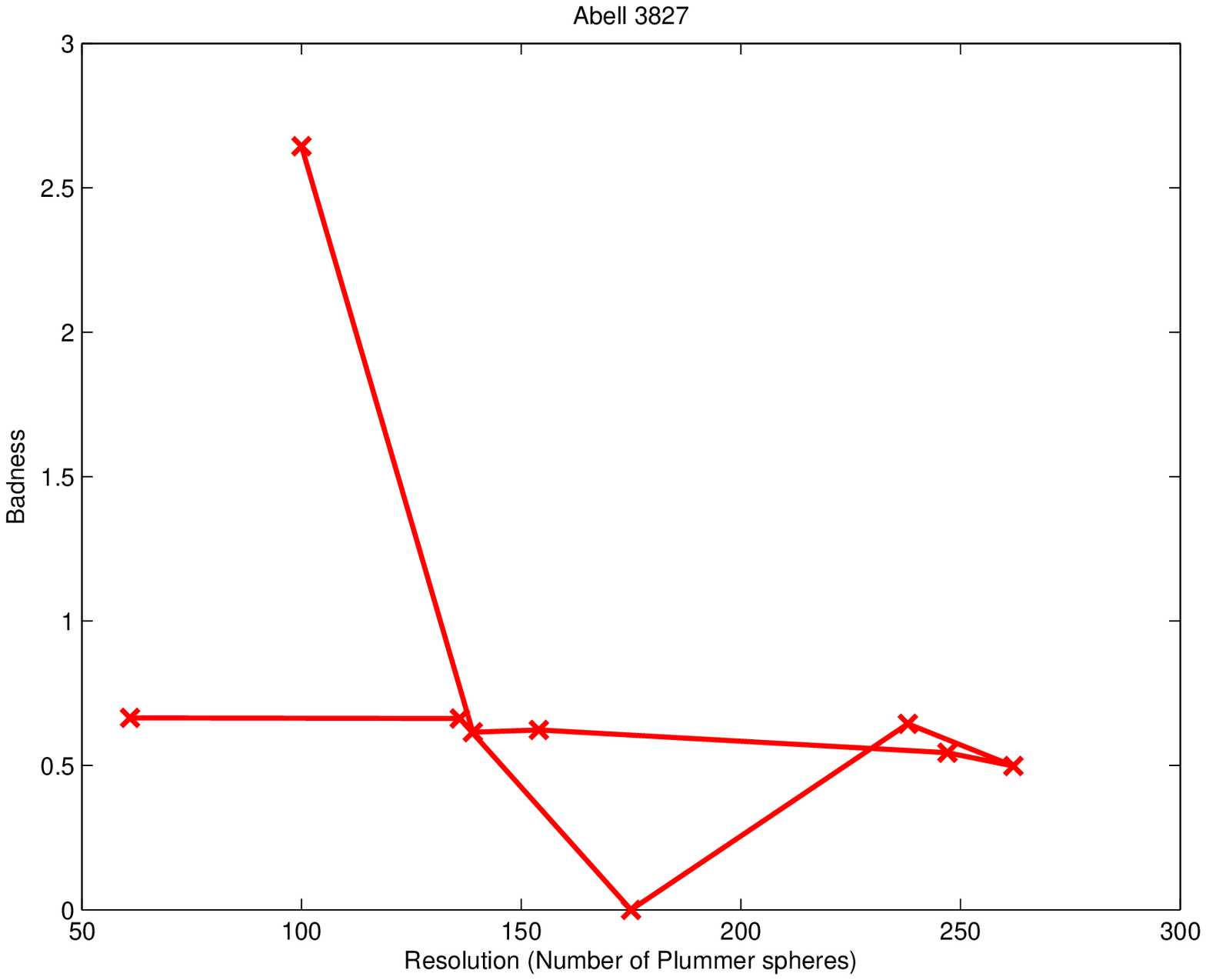}

\caption{Reconstruction of the lens in
  Figure~\ref{fig:plummerin} from the data in that figure.  The
  badness curve (bottom panel) shows that the best
  model is the third one (top right map in the grid of nine.) The dashed
  circle in each map delineates the modeled region. \label{fig:plummerout}.
The sequence of mass maps is in reading order (from top-left to bottom-right).}
\end{figure}

A Plummer lens of mass $10^{14} $M$_\odot$ was generated at redshift
0.1.  Six sources were put at different redshifts (one at 0.15, two
at 0.2, two at 0.4 and one at 1.0).  The mass profile and image plane
are shown in Figure \ref{fig:plummerin}.  The images and source
redshifts were given to the inversion module of GRALE.
Figure~\ref{fig:plummerout} shows the reconstructed masses at
different resolutions and the badness values.

When reconstructing the lens, GRALE did not have the information that
in fact it had a simple parametric form, without substructures.  The
reconstructions do have some substructure, as well as small offsets
from the centre.  Such spurious features increase with resolution.
The least-badness criterion, however, favours a model with relatively
little substructure.

\subsection{A more complex lens}

We now increase the complexity, both of the input lens and of the
reconstruction procedure.  For each data set, from now on we will
present a mean map $\Sigma$ and a fraction rms-deviation map
$\delta\Sigma/\Sigma$, obtained as follows.  From the images, we first
let GRALE construct a sequence of maps at nine different resolutions
(as with the simple lens), and then select the one at minimum badness.
This whole procedure is repeated 10 times, to obtain an ensemble of
reconstructions.  The mean and rms deviation refer to such an
ensemble, as
\begin{equation}
\delta\Sigma = \left(\langle\Sigma^2\rangle - 
                     \langle\Sigma\rangle^2\right)^{\frac12} \,.
\end{equation}
Each map of $\Sigma$ and $\delta\Sigma/\Sigma$ comes out of 90 separate
reconstructions at different resolutions.  The typical computational
requirement is 50~hours $\times$ 16 cores.

\begin{figure}
  \includegraphics[width=0.24\hsize]{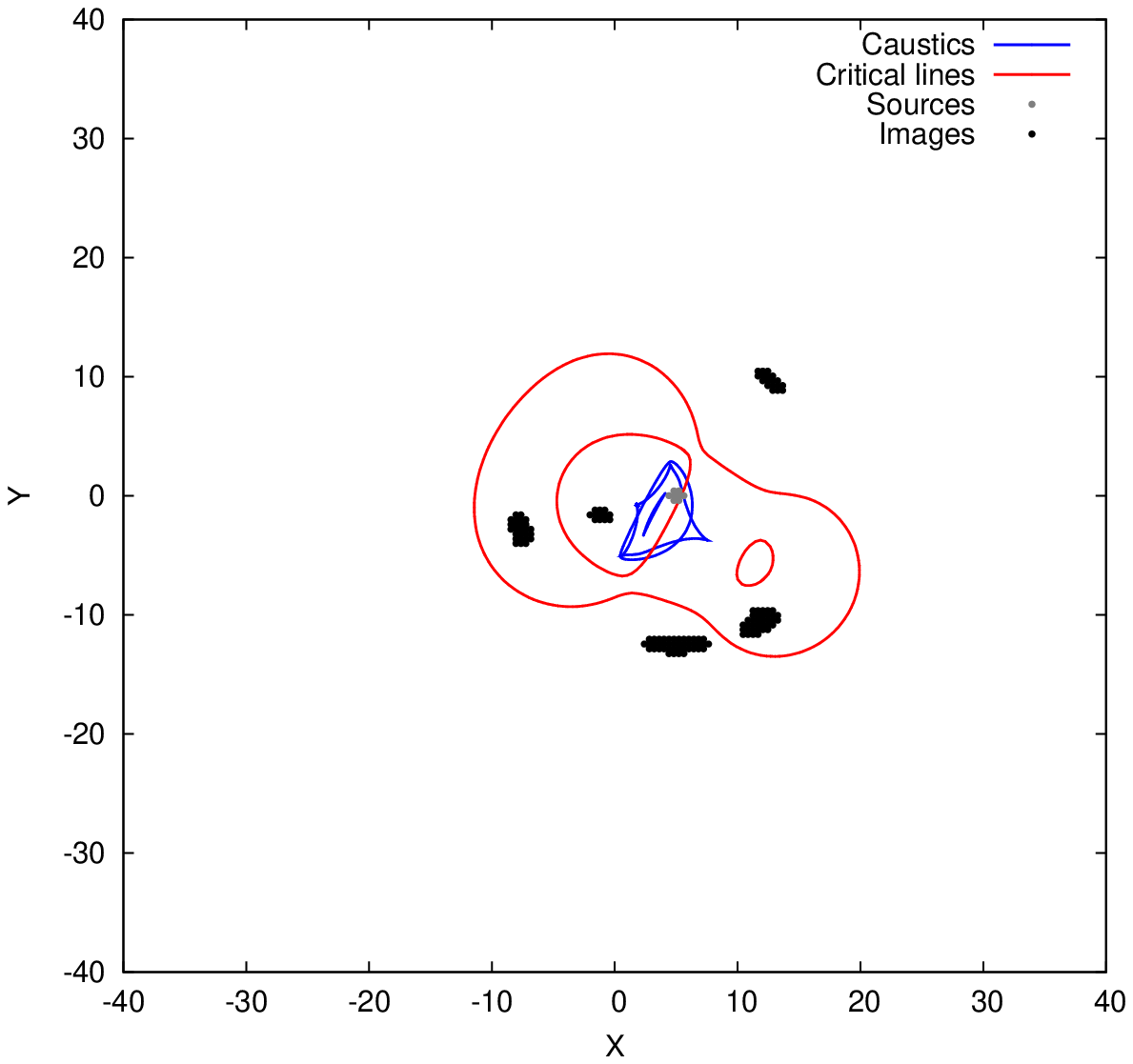}
  \hfil
  \includegraphics[width=0.3\hsize]{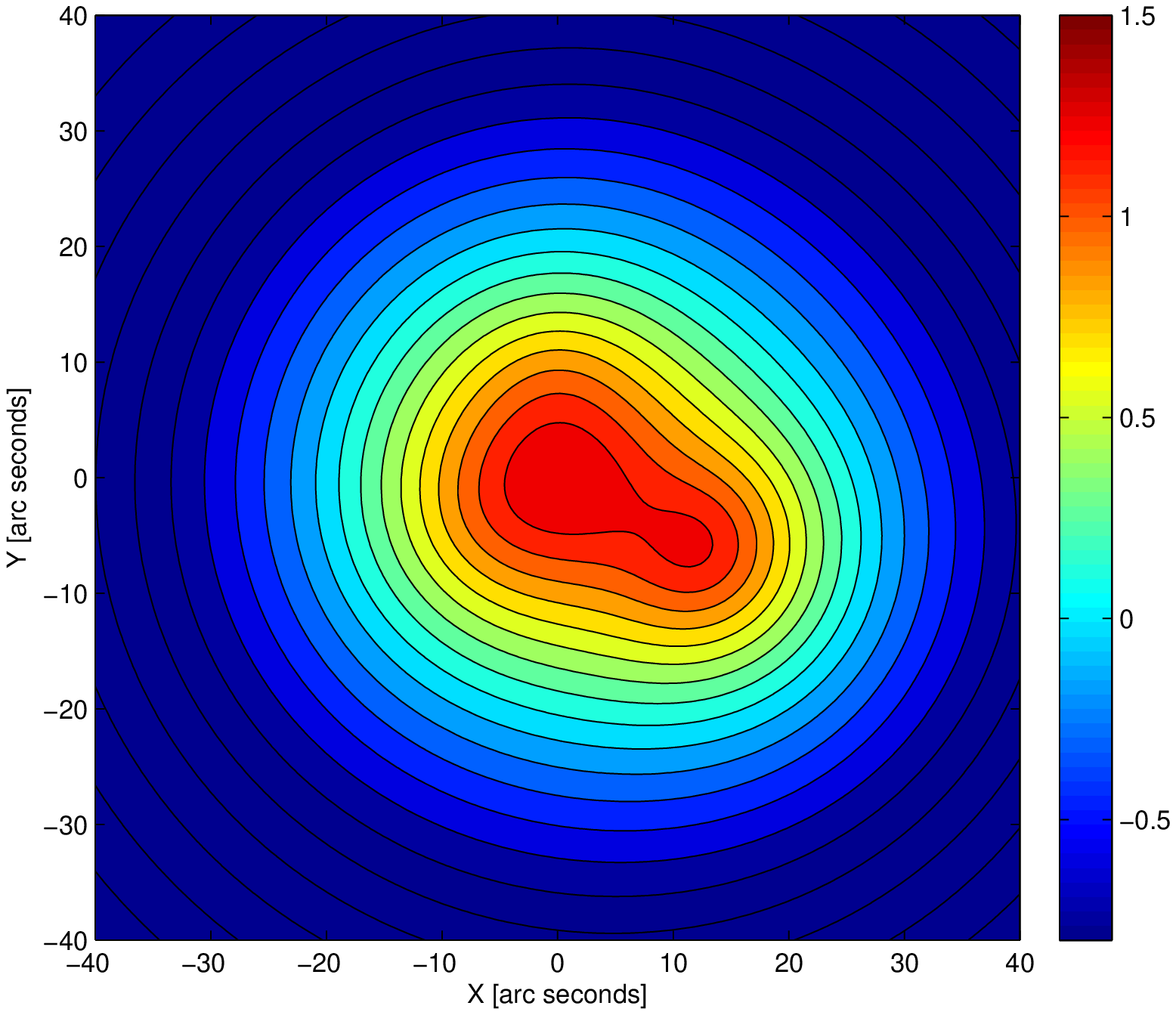}
  \hfil
  \includegraphics[width=0.24\hsize]{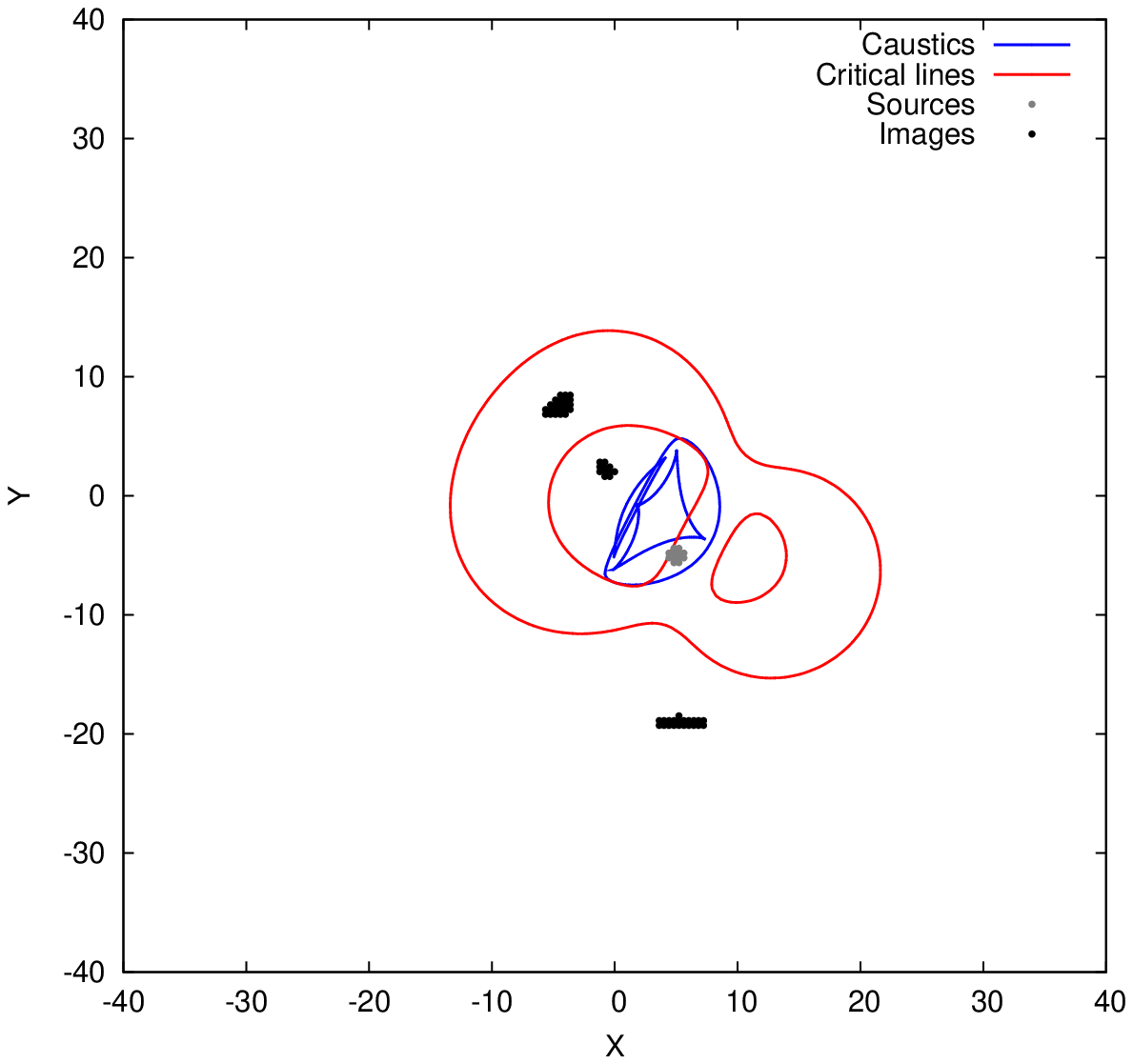}\\
  \includegraphics[width=0.24\hsize]{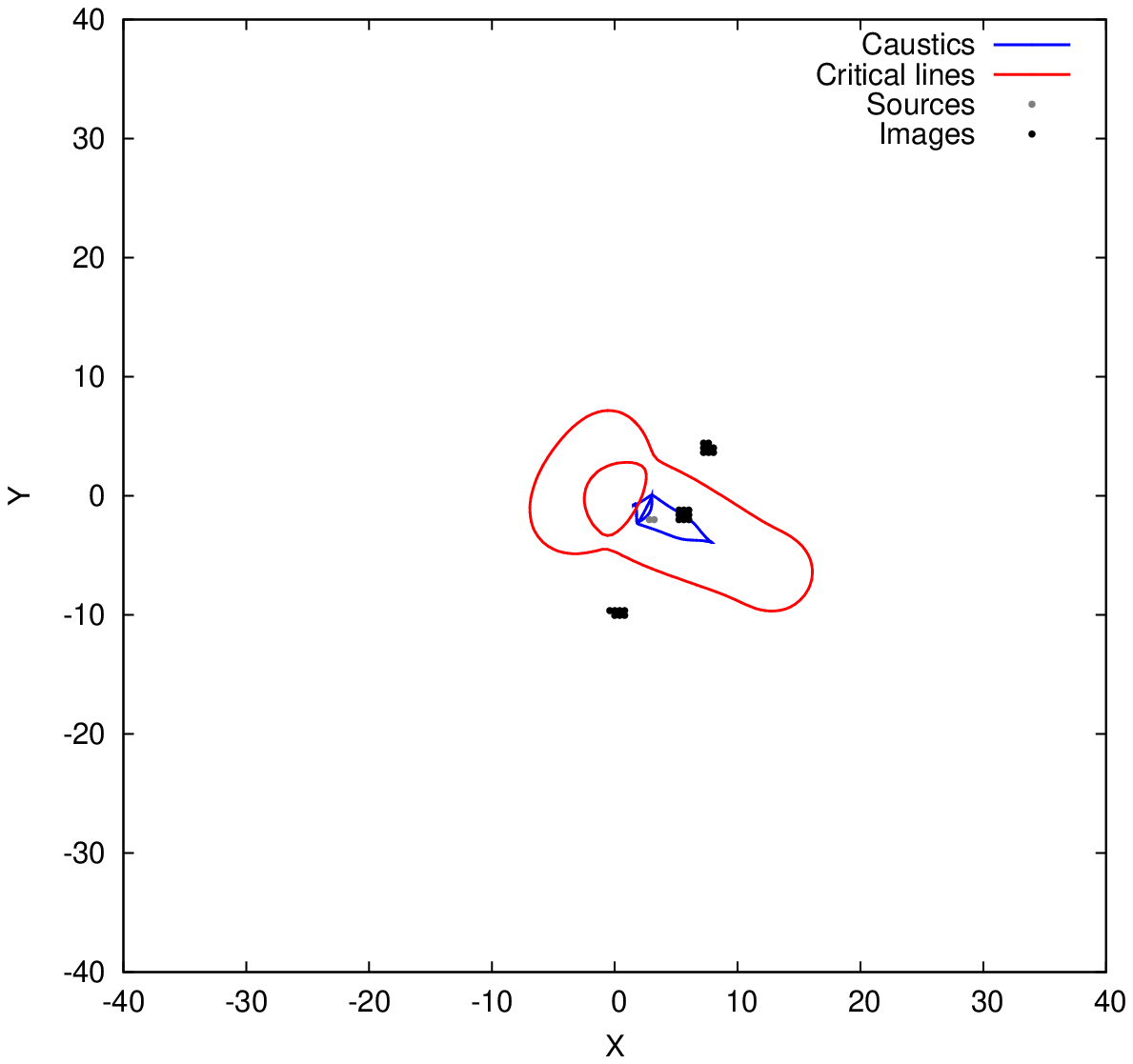}
  \includegraphics[width=0.24\hsize]{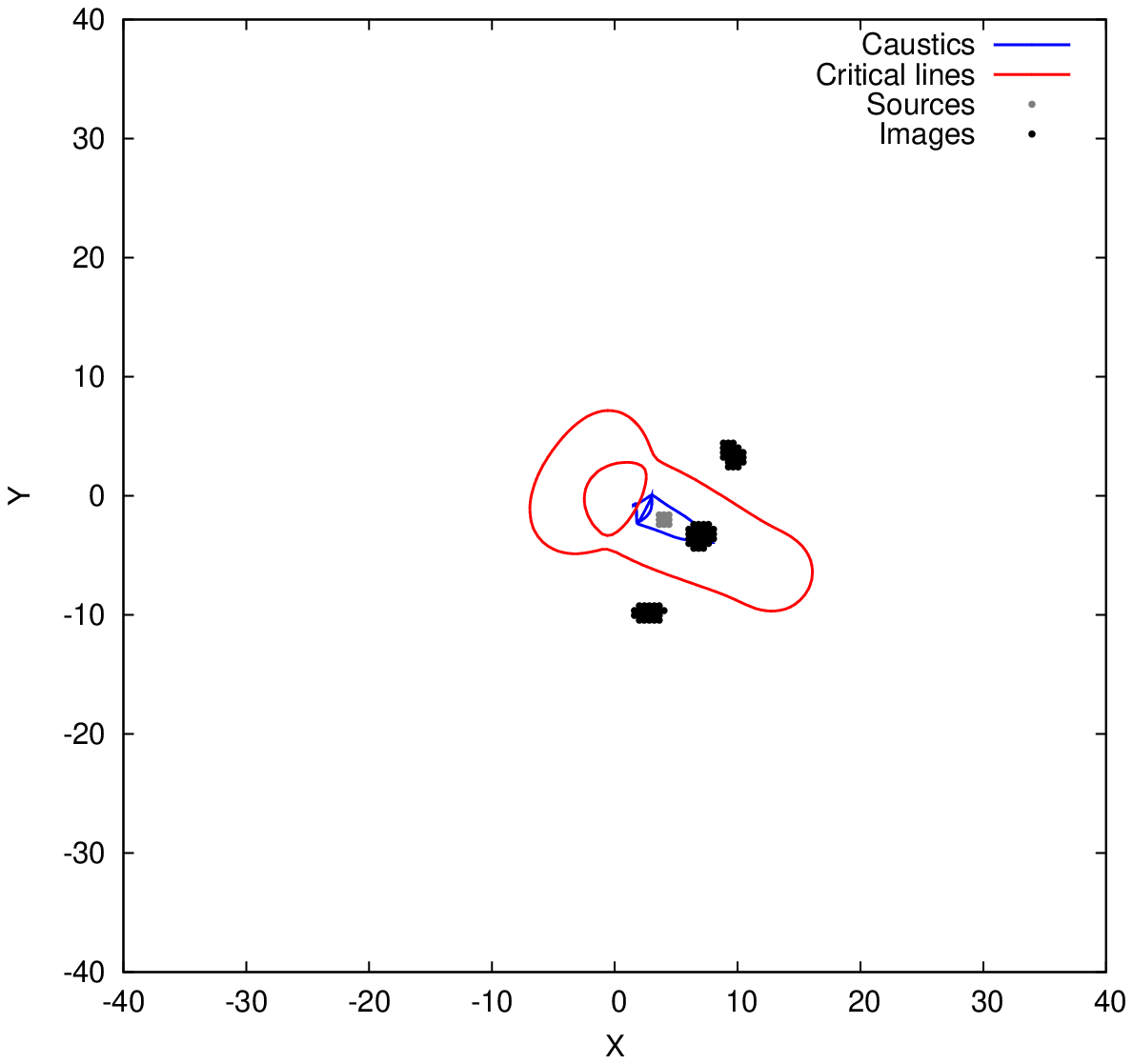}
  \includegraphics[width=0.24\hsize]{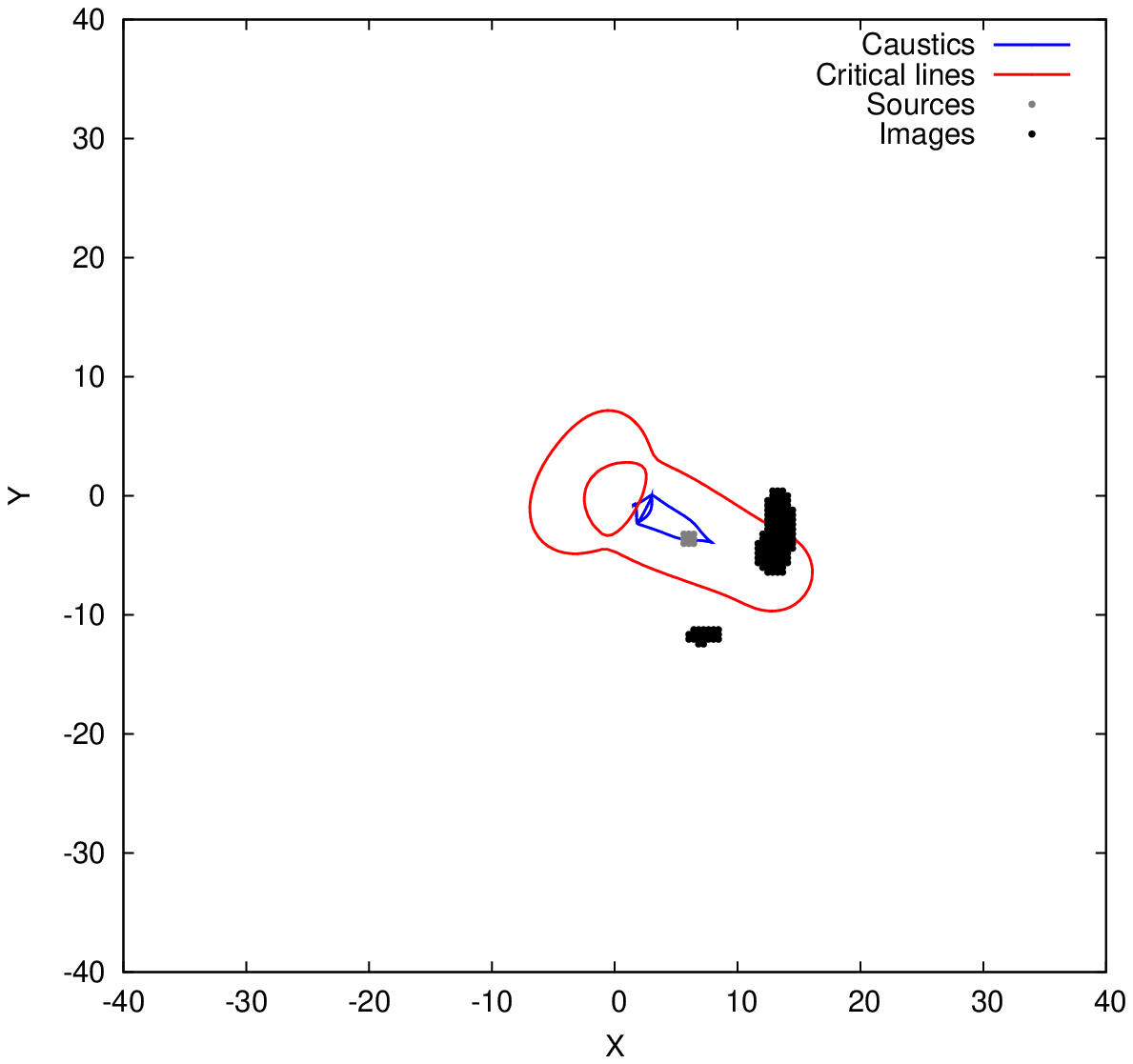}
\caption{\label{fig:plummer5in} A synthetic lens with a main
  mass concentration and a nearby secondary mass peak. Five projected Plummer
  spheres are used to construct this lens. Image systems from five
  sources at different redshifts are shown in separate panels.}
\end{figure}

\begin{figure}
  \includegraphics[width=0.32\hsize]{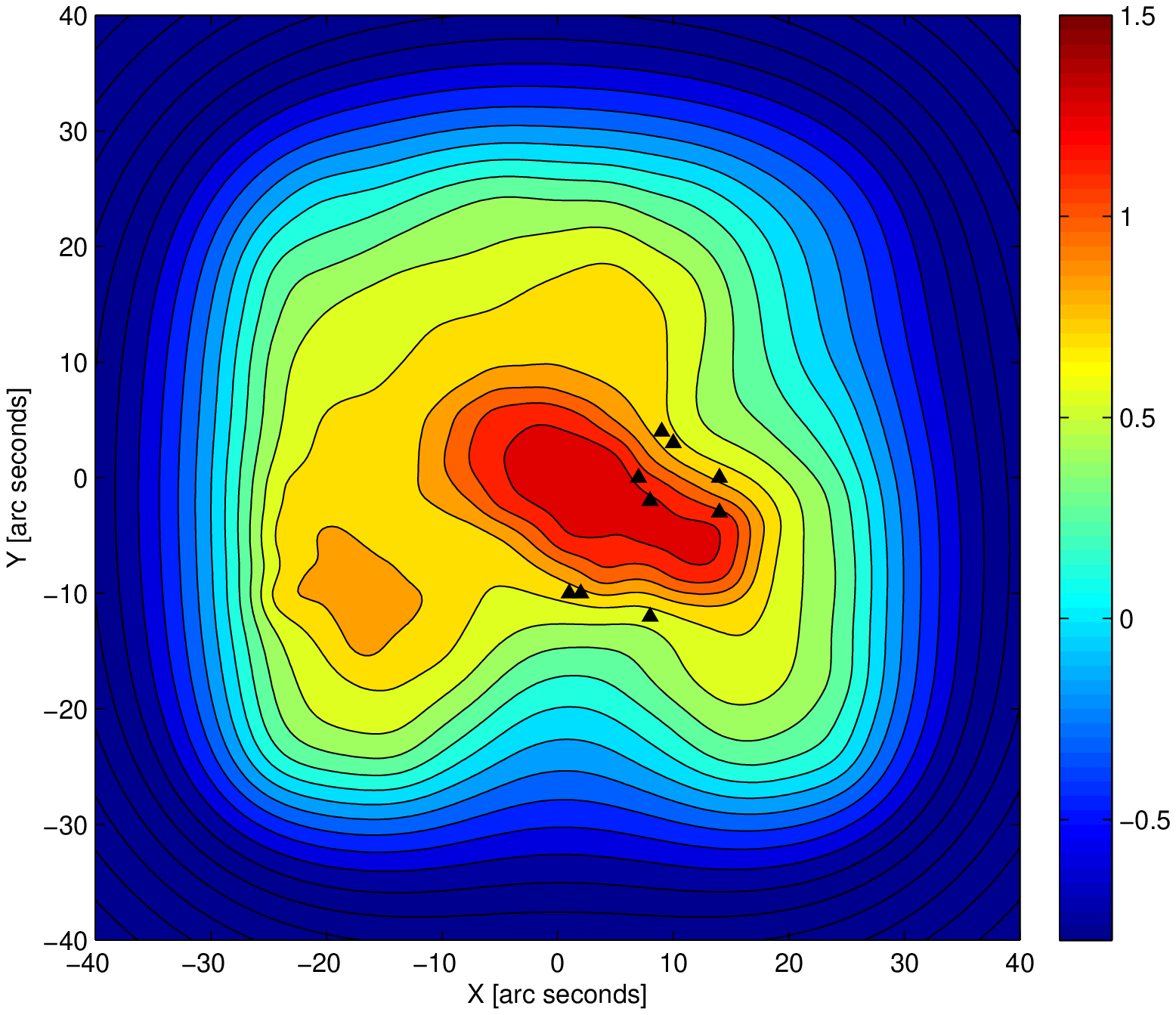}
  \includegraphics[width=0.32\hsize]{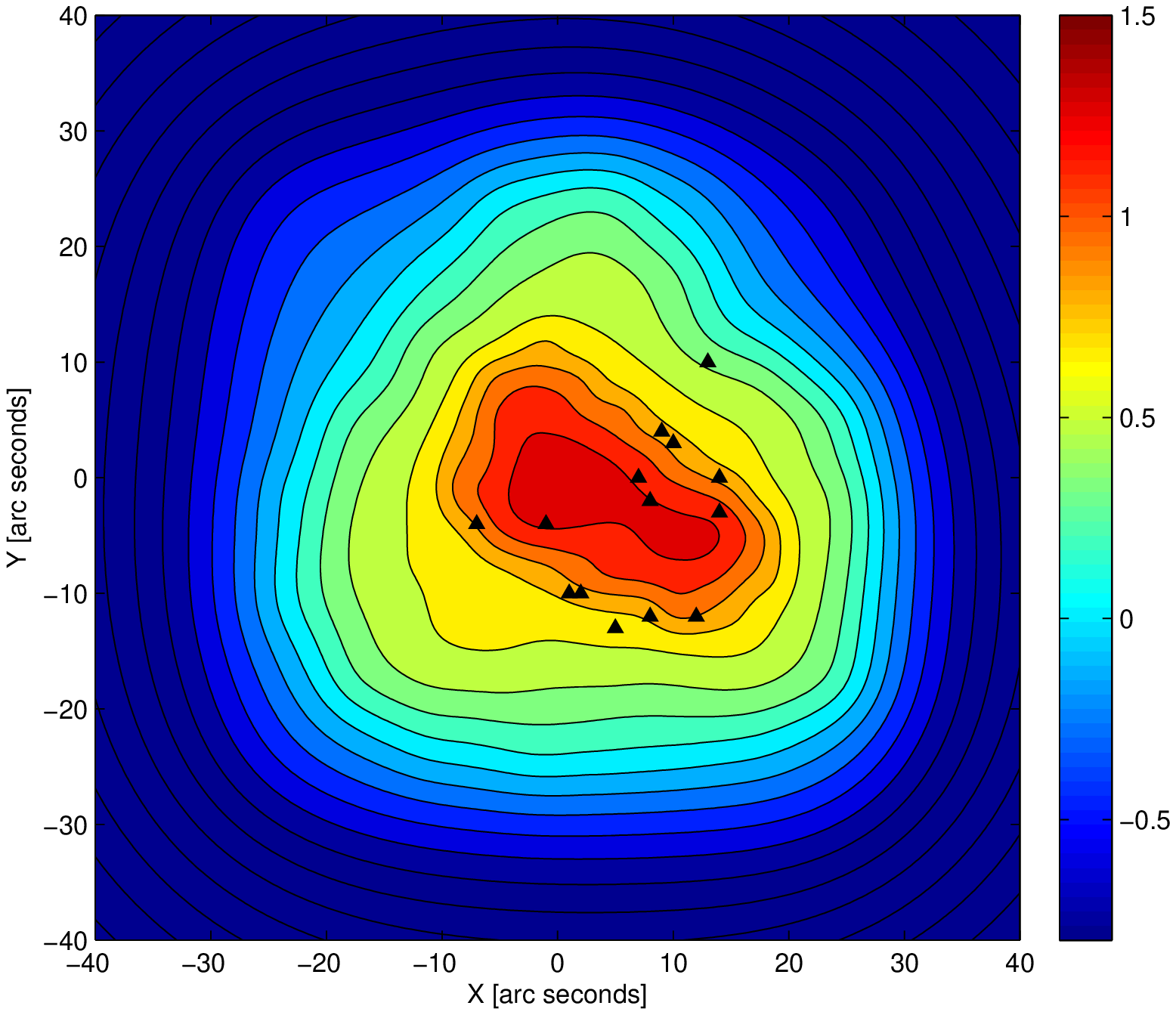}
  \includegraphics[width=0.32\hsize]{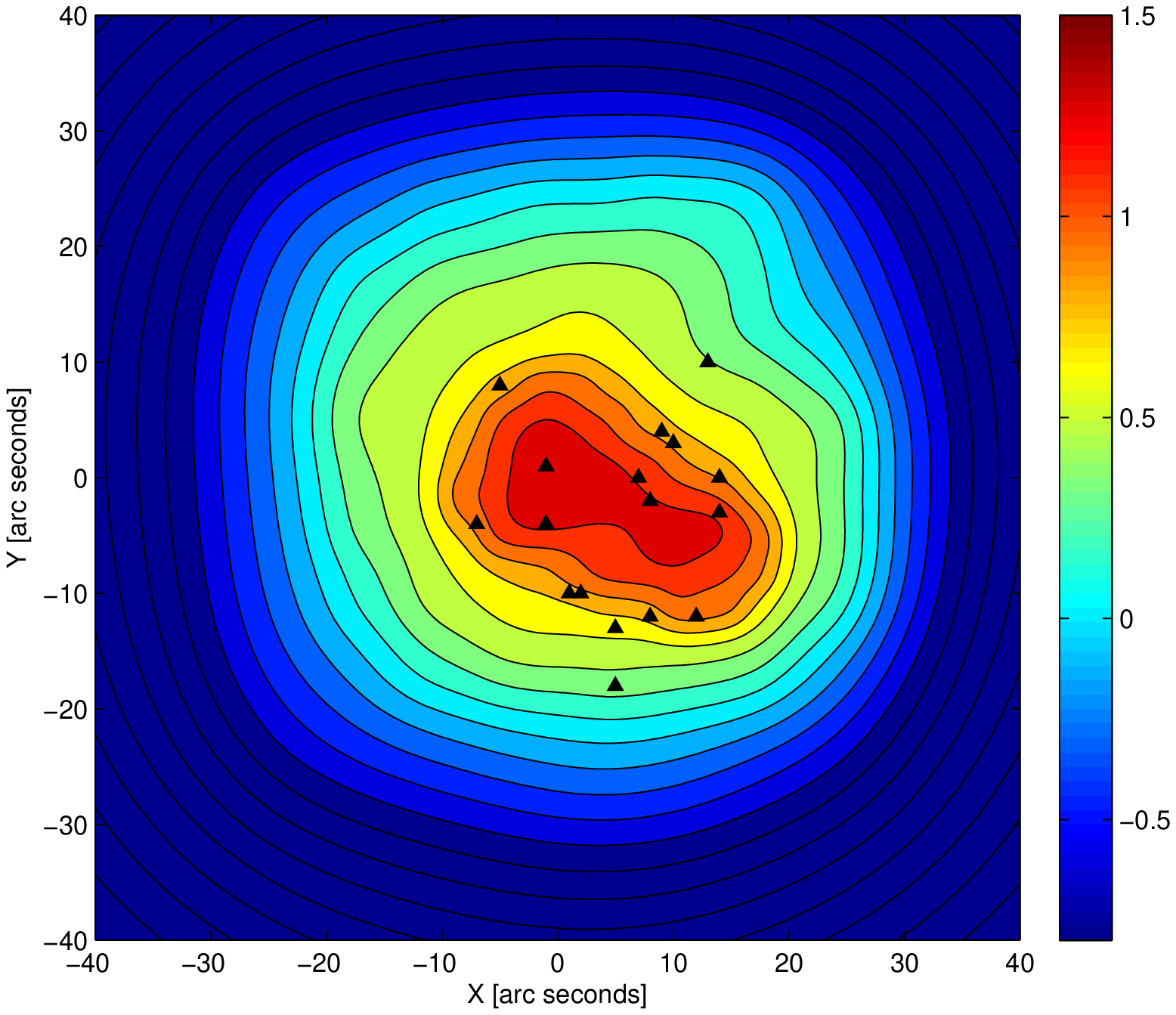}\\

  \includegraphics[width=0.32\hsize]{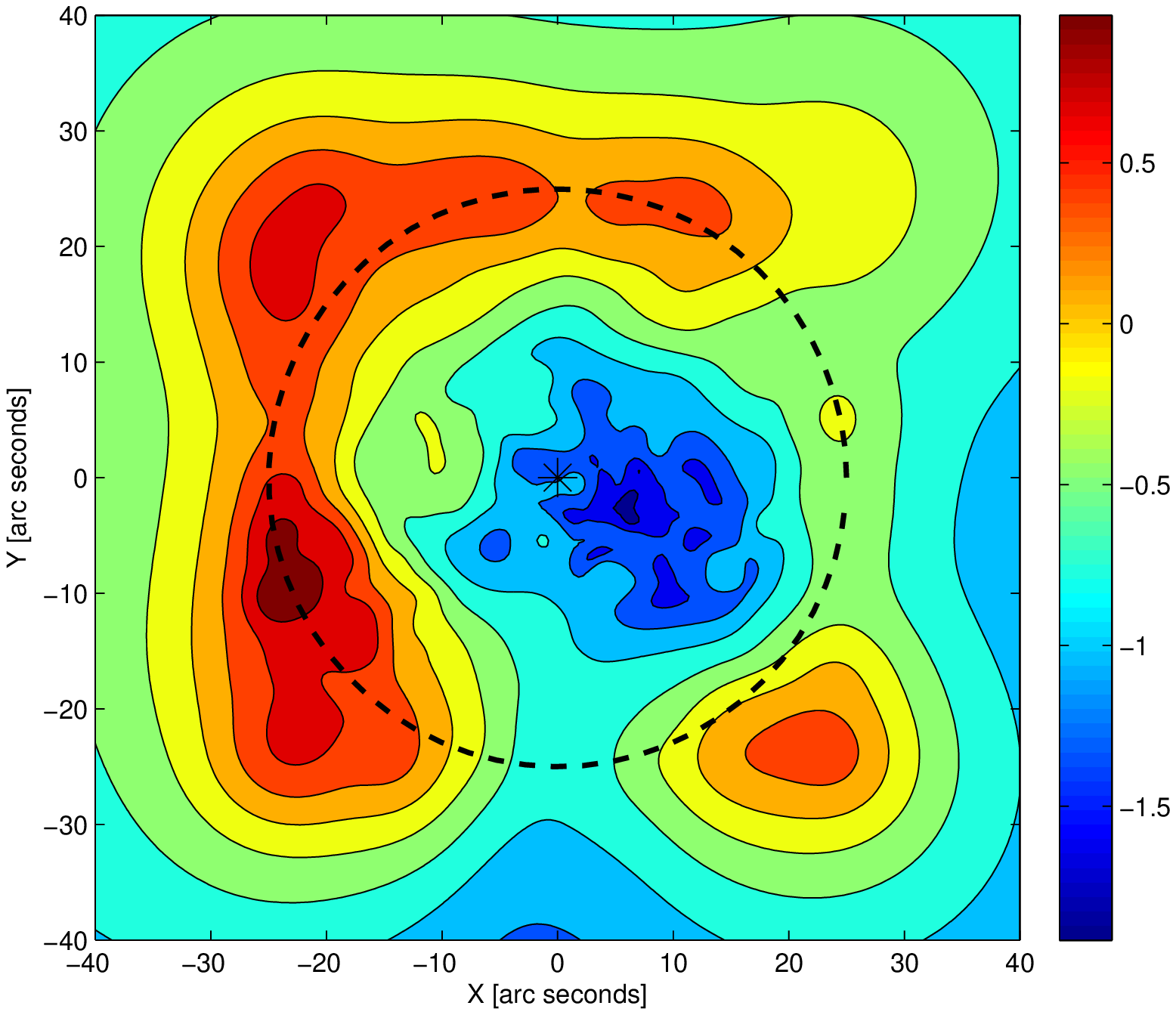}
  \includegraphics[width=0.32\hsize]{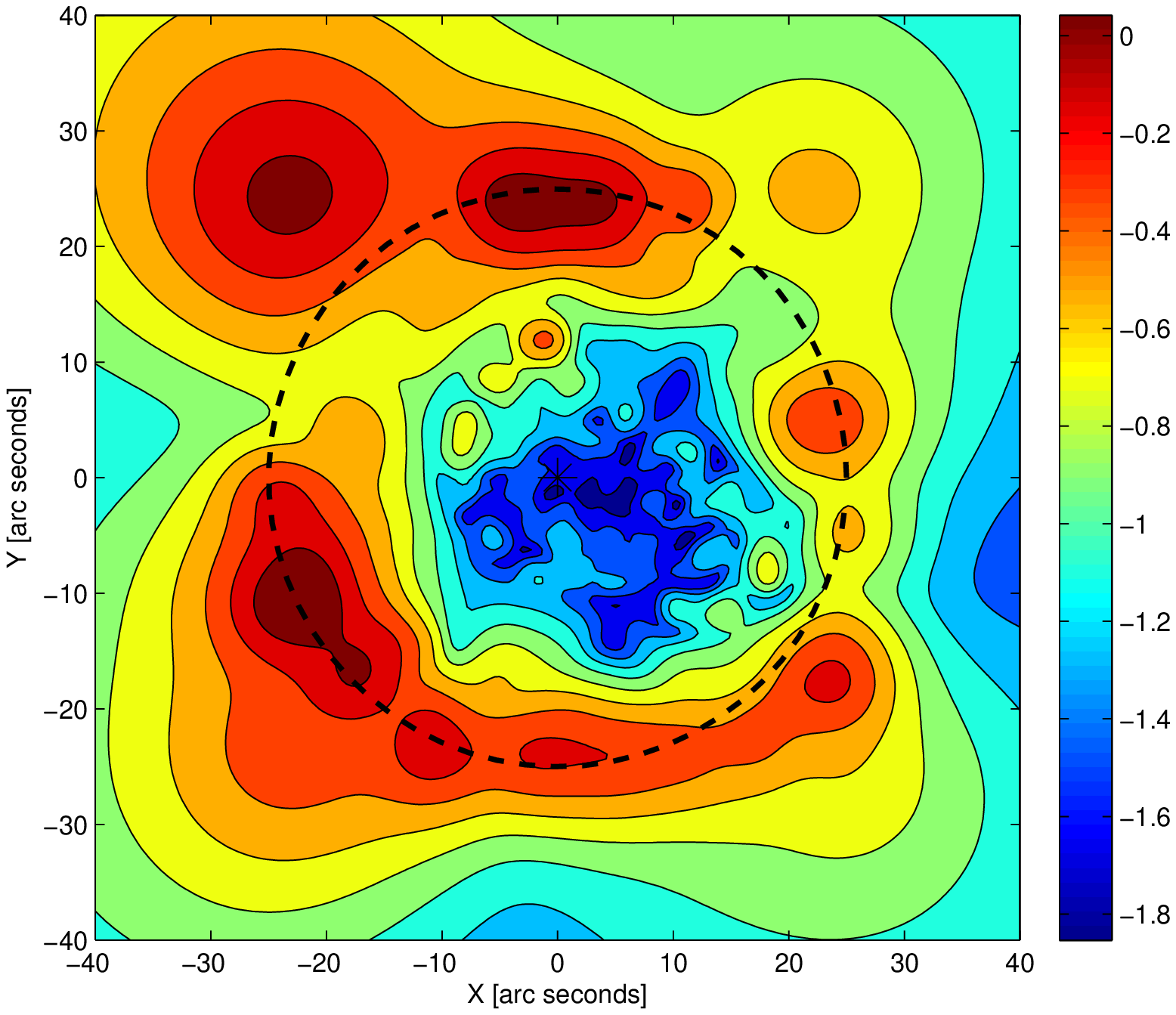}
  \includegraphics[width=0.32\hsize]{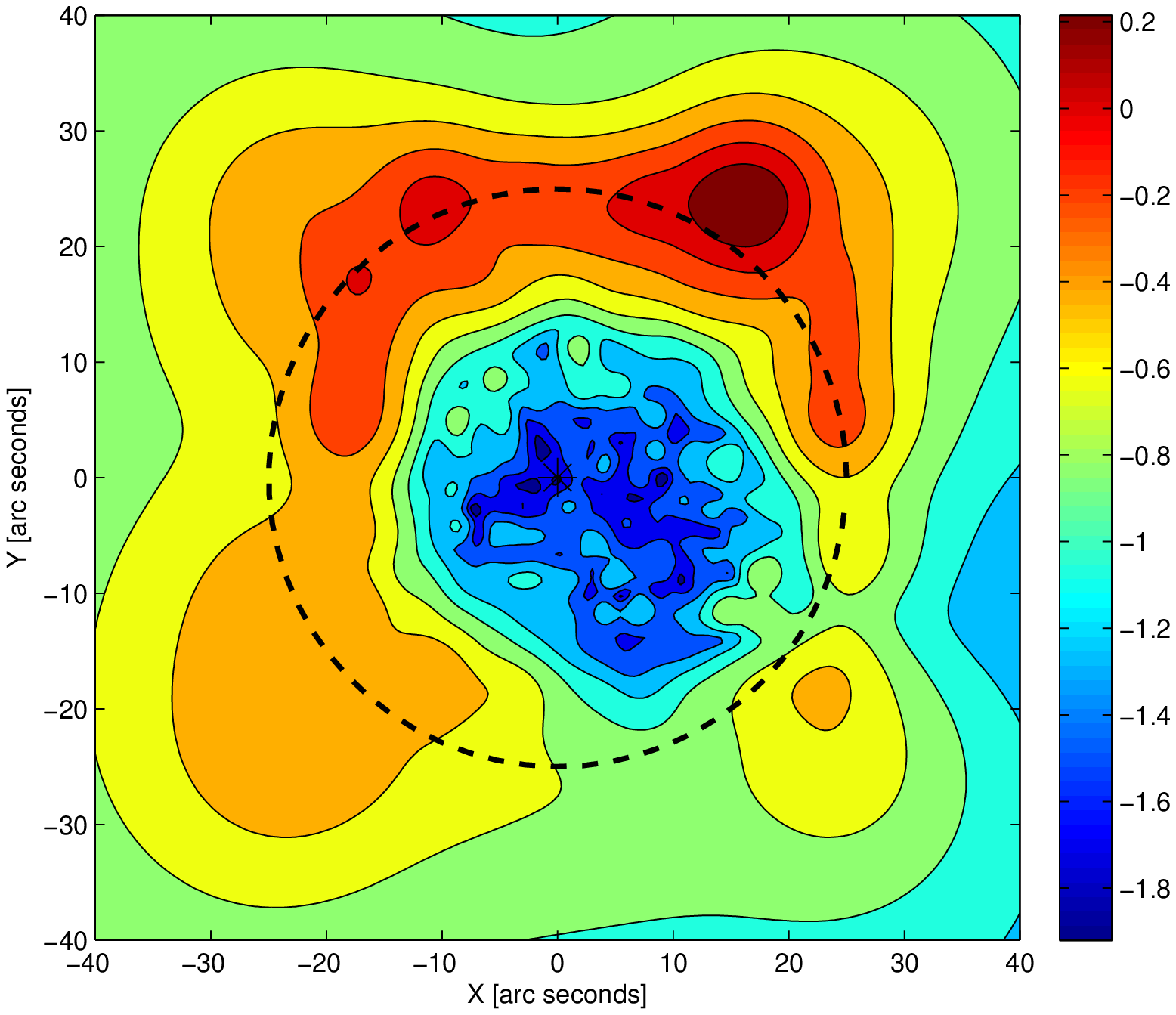}\\

  \includegraphics[width=0.32\hsize]{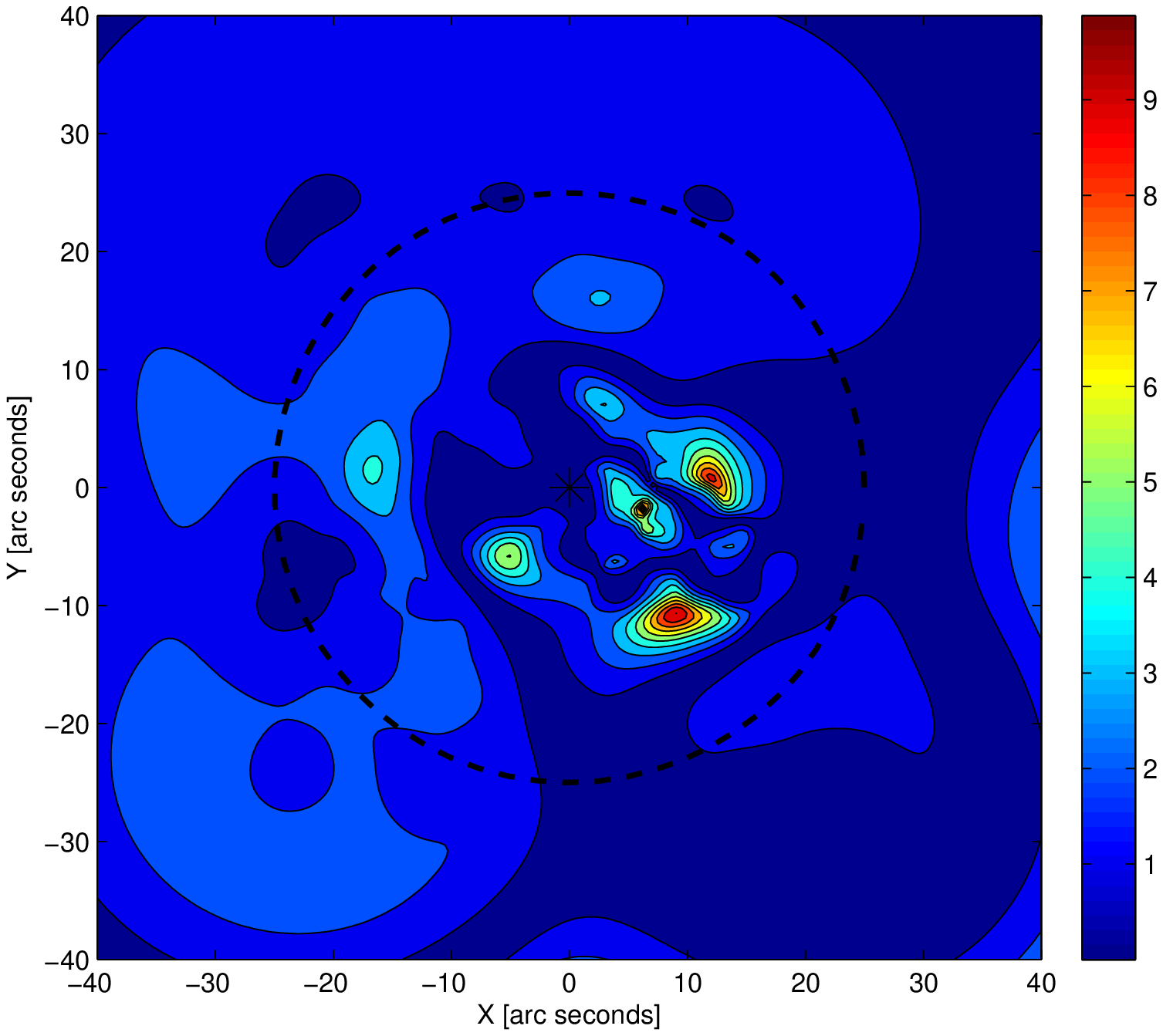}
  \includegraphics[width=0.32\hsize]{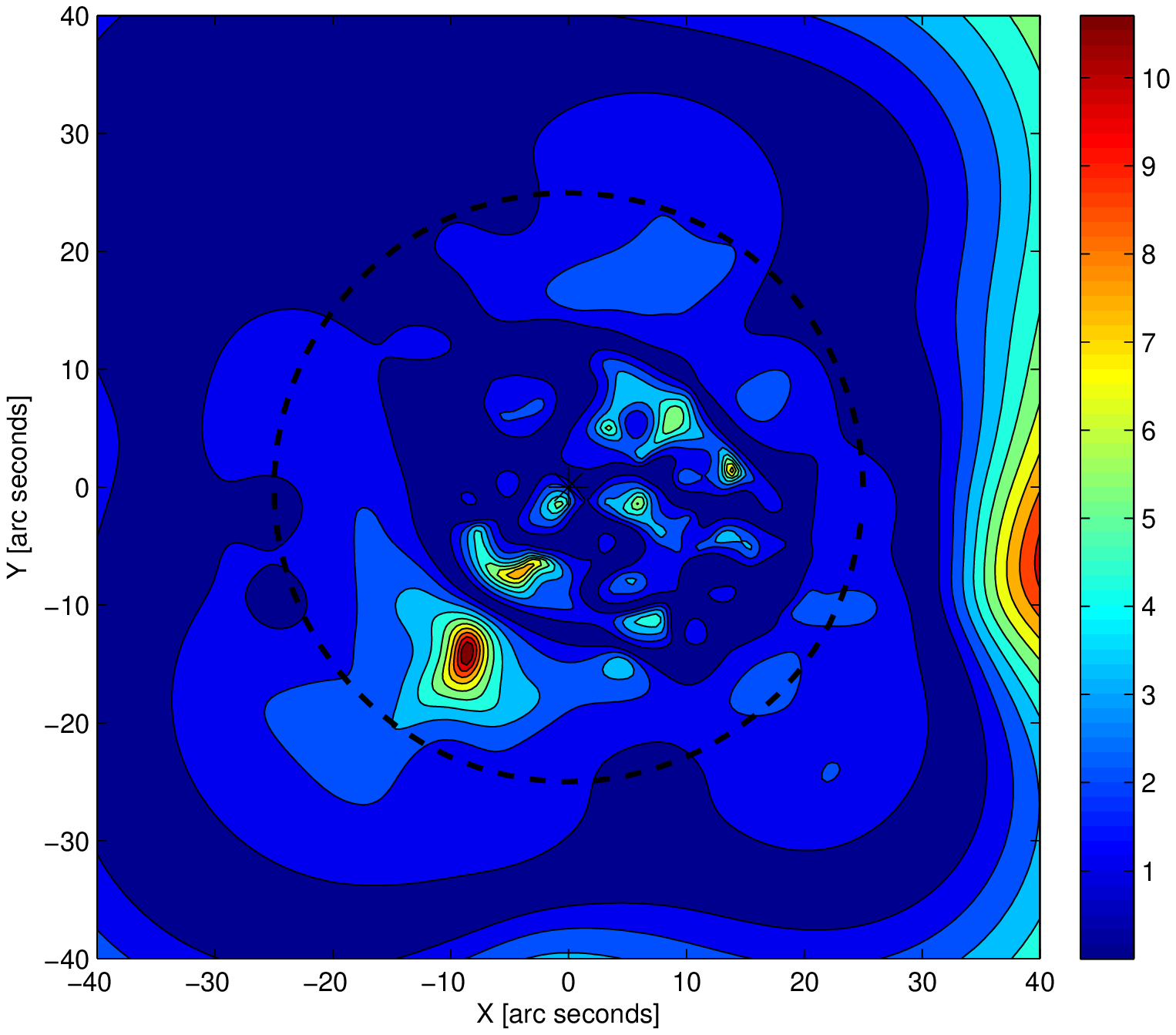}
  \includegraphics[width=0.32\hsize]{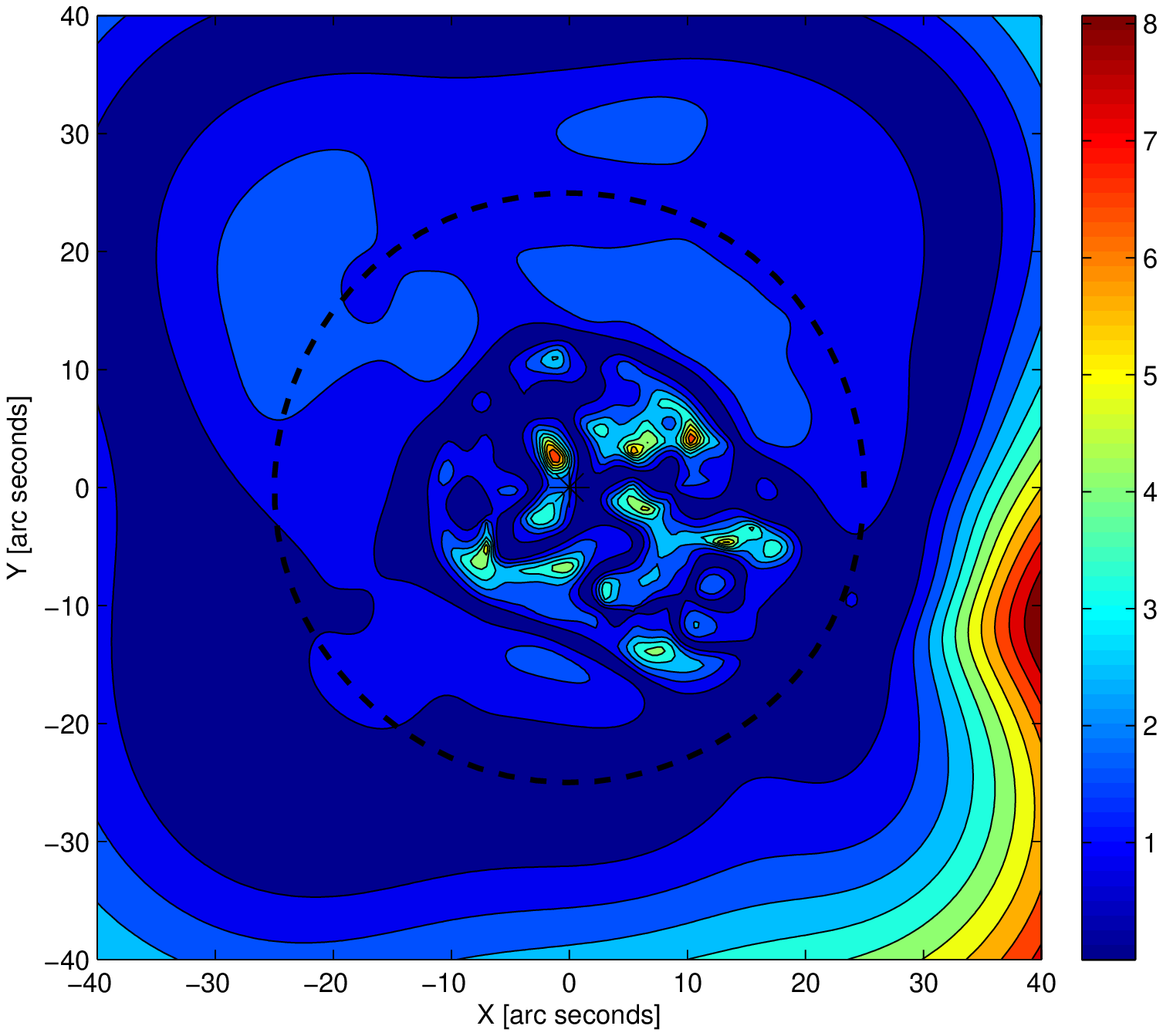}\\

  \includegraphics[width=0.32\hsize]{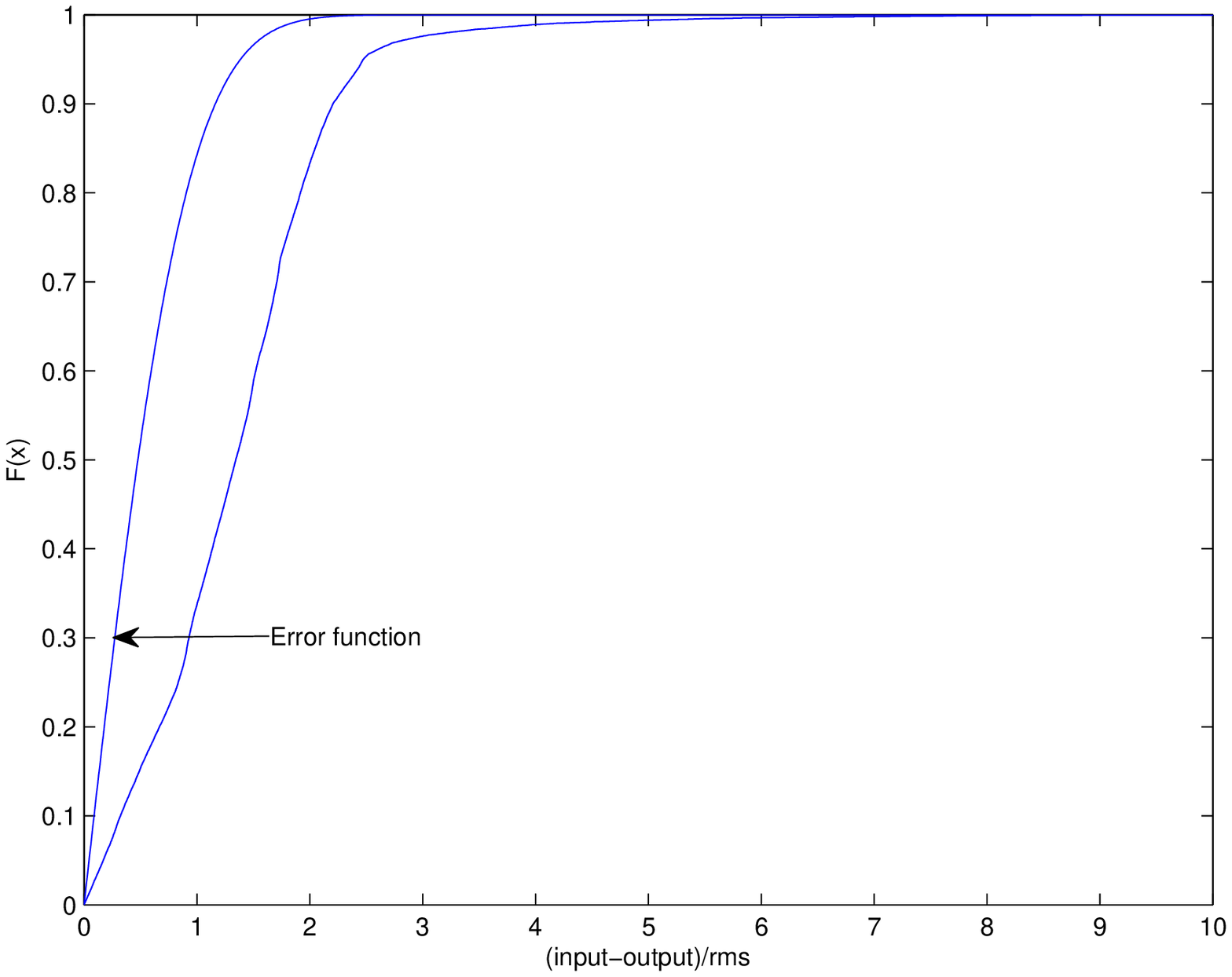}
  \includegraphics[width=0.32\hsize]{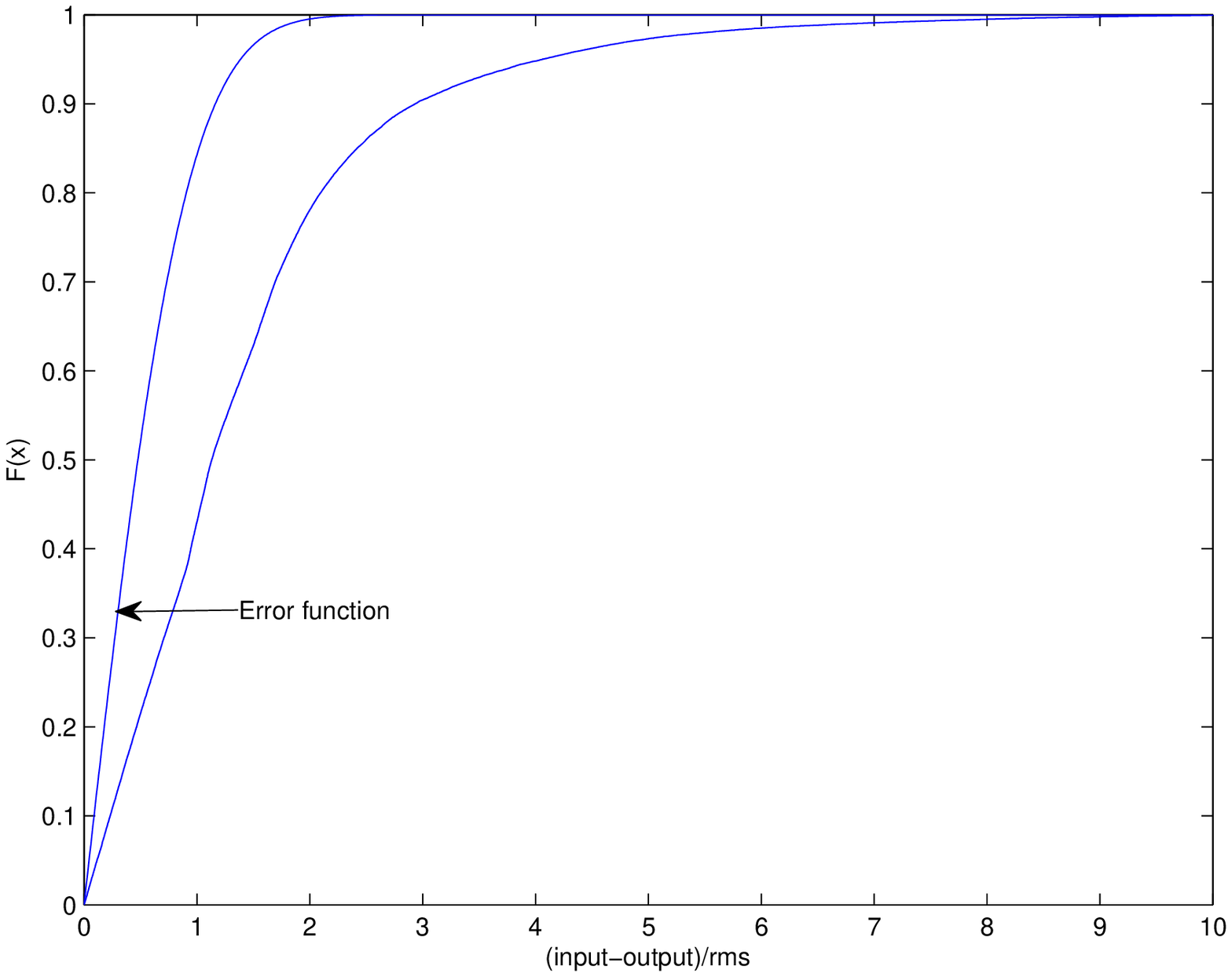}
  \includegraphics[width=0.32\hsize]{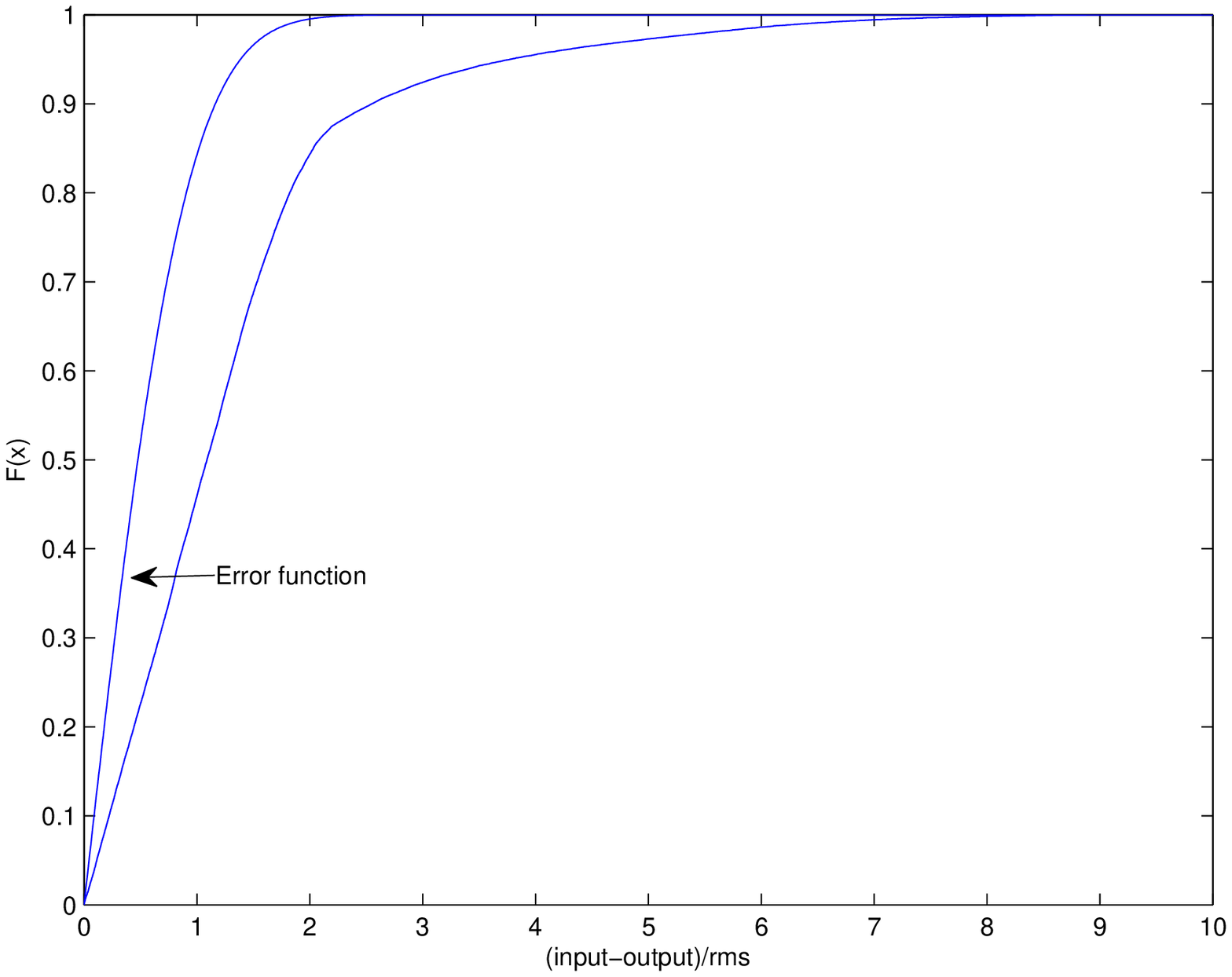}\\
\caption{\label{fig:plummer5out} Reconstruction of the lens in
  Figure~\ref{fig:plummer5in}. Column~1: using three sources only,
  with the corresponding images shown as black triangles; column 2:
  using four sources; column 3: using all five sources.  The top row
  shows average surface mass density $\Sigma$; units are same as in
  Figure~\ref{fig:plummerin}.  The second row shows the fractional rms deviation
 of ten reconstructions, $\delta\Sigma/\Sigma$.  The
  third row contains $\Delta\Sigma/\delta\Sigma$ where
  $\Delta\Sigma$ is the pixelwise difference between the true map and
  the average reconstructed map.  The bottom row shows the cumulative
  $\Delta\Sigma/\delta\Sigma$, along with the corresponding curve
  (marked `error function') for Gaussian errors with dispersion
  $\delta\Sigma$.  We conclude that the error estimate $\delta\Sigma$
  needs to be multiplied by $\sim 2$ (or increased by 0.30 on a
  $\log_{10}$ scale).  The worst cases are some very small regions
  (red in the lower panels) where $\log_{10} \Delta\Sigma$ should be
  increased by $\sim+1$.}
\end{figure}

A simulated lens at redshift 0.1 was next created with five Plummers
positioned such that the configuration resembles the inner region of
Abell 3827.  Sources were put at different redshifts, as
follows.
\begin{enumerate}
\item Three-source case: three sources at $z=0.2$ were were given as input.
\item Four-source case: a fourth source at $z=0.4$ was added.
\item Five-source cases: a fifth source at $z=1.0$ was added.
\end{enumerate}
The resulting images, along with caustics and critical curves, is
shown in Figure \ref{fig:plummer5in}).  Results from these are shown
in Figure \ref{fig:plummer5out}.  The top row of the figure shows the
mass maps $\Sigma$.  The second row shows $\delta\Sigma/\Sigma$, or
the fractional rms deviation.  The third row shows
$\Delta\Sigma/\delta\Sigma$ where $\Delta\Sigma$ is the (absolute)
actual deviation of the reconstructed mass map from the real mass map.  
If $\delta\Sigma$ were close to $\Delta\Sigma$, we
could simply take the rms deviation as the uncertainty.  In fact the
rms deviation under-estimates the true error by about a factor of two.
That can be read off the bottom row of Figure \ref{fig:plummer5out},
which plots the cumulative distribution of
$\Delta\Sigma/\delta\Sigma$.

The main result from this test is that the rms deviation times two is
a reasonable approximation of the errors.  In addition, we can also read off some
qualitative features from Figure \ref{fig:plummer5out}. First, the
spur or handle-like feature to the lower right is recovered in the
lens reconstruction in all cases, even if not perfectly reproduced.
Second, the maps get more accurate as more sources, especially at
different redshifts, are introduced.

We conclude that GRALE is able to find offsets as well as extended
structures (if any) in lenses.

%%%%%%%%%%%%%%%%%%%%%%%%%%%%%%%%%%%%%%%%%%%%%%%%%%%%%%%%%%%%%%%%%%%%%%%%%%

\section{Reconstruction of three real clusters}

In this Section we do mass reconstructions of three galaxy-clusters,
and present these with their accompanying mass error maps. The two
sets of maps for each cluster allow us to judge whether
light-follows-mass (LFM) is a good assumption. We defer the discussion
of the implications of the deviations from LFM to Section~\ref{disc}.

\subsection{Abell 3827}

Abell 3827 is a lensing cluster at redshift $0.099$.  Three multiply
lensed image systems have been identified \citep{2010ApJ...715L.160C}
belonging to three sources at redshift $0.204$, most probably different
parts of the same source. Another big arc is identified belonging to a
source at redshift $0.408$, but its multiply imaged counterpart has not
yet been identified.  A mass map based on these images
\citep{2011MNRAS.415..448W} indicates a dark extended clump, offset by
$\sim 6\rm\,kpc$ from the brightest of the four or five ellipticals in
the cluster core. This offset, if confirmed, would afford us a unique
opportunity to examine and understand the dynamics in dense regions of
clusters.  One of the primary goals of this paper is to assess the
reality of this offset and estimate its statistical
significance. GRALE is a very different lens mass reconstruction
method from the one used in \cite{2011MNRAS.415..448W}, so detecting
the offset with GRALE will lend credence to its reality.

\begin{figure}
  \includegraphics[width=0.33\hsize]{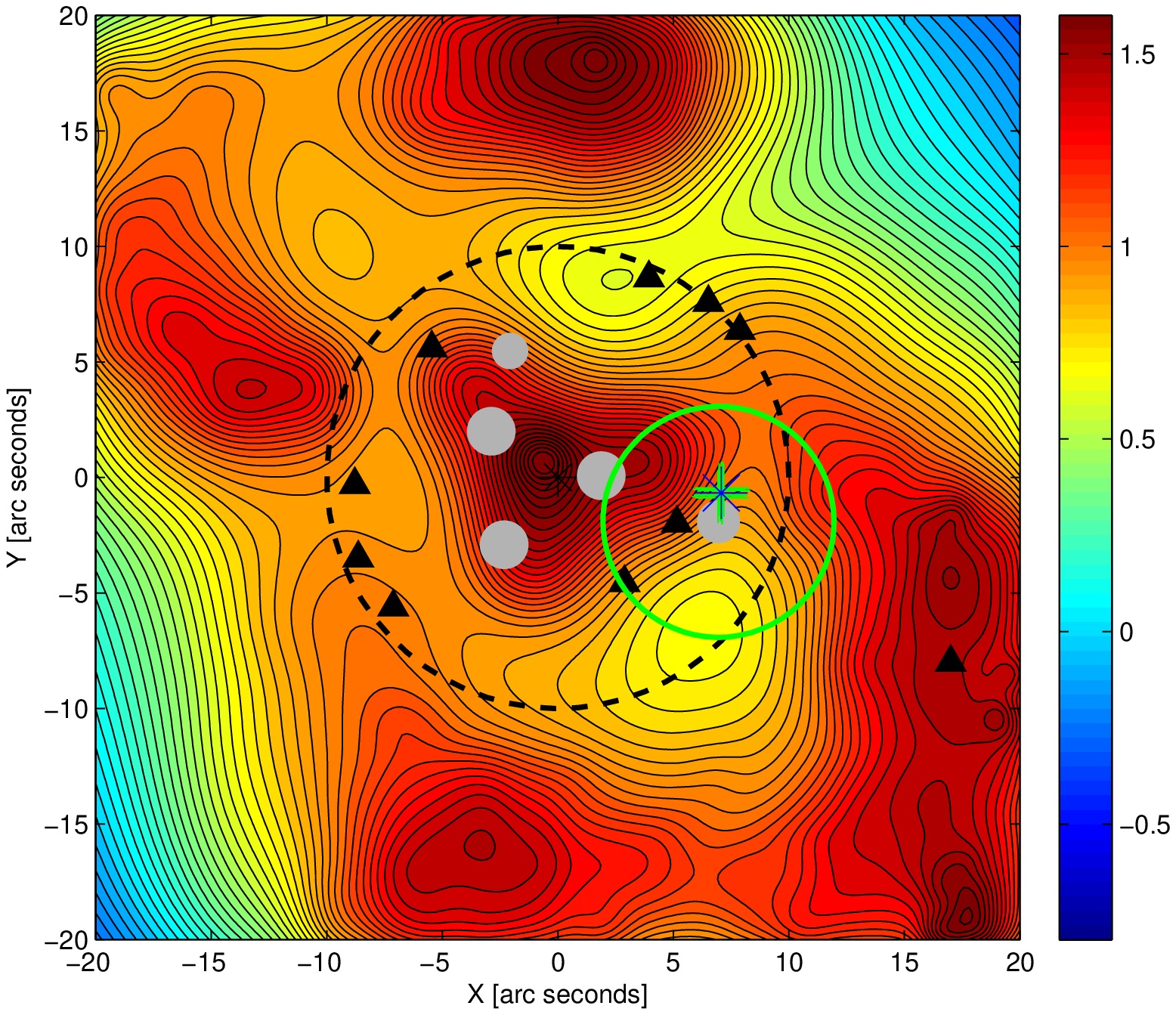}
  \includegraphics[width=0.33\hsize]{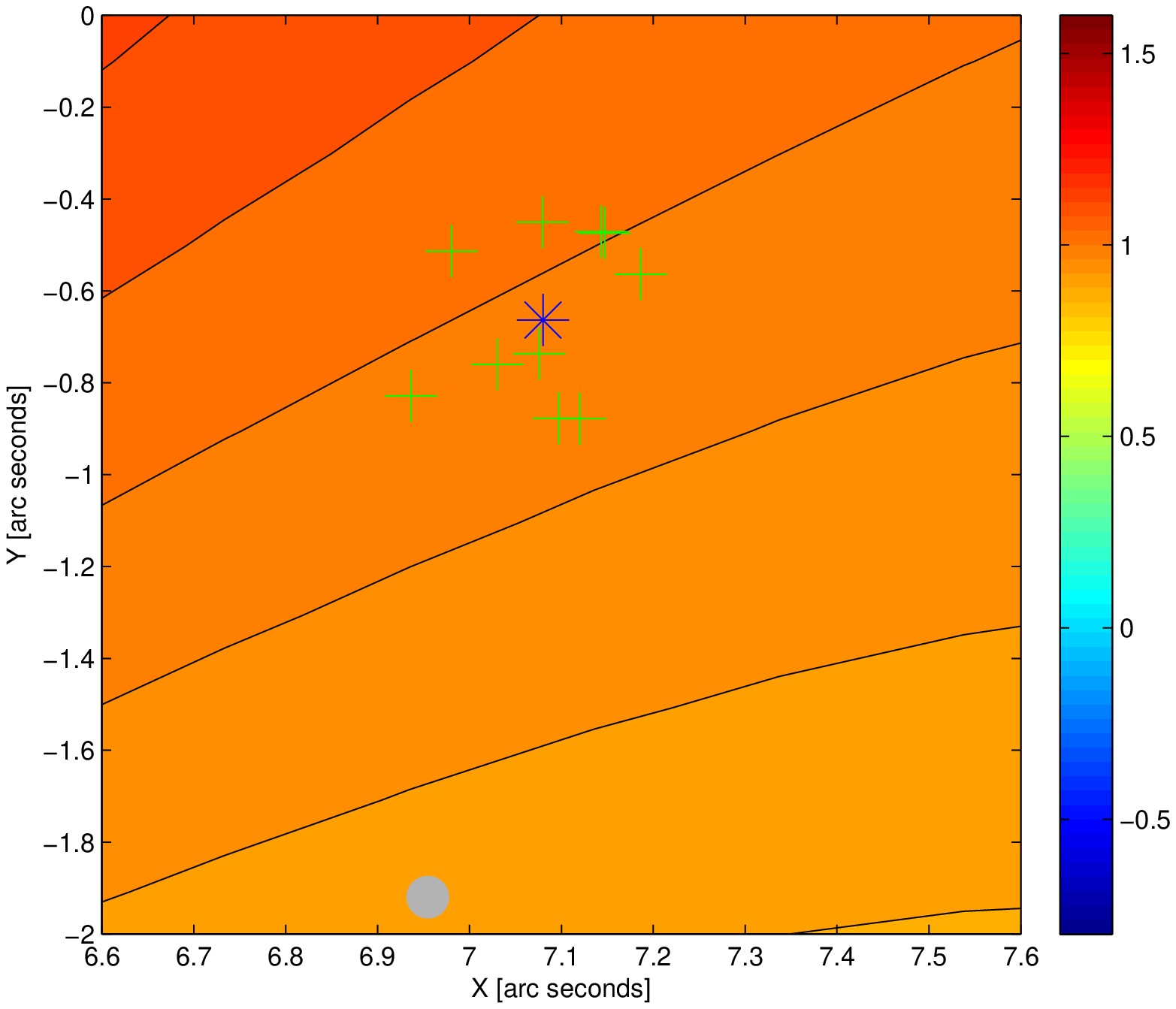}
  \includegraphics[width=0.33\hsize]{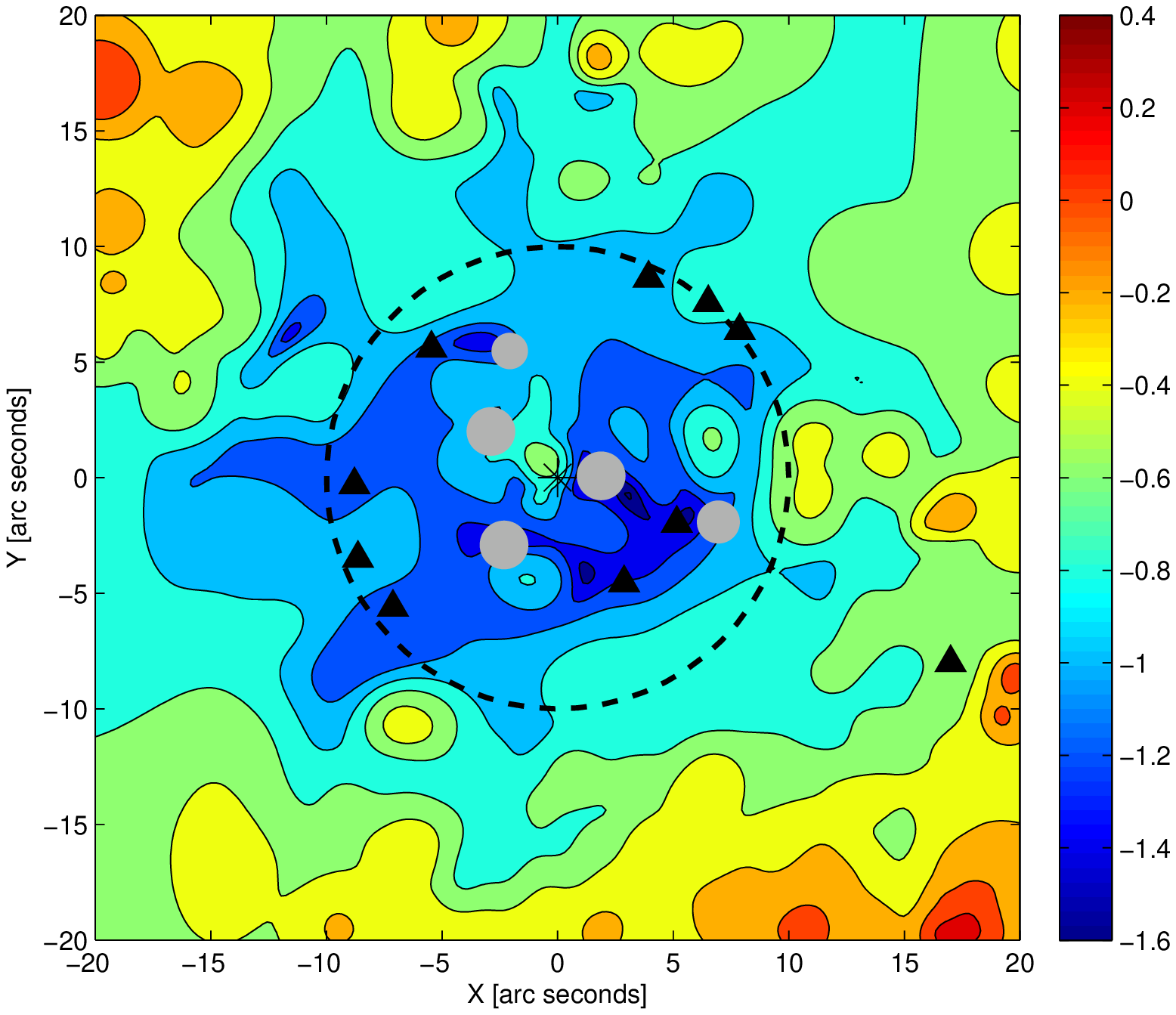}\\
  \includegraphics[width=0.33\hsize]{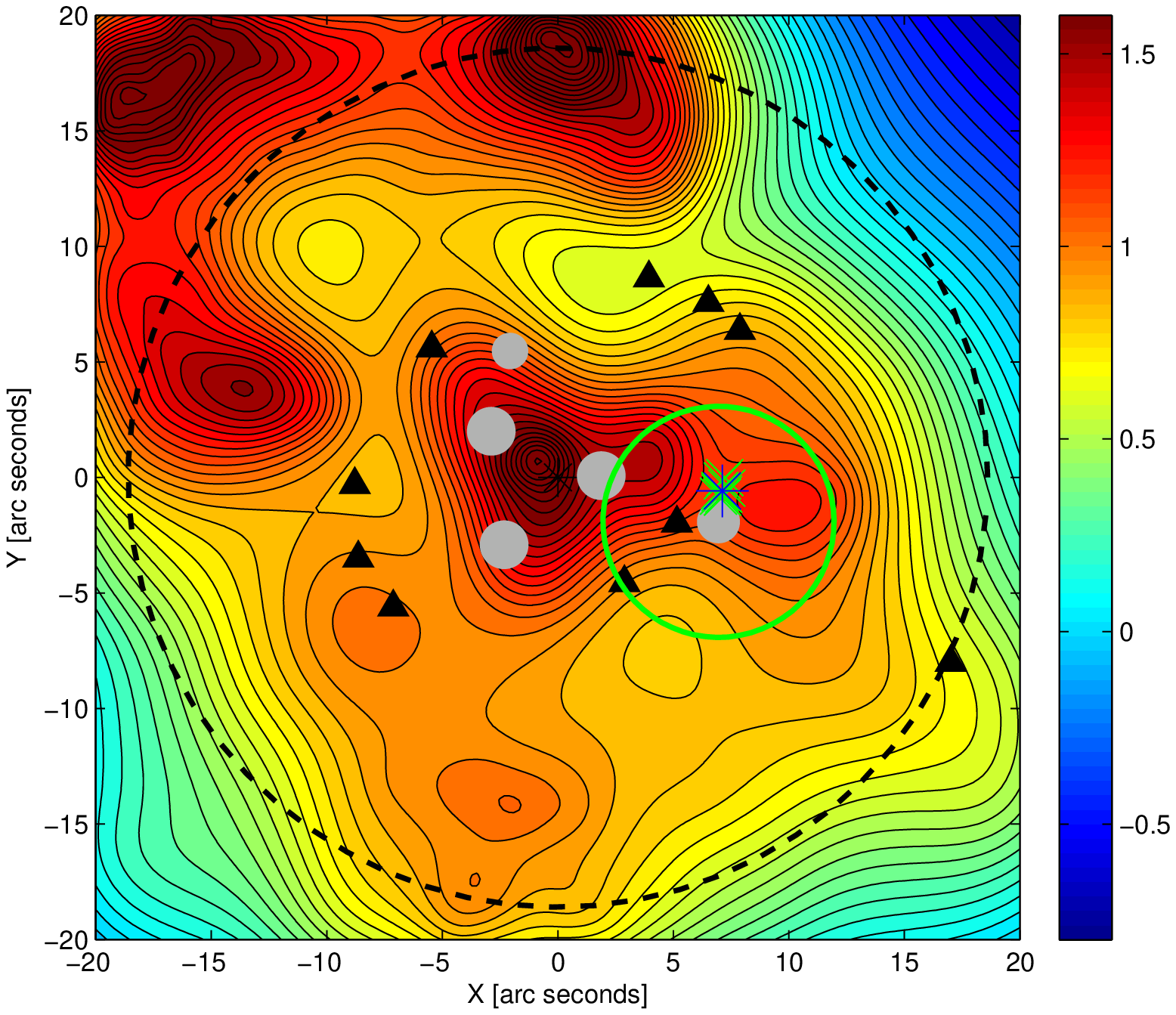}
  \includegraphics[width=0.33\hsize]{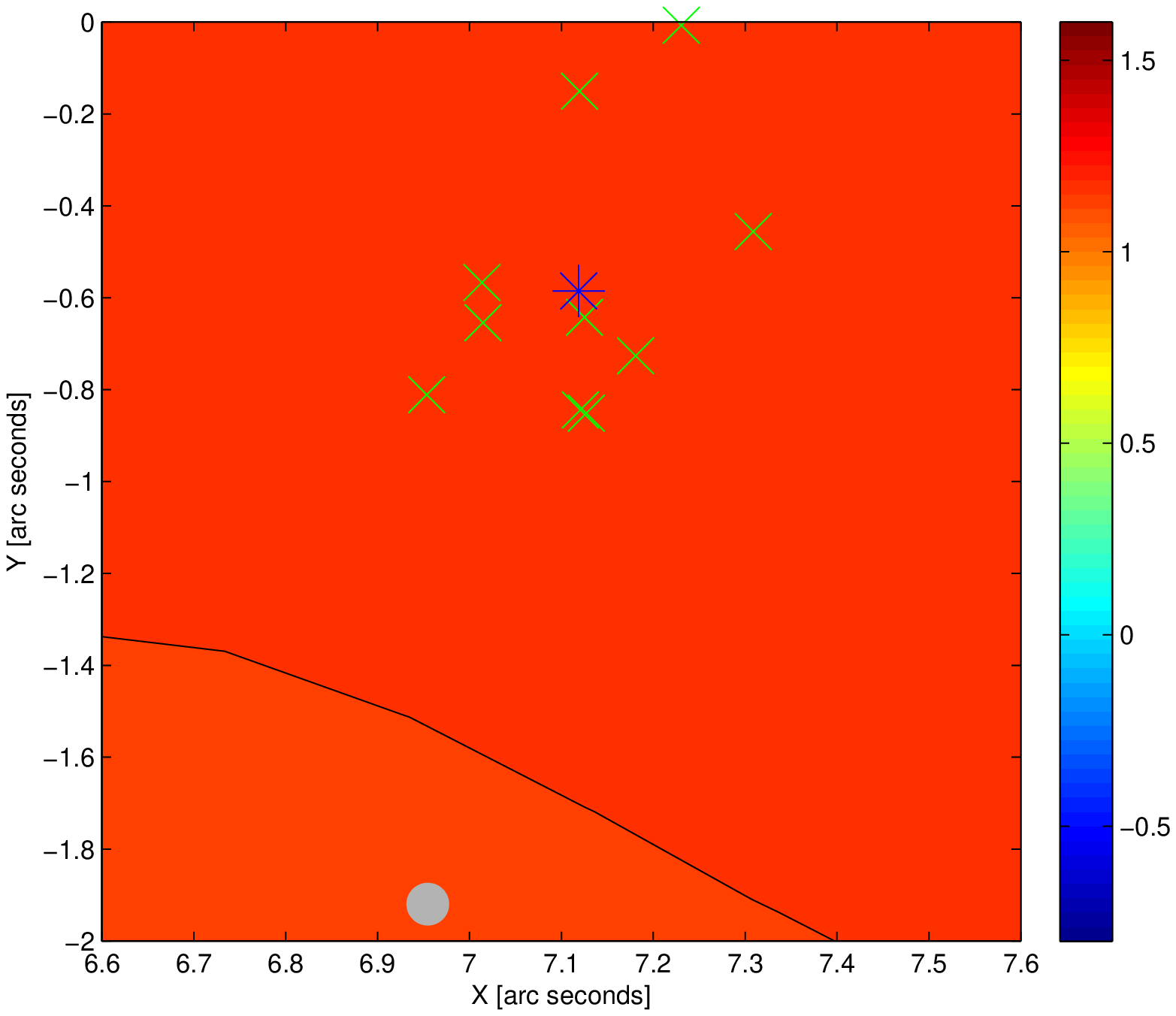}
  \includegraphics[width=0.33\hsize]{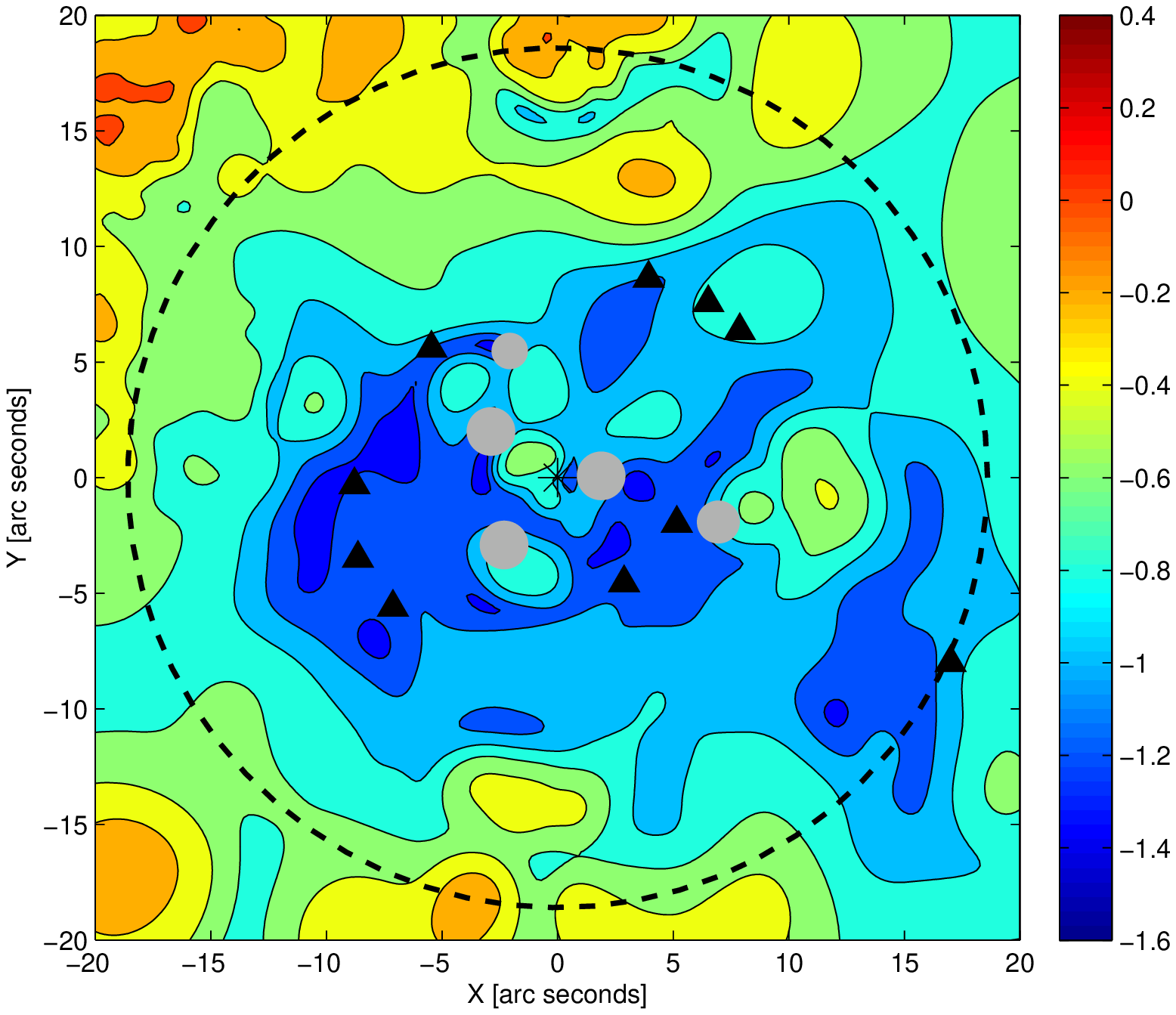}\\
  \includegraphics[width=0.33\hsize]{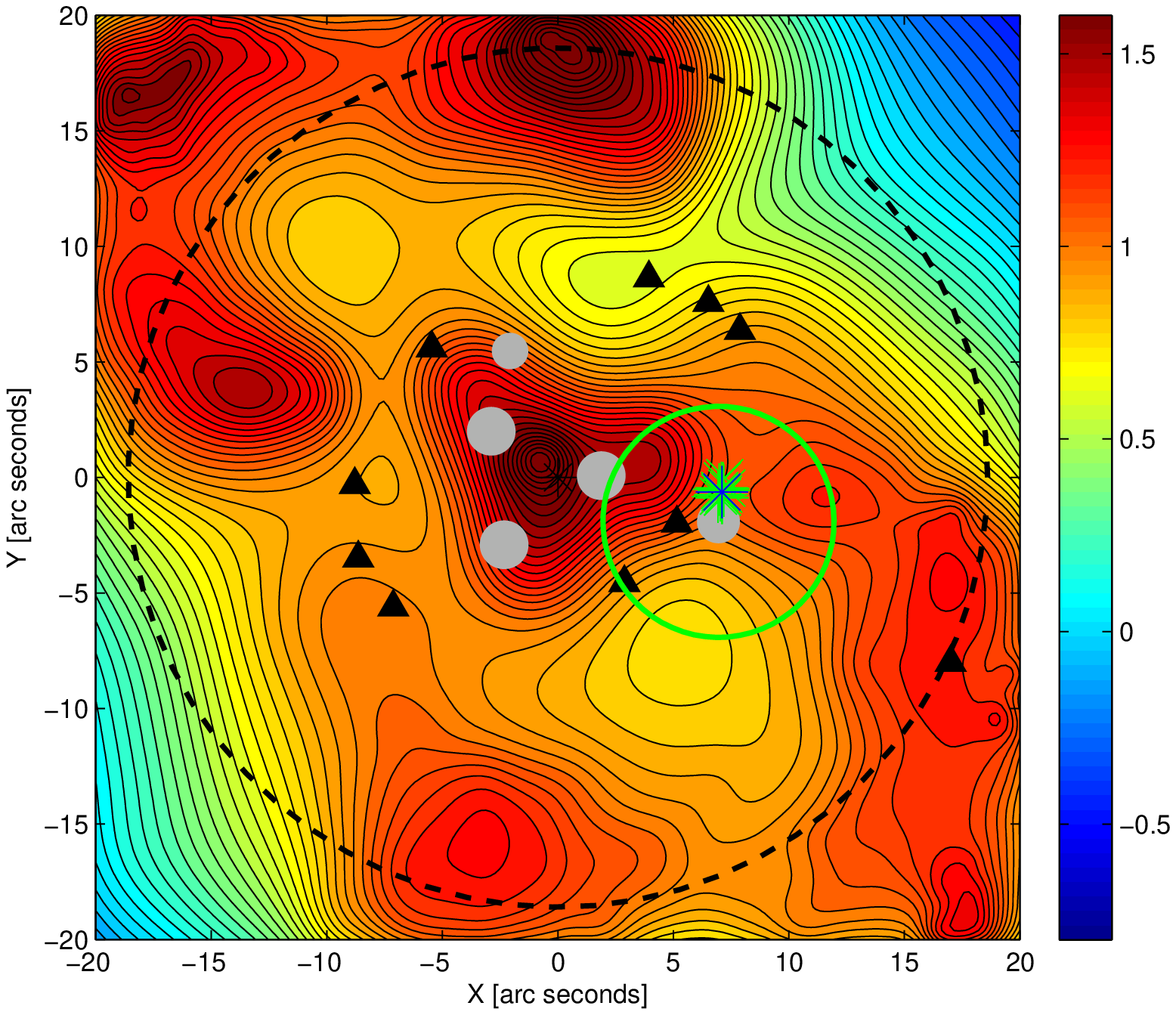}
  \includegraphics[width=0.33\hsize]{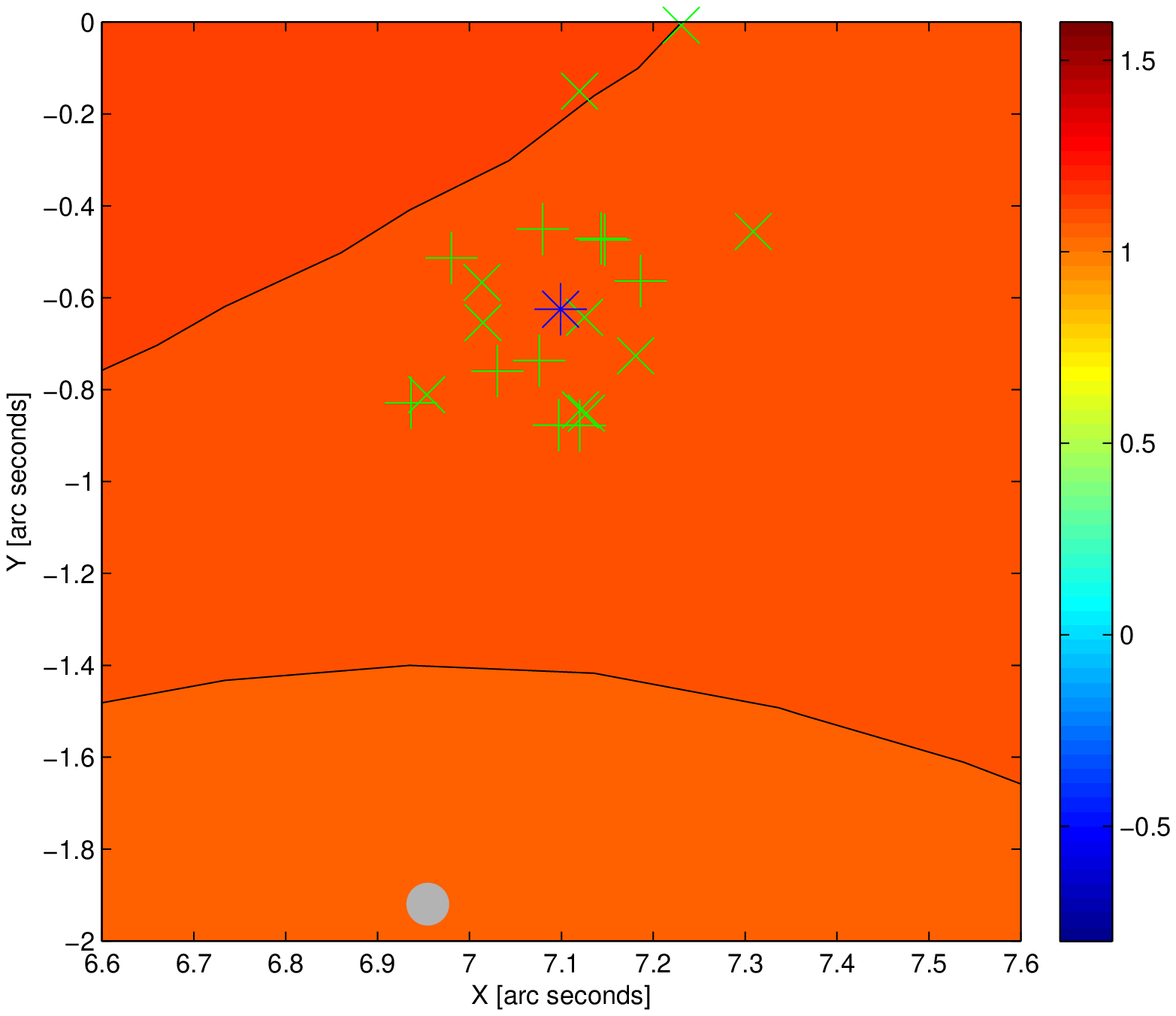}
  \includegraphics[width=0.33\hsize]{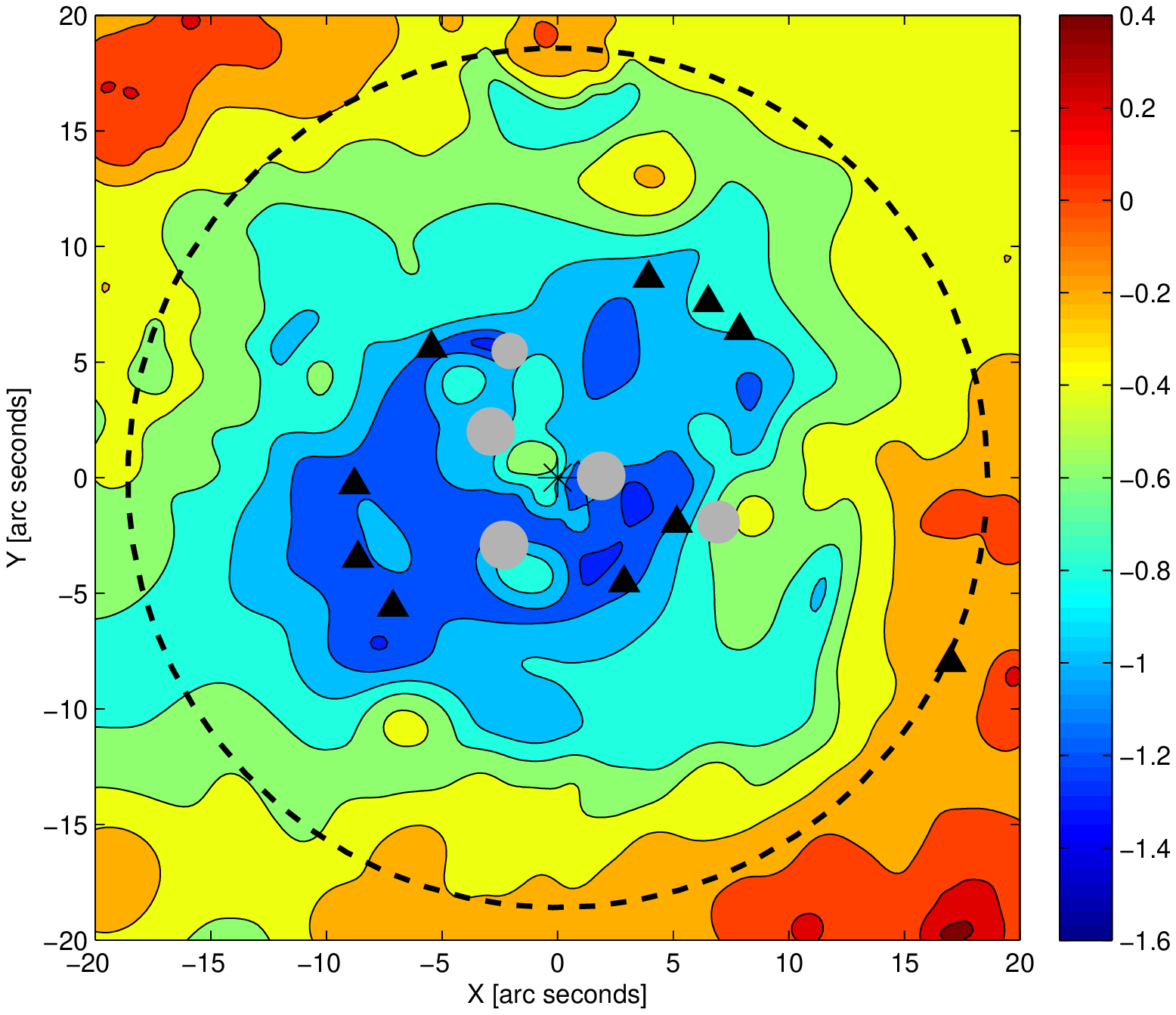}
\caption{\label{fig:a3827} Mass reconstructions of A3827. North is up
  and East to the right. The scale is 1.82~kpc/arcsec. The upper row
  maps are an ensemble of ten maps, each obtained using only the nine
  images of the source at $z_s=0.2$.  The middle row shows an ensemble
  of ten maps, using nine images of the $z_s=0.2$ source and the
  single image at $z_s=0.4$.  The bottom row combines both ensembles.
  The left column presents the average of the ten mass maps. The
  middle column is a zoom centered on the most luminous elliptical
  N1. The ten green `$+$' signs (top row) and `$\times$' signs (middle
  row) represent centroids from ten individual maps of the mass within
  the green circle shown in the left column. The grey dot towards the
  bottom of the plots (in the middle column) is N1.  The blue asterisk is the centroid of the
  average of the ten realisations.  The right column shows the fractional rms deviation
 between the ten maps, $\delta\Sigma/\Sigma$.}
\end{figure}

Using the identified images we reconstructed the mass distribution in
two ways, and then combined the results.  These are displayed in the
three rows of Figure \ref{fig:a3827}.

First, we used the three image systems belonging to the sources at
redshift $0.2$.  The first panel of the top row of Figure
\ref{fig:a3827} shows a spur in the mass map, which is offset from the
nearby elliptical galaxy (the right most of the five grey dots). The
spur's location is similar to the location of the local overdensity
reported in \cite{2011MNRAS.415..448W}, so the offset is similar in
both reconstructions. From the map of fractional rms deviation
$\delta\Sigma/\Sigma$ (right panel of the first row) the spur appears
to be significant; the rms deviation in that region is about $0.1$ kg
m$^{-2}$, and so the fractional error is about 10\%. Since the
structure appears to be extended and not a single clump, it is not
obvious how to quantify it.  We can nonetheless test its significance.
We chose a circle of radius $5''$ (green circle) around the nearby
elliptical.  (The choice of size is somewhat arbitrary; other choices
would also serve our purpose.)  We then calculate the centre of mass
within this circle, for each mass map within the ensemble, and mark
them with green `$+$' signs in the middle panel of top row, which is a
zoom on to the relevant region.  All ten centroids are consistently
displaced from the nearby galaxy (grey circle), by about $1.2''$. The
average of the ten centroids is marked with a blue star symbol.  We
may interpret these results as a hypothesis test.  The null hypothesis
is that the cluster has no mass/galaxy offset, and the mass is centred
on the galaxy light.  A mass reconstruction could nonetheless put the
aperture centroid displaced from the galaxy, simply from the
stochastic element in the genetic algorithm --- note that the mass
reconstructions are not given any information about the cluster
galaxies.  If there is no mass offset, the model offsets would be
random, and the change of all 10 mass reconstructions having an offset
in the same direction would be only 10\%.  But the aperture centroids
are consistently offset in the same region.  Hence there does appear
to be an offset, significant at 90\% confidence, between the mass spur
and the galaxy.

Second, we used all four image systems: three belonging to the sources
at redshift $0.2$ and one with source redshift $0.4$. As mentioned
before, no image counterpart of the latter has been identified, but
there is a possibility of such a counter-image near the centre of the
cluster.  Accordingly, we allowed GRALE to produce extra images in
that region.  The corresponding mass maps are shown in the second row
of Figure~\ref{fig:a3827}. This time the extent of the image region is
larger, and the fraction rms between reconstructions (right panel) is smaller
in the general region of the image at $z_s=0.4$.  A clear mass subpeak
is seen near the elliptical, offset from it by $\sim4''$ or $\sim 7
\rm\,kpc$.  To be consistent with the previous case, we again
calculate the centre of mass, or centroid, in a circular region of
radius $5''$.  Individual centroids are marked with green `$\times$'
signs, and their average is the blue star. Again the offset is detected
at a significance similar to the one above.

Finally, we then combined the two sets of ensembles described above,
for a total of twenty individual maps.  The bottom row of
Figure~\ref{fig:a3827} shows the average mass map, and the map of
$\delta\Sigma/\Sigma$ for the combined ensemble.  The conclusion
remains unchanged.

\subsection{Abell 2218}

Abell 2218 is a well known and much studied lensing cluster
\citep[e.g.,][]{1998AJ....116.1541A}.  Like other rich clusters, it
has been used in the recent years as a cosmic telescope
\citep{2010A&A...518L..17A,2010ApJ...716L..45H,2010ApJ...709..210K} to
get a better view of distant or faint galaxies.  The strong lensing
region is somewhat larger on the sky than in Abell 3827, and the
greater redshift, $z_l=0.175$, implies a larger physical scale, $3$
kpc arcsec$^{-1}$.

\begin{figure}
  \includegraphics[width=0.32\hsize]{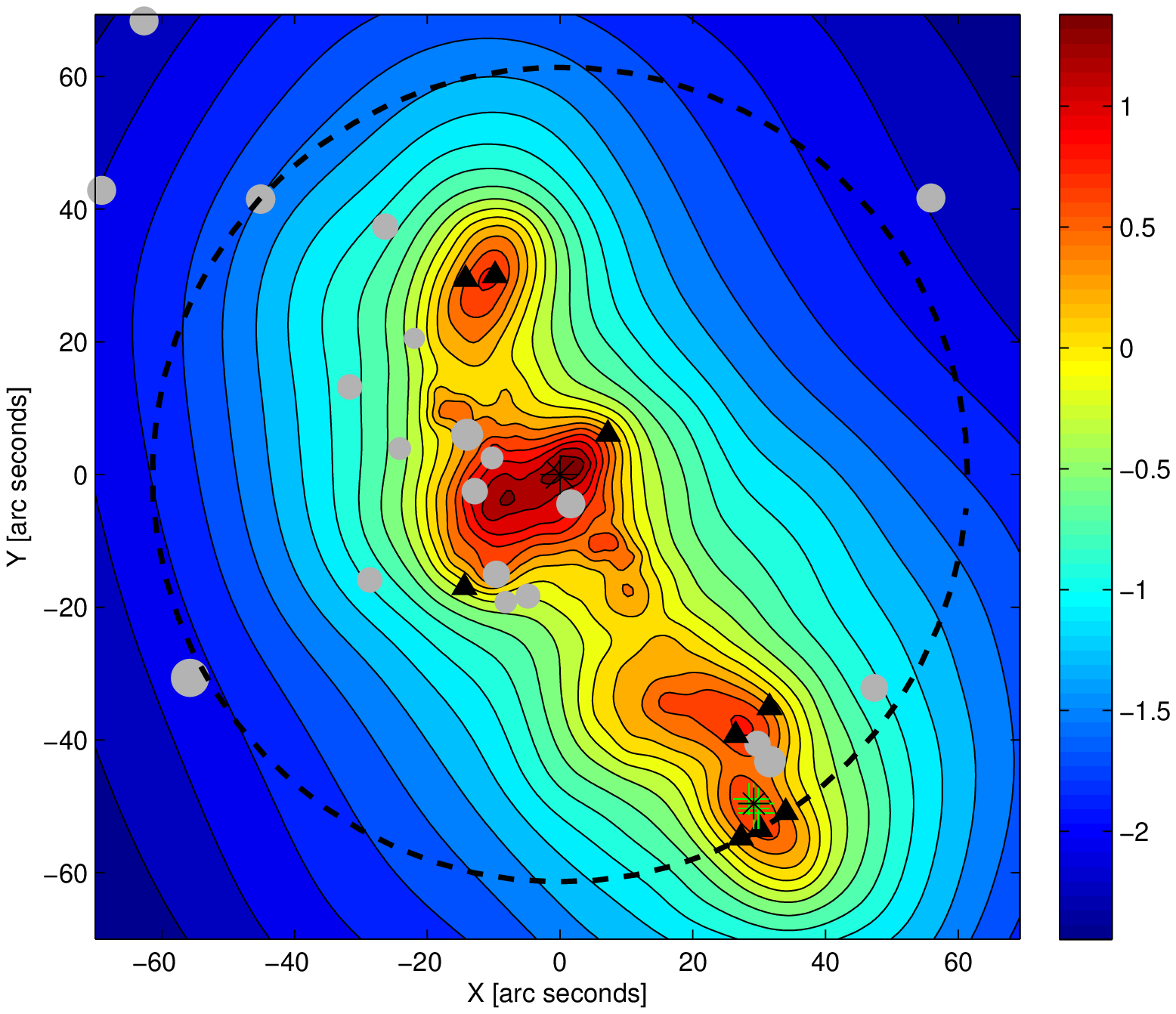}
  \includegraphics[width=0.32\hsize]{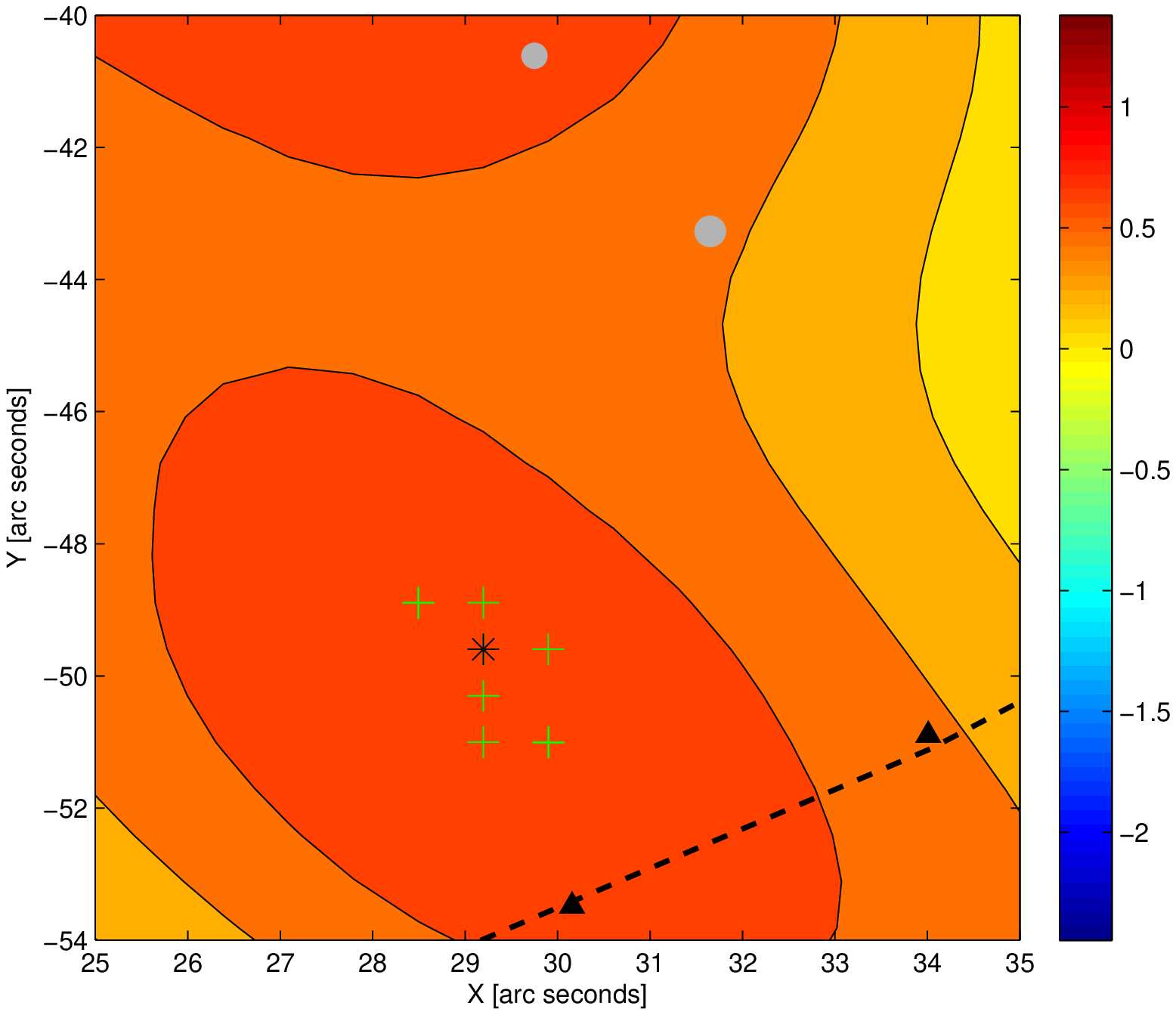}
  \includegraphics[width=0.32\hsize]{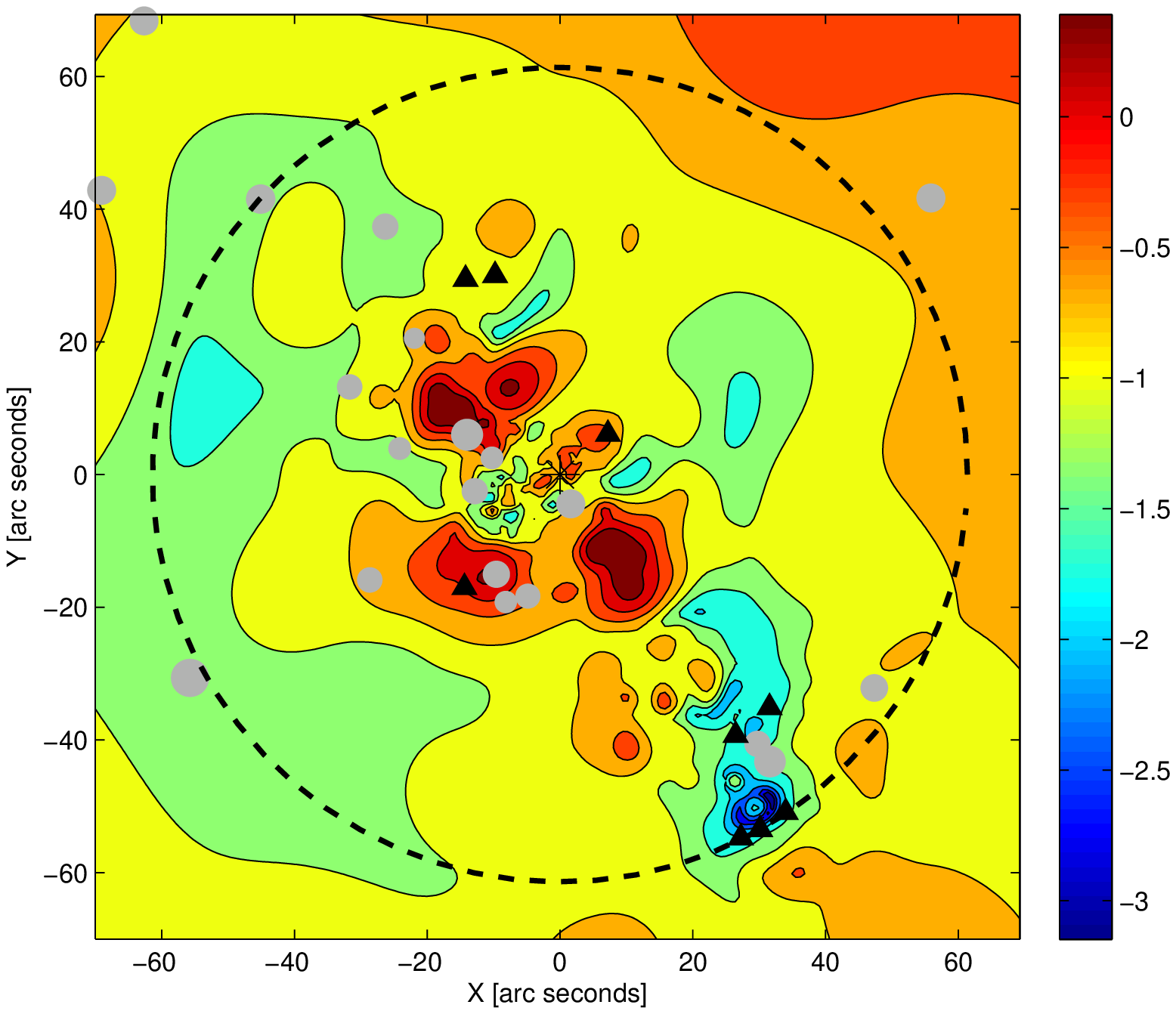}
\caption{\label{fig:a2218} Mass map of A2218.  North is up and East to
  the right.  The average mass map (left column) and fractional rms (right column) are
  based on ten realisations. The central column shows the zoom of the
  region with mass-light offsets, and the green '+' signs are the
  local mass peaks from individual reconstructions. The scale is
  3~kpc/arcsec.  Galaxies with $R<20$ \citep{1992A&A...266....6P} are
  marked with grey dots.}
\end{figure}

We reconstructed the cluster using the four most secure strong lensing
systems.  Figure~\ref{fig:a2218} shows the mass map (left panel) and
fraction rms dispersion between the ten individual maps of the ensemble 
(right panel).  While apparent offsets are visible between galaxies 
(grey dots) and mass in the central region of the cluster, these are not
significant, because rms in that region is comparable to the typical
value of the surface mass density.  Significant offsets are seen
around the lower right mass clump, where the rms dispersion between
mass maps is low. In the central panel we show a zoom of that region,
similar to that in the middle panel of Figure~\ref{fig:a3827}.  The
green `$+$' signs represent the local mass peaks (not centroids as in
the case of A3827) of individual reconstructions, which are displaced
from the nearest cluster galaxies, represented by grey dots in the
upper right of that panel.

\subsection{Abell 1689}

\begin{figure}
  \includegraphics[width=0.32\hsize]{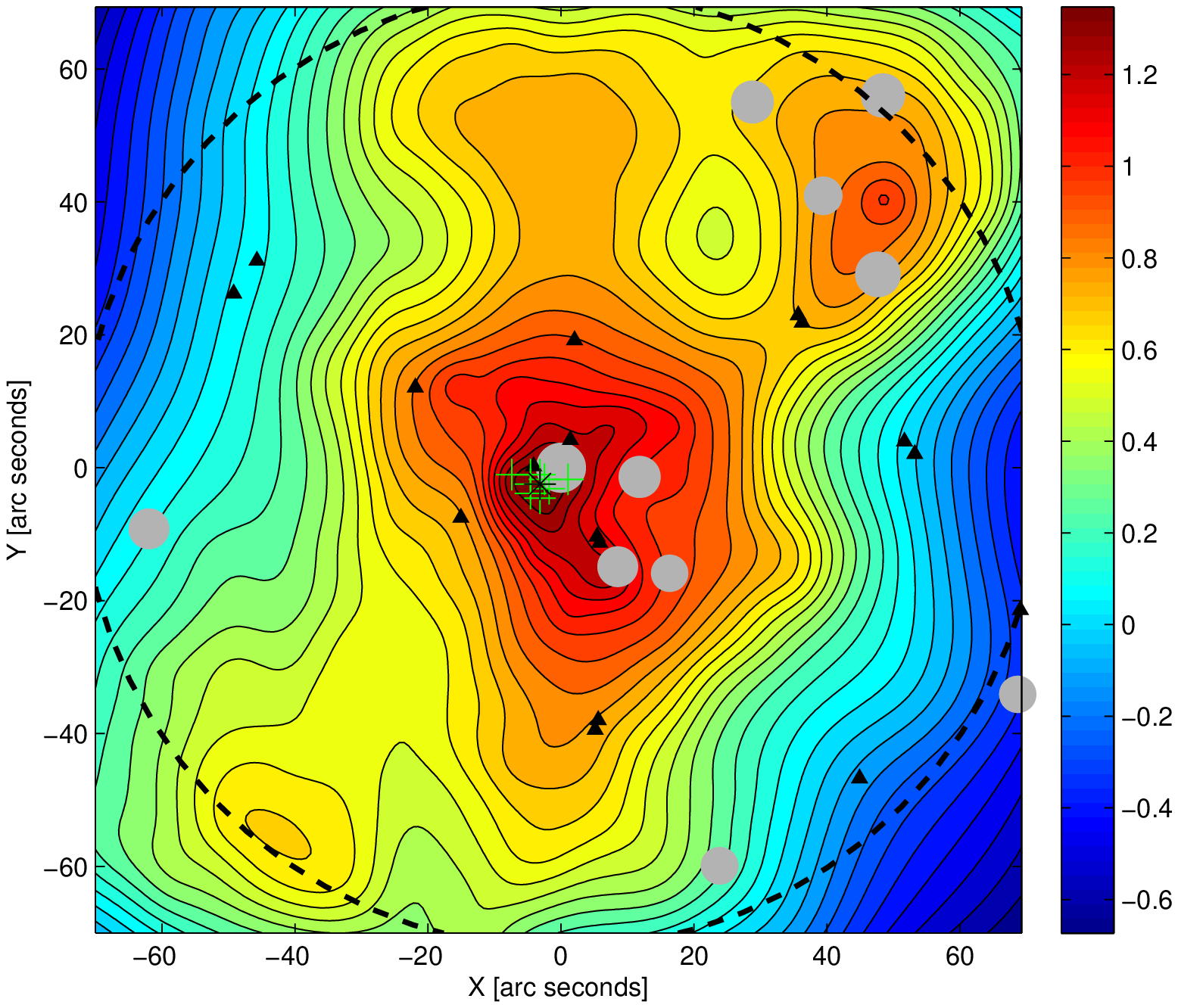}
  \includegraphics[width=0.32\hsize]{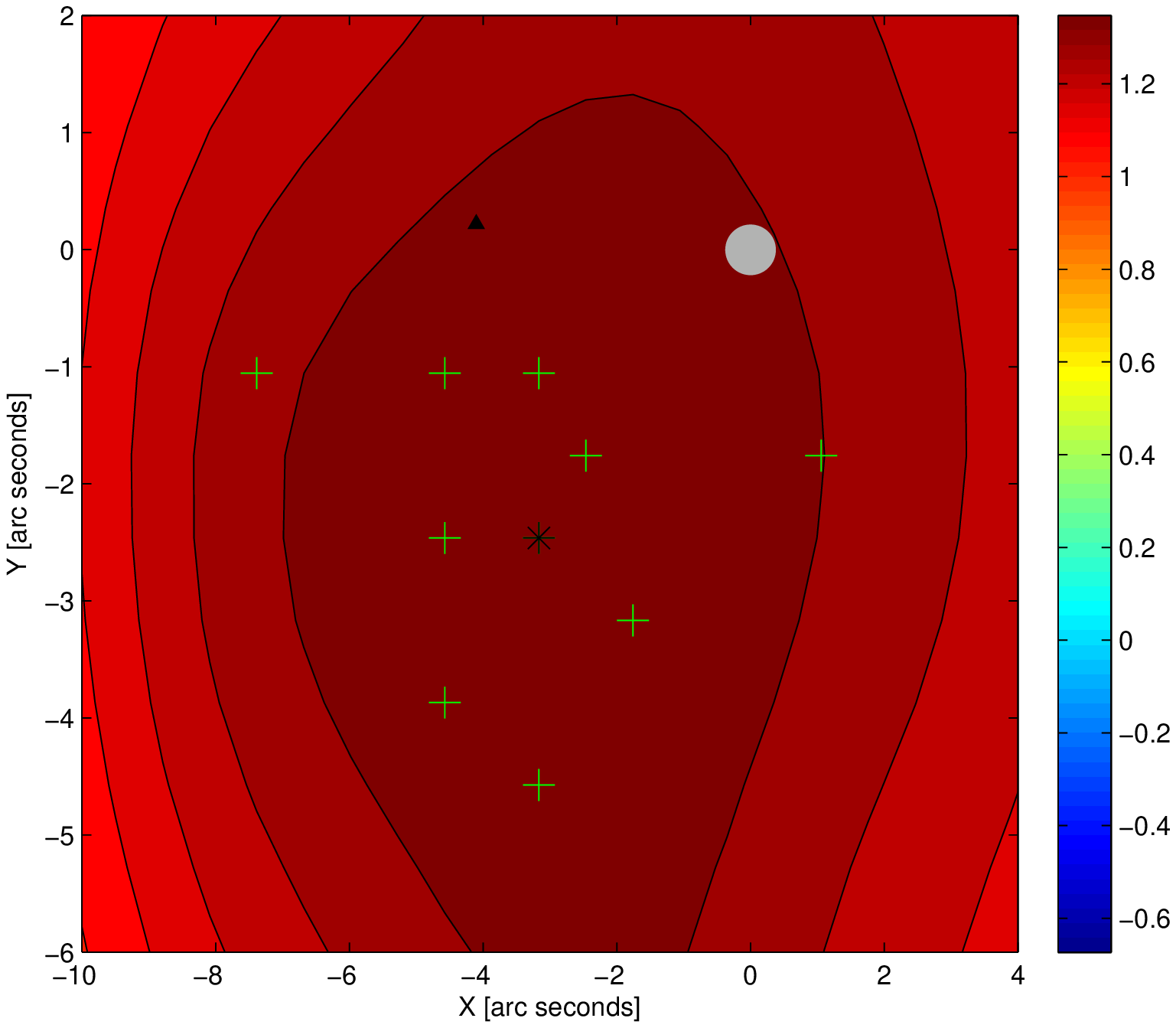}
  \includegraphics[width=0.32\hsize]{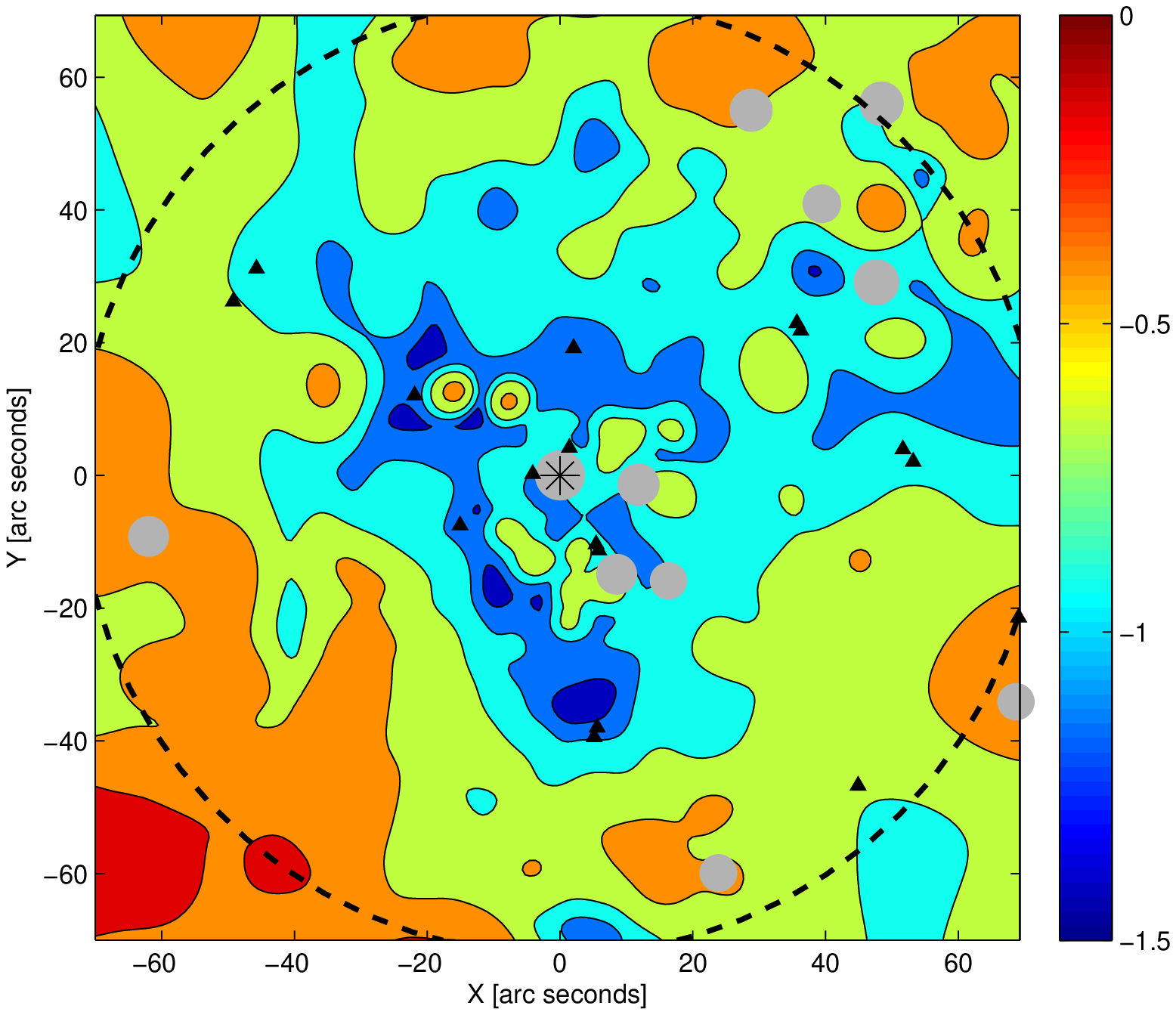}
 \caption{\label{fig:a1689} Mass maps of A1689.  North is up and East
   to the right. The columns are similar to those in
   Fig.~\ref{fig:a2218}. Galaxy positions \citep{2002A&A...382...60D}
   also marked.}
 \end{figure}

Abell 1689, at redshift $0.183$, is perhaps the best known lensing
cluster, containing over a hundred lensed images from at least thirty
background sources extending to high redshifts
\citep{2005ApJ...621...53B}.  Our reconstruction of its mass is shown
in Figure \ref{fig:a1689}.  As with Abell 2218, the mass map and the
rms maps are in the left and right panels. There are no significant
mass/light offsets in this cluster. To illustrate that, in the central
panel we show a zoom into the central region, where the mass peaks of
the ten individual maps are shown as green '+' symbols. Their
distribution with respect to the central cluster galaxy (grey dot) is
consistent with the two being coincident.

\begin{figure}
  \includegraphics[width=0.32\hsize]{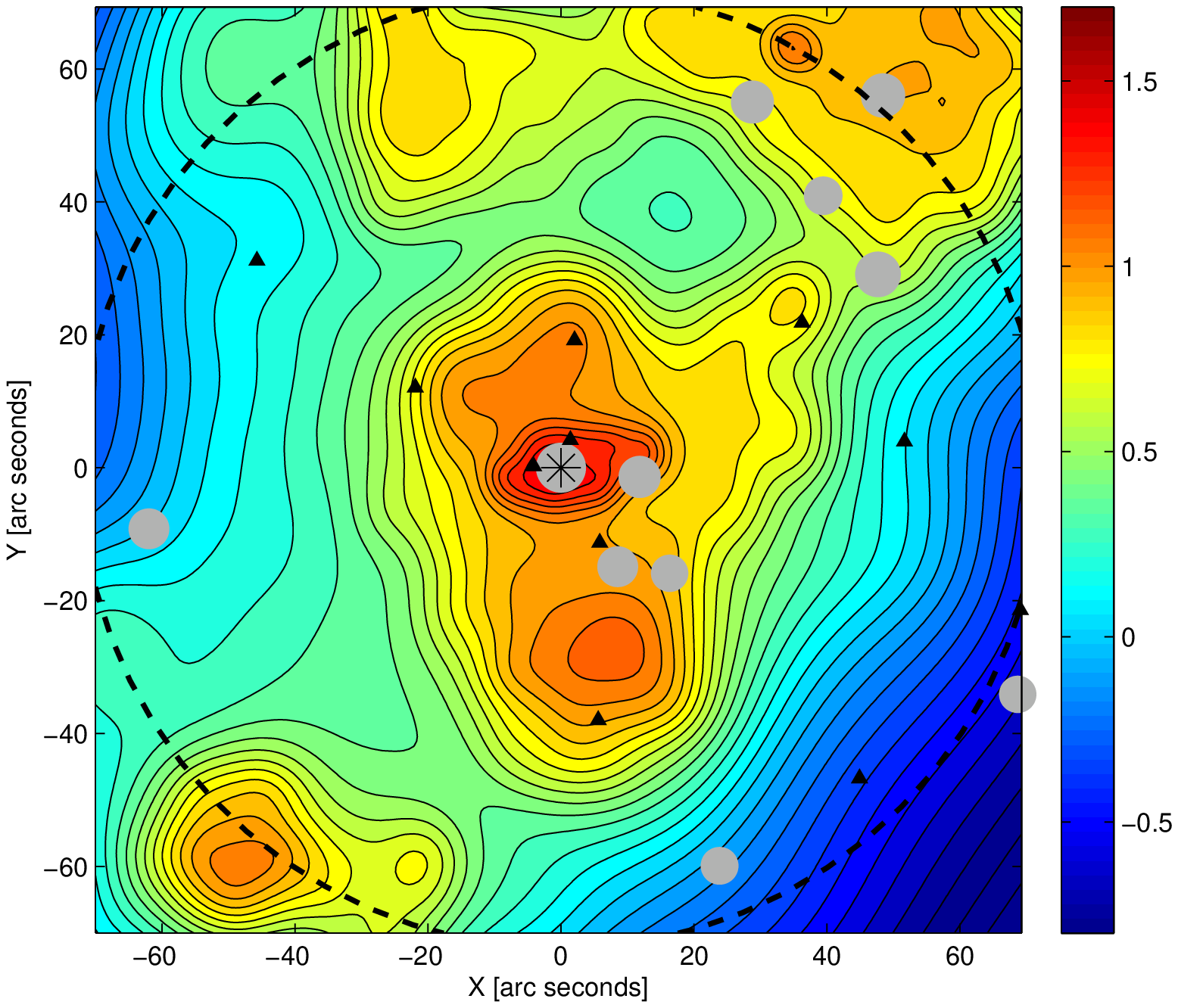}
  \includegraphics[width=0.32\hsize]{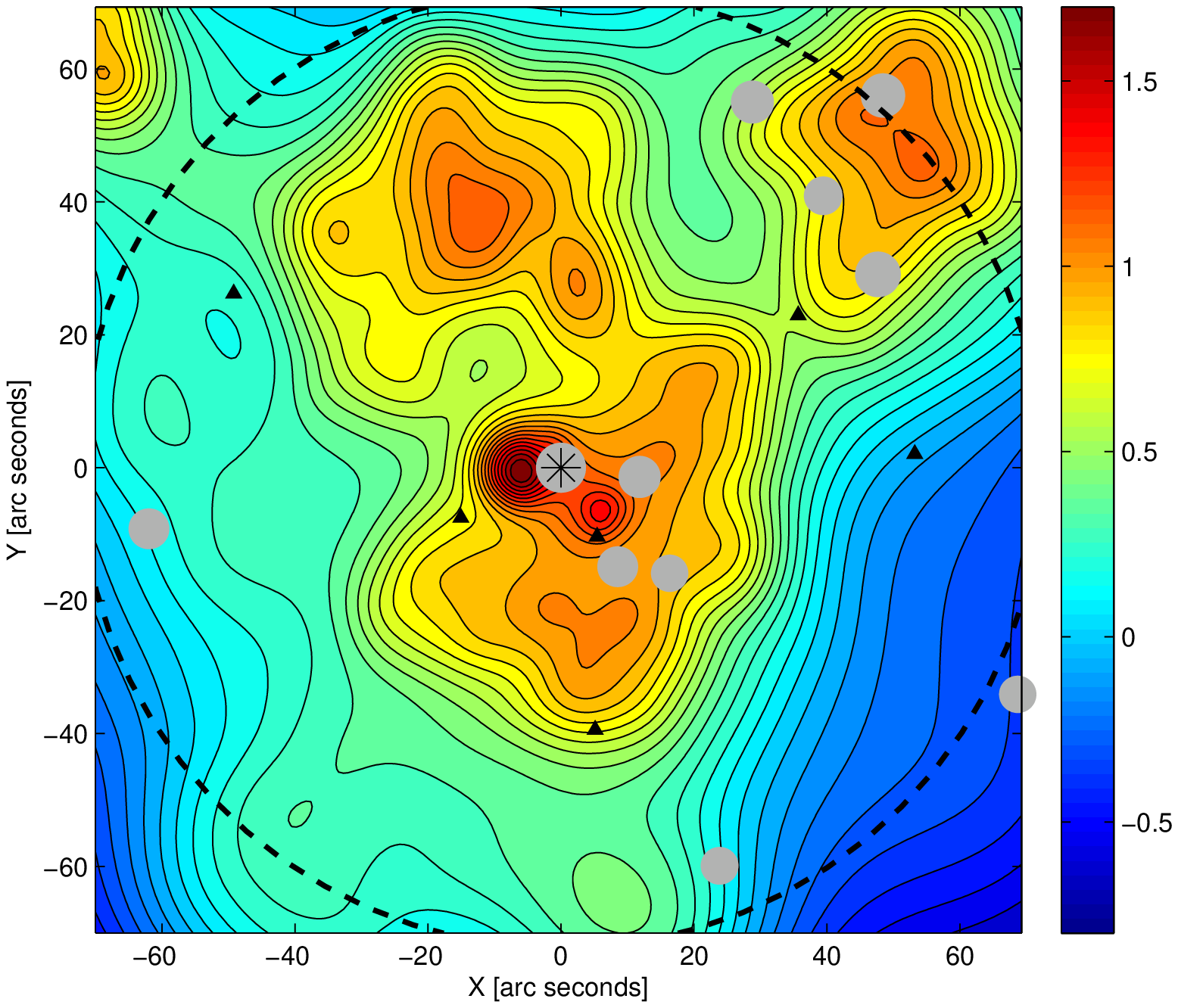}
  \includegraphics[width=0.32\hsize]{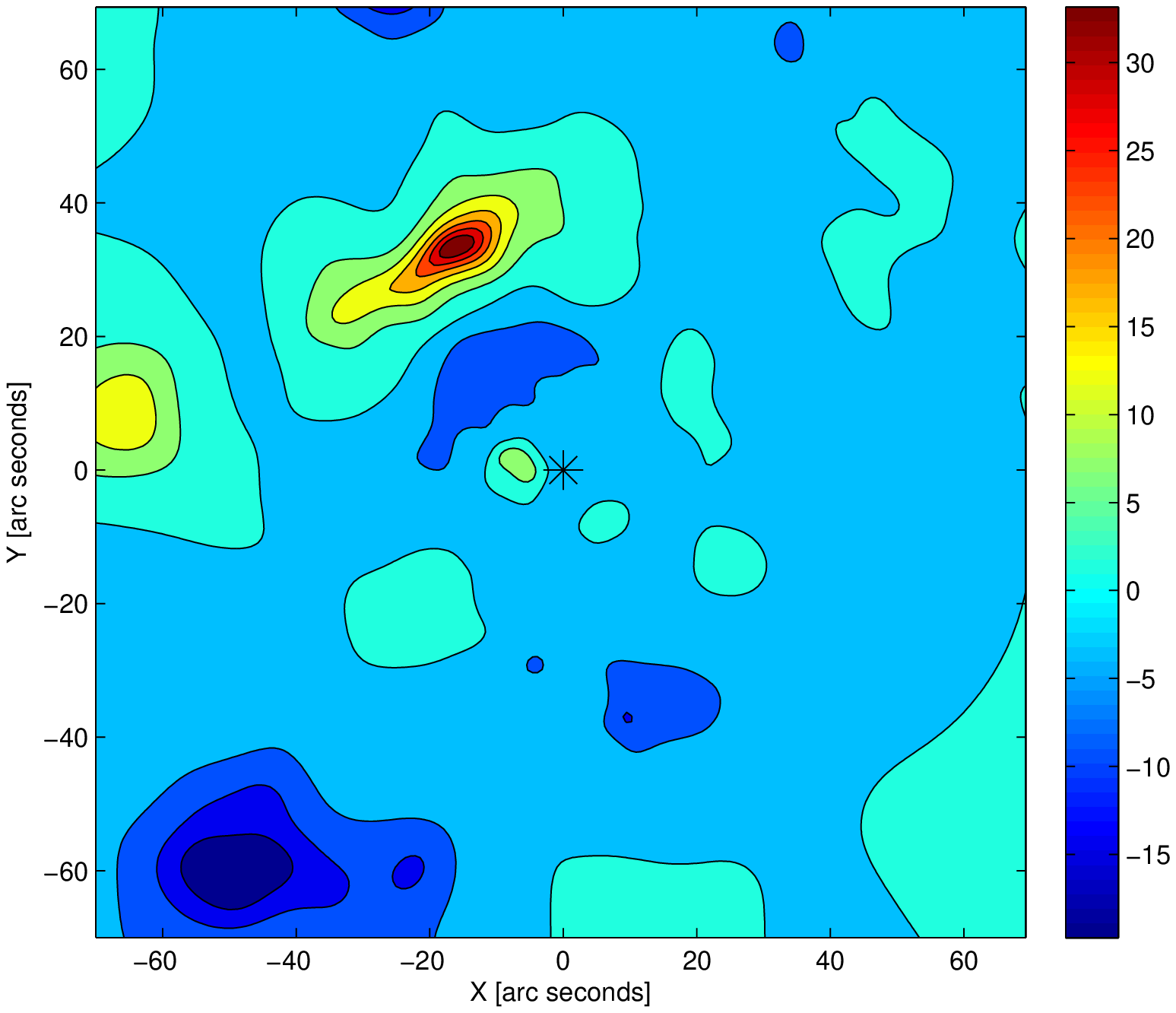}\\
  \includegraphics[width=0.32\hsize]{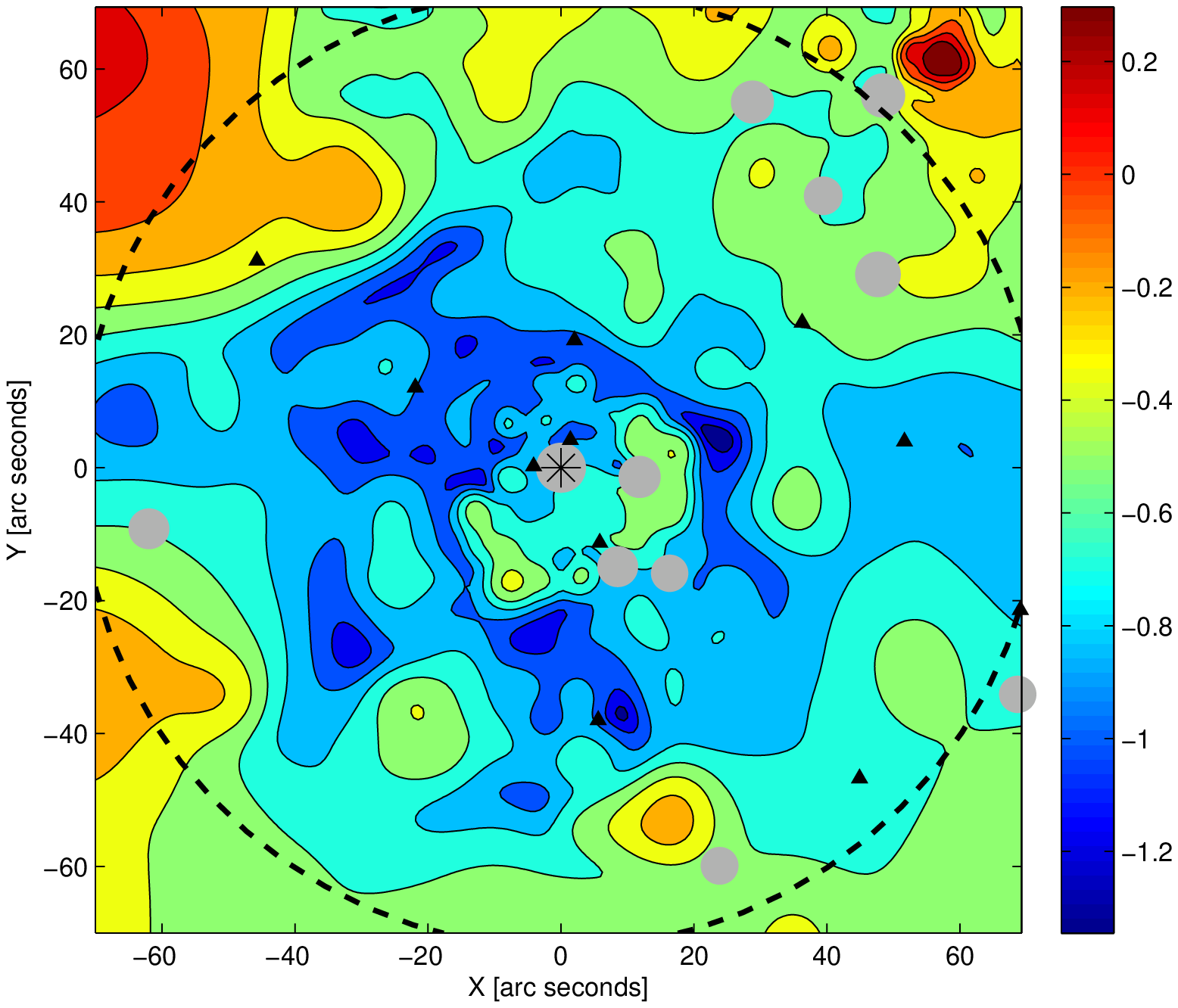}
  \includegraphics[width=0.32\hsize]{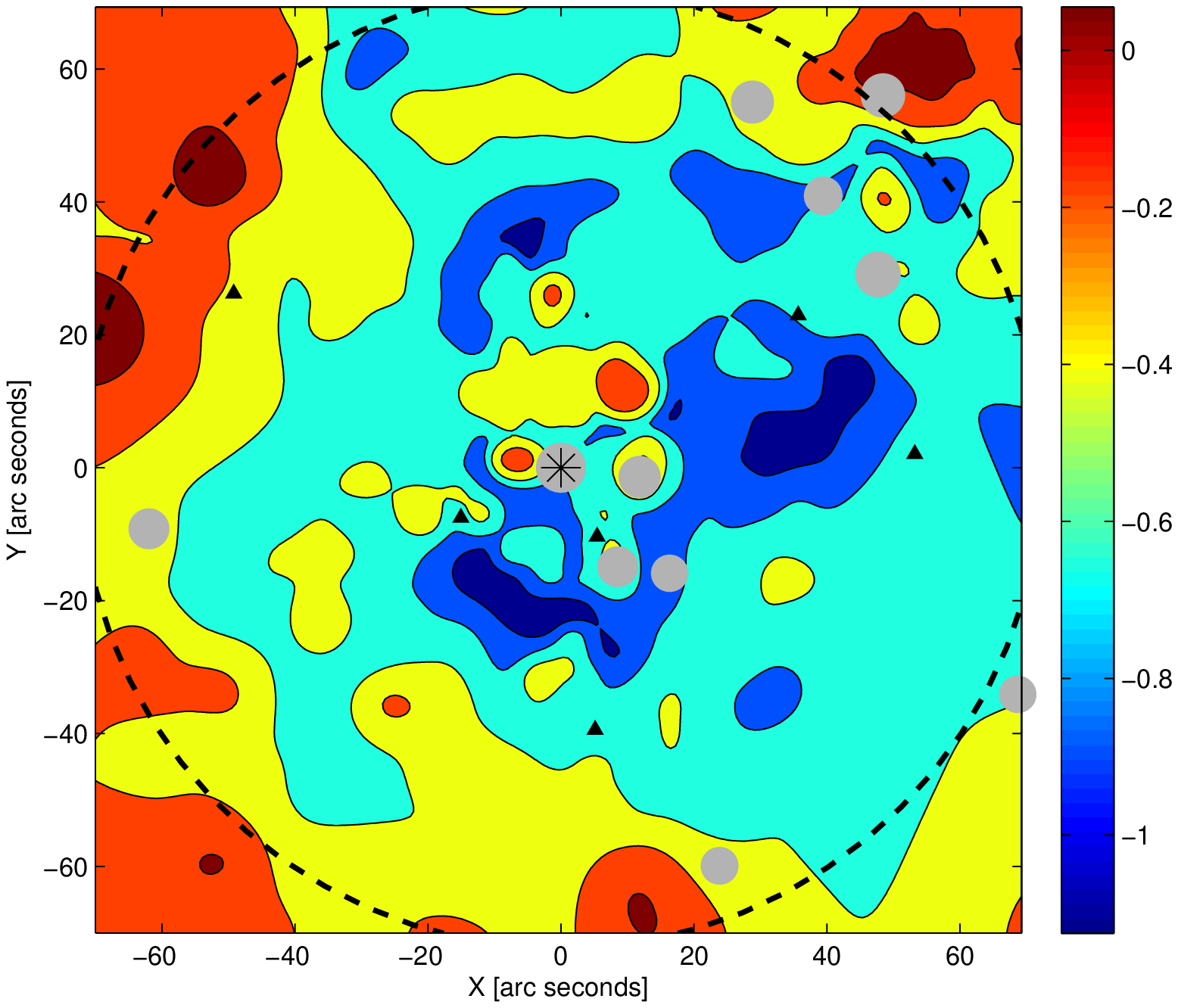}
  \includegraphics[width=0.32\hsize]{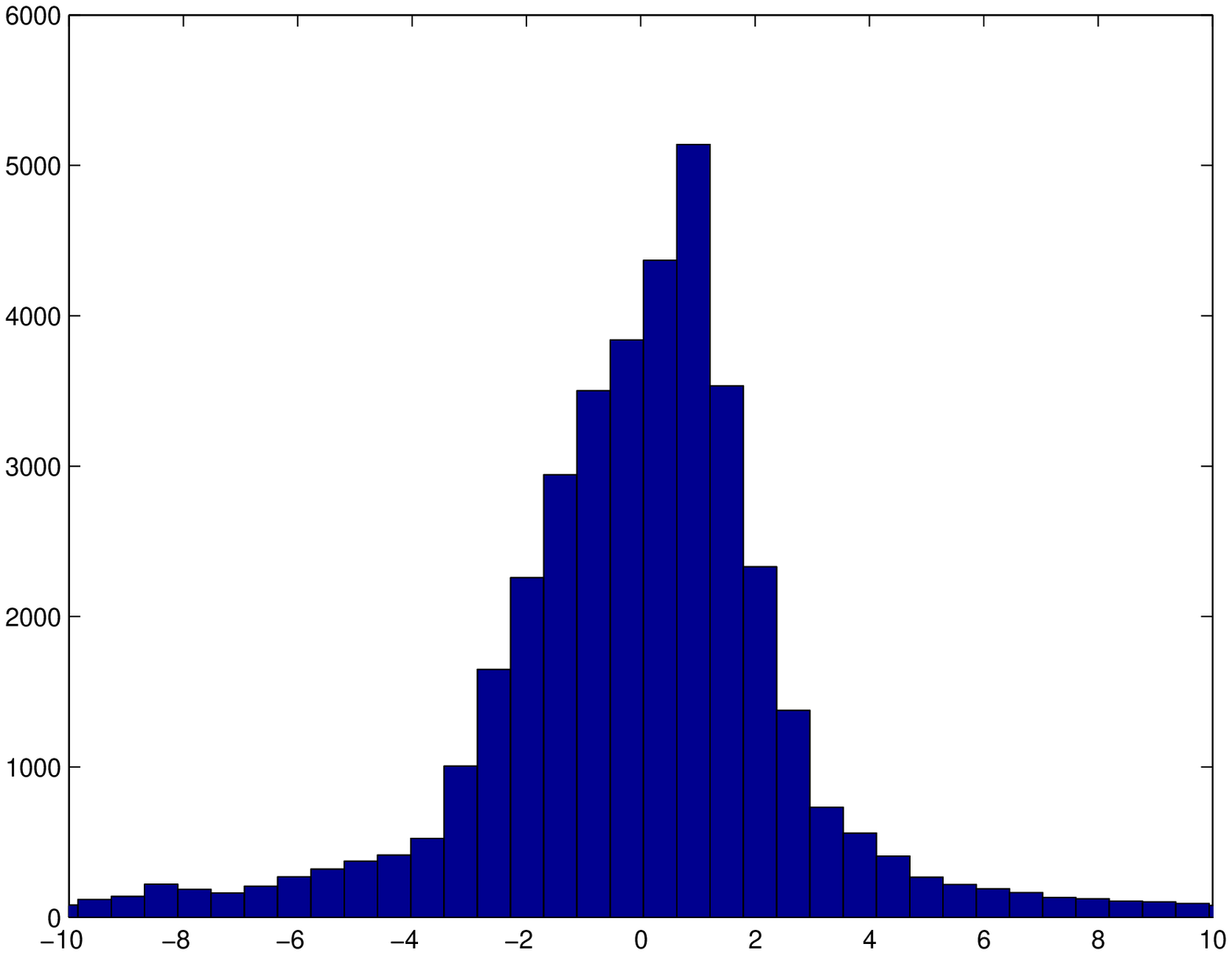}
\caption{\label{fig:a1689_los} Test for the line of sight structure in
  A1689.  Upper left and upper middle panels are the mass maps
  obtained using two separate sets of sources: at low and high
  redshifts respectively. Lower left and lower middle panels are the
  corresponding fractional rms maps. Upper right is the difference between the
  high-$z$ (HRS) and the low-$z$ (LRS) maps divided by the rms
 of the low-$z$ maps  (i.e., $\Delta\Sigma/\delta\Sigma$, which is dimensionless); the scale is linear. Note the apparent structure at higher $z$,
  near $(-20'',\,35'')$.  Lower right is the histogram of the map above it (pixelwise) $\Delta\Sigma / \delta\Sigma$.}
\end{figure}

Because the cluster has many multiply imaged systems spanning a wide
range of redshifts it is possible to test if there are line of sight
(los) structures that have affected the positions of images. We
divided the multiply lensed sources into two groups, the low redshift
system (LRS) and high redshift system (HRS). LRS consists of a total
of three multiply imaged systems with five, three and three (total of
eleven) images at redshifts 2.54, 1.99 and 1.98, respectively.  HRS
consists of a total of two multiply imaged systems with two and five
(total of seven) images at redshifts 4.53 and 2.99, respectively. We
then carried out mass reconstruction for A1689 using LRS and HRS
separately.  The two mass maps are shown in
Figure~\ref{fig:a1689_los}, in the upper left and upper middle
panels, respectively. The corresponding fraction rms distributions are shown
below each map.  The upper right panel is the difference between HRS
and LRS maps divided by the rms of the LRS maps ($\Delta \Sigma/\delta\Sigma$).  
Most of this map is consistent with a uniform surface
mass density of low amplitude, about a factor of ten below the
critical surface mass density. This could be due to steepness, or mass
sheet degeneracy which affected one map more than the other. The only
prominent feature is a mass excess in the HRS map, compared to the LRS
map, centred at around $(-20'',\,35'')$.  The $\delta\Sigma$ maps for both 
HRS and LRS are both low in that region, suggesting that the structure is
real. We interpret this feature as a los structure, probably in the
redshift range 2--3.  Another test of the structure's significance is shown
in the lower right, which contains a histogram of the upper right plot 
$\Delta\Sigma /\delta\Sigma$ (pixelwise). 
The putative los structure contributes to the tail extending beyond
the right edge of the distribution.
The corresponding lensing mass would be $\sim10^{13}M_\odot$ if the
structure were at the same same redshift at A1689, but since the
structure can only be at $z>2.5$, the critical density and hence the
lensing mass are much lower --- a few times $10^{12}M_\odot$ ---
amounting to a modest galaxy group.  There is another feature at
$(-50'',\,-60'')$, but it is outside the image circle, and the $\delta\Sigma$
in that region says that it is not significant.

%%%%%%%%%%%%%%%%%%%%%%%%%%%%%%%%%%%%%%%%%%%%%%%%%%%%%%%%%%%%%%%%%%%%%%%%%%

\section{Discussion}\label{disc}

Gravitational lensing offers a unique opportunity to study the
distribution of matter in clusters of galaxies. Free-form
reconstruction methods take full advantage of this.  Our synthetic
tests show that GRALE recovers the mass distribution well, and the
concomitant errors provide a reliable guide to assessing the
significance of various mass features. The test case in Figure
\ref{fig:plummerout} and \ref{fig:plummer5out} shows no spurious offsets in the mass maps.

Reconstructions of the three real lensing clusters indicate some
curious features.  In two clusters we see offsets between the optical
light and the nearest mass concentrations.  The form of the offsets is
not resolved: they could be distinct peaks in the projected mass
distribution; or they could be spurs that extend from a peak that
itself coincides with the galaxy light; or the offsets could very
lopsided dark halos around galaxies.  (We emphasize that not all
offsets seen in the reconstructed mass maps are significant, but only
those that pass the statistical significance tests.)  A caveat to bear
in mind is the assumption that the observed image positions are
accurate. Because lensed images are often faint, have low surface
brightness and are superimposed on brighter cluster galaxies, image
identification is not always straightforward. It is thus conceivable
that some images have been misidentified.  But assuming the image
identifications are all valid, confirmation by independent techniques
is desirable.  Lens reconstruction methods not assuming light traces
mass in some way include Lensview \citep{2006MNRAS.372.1187W},
LensPerfect \citep{2008ApJ...681..814C} and PBL
\citep{2008ApJ...687...39D} and any of these would be suitable.  If
the mass/galaxy offsets are confirmed, they would lead to interesting
conclusions about the nature of clusters and dark-matter.

In general, several reasons for offsets are possible. Superimposed, but dynamically unrelated 
line of sight structures could contribute lensing mass, with no apparent associated light, 
especially if the structures are considerably further away from us than the main lensing cluster.  
However, we argue that the offset in A3827,  is not due to the line of sight structure
because of the very low redshifts of the sources. In A2218 line of sight structures are also 
unlikely to be the cause because only a very concentrated and massive los structure can 
contribute significantly in the vicinity of a massive clump within a cluster. Such chance
superposition are expected to be rare. 

\begin{figure}
\includegraphics[width=1.0\hsize]{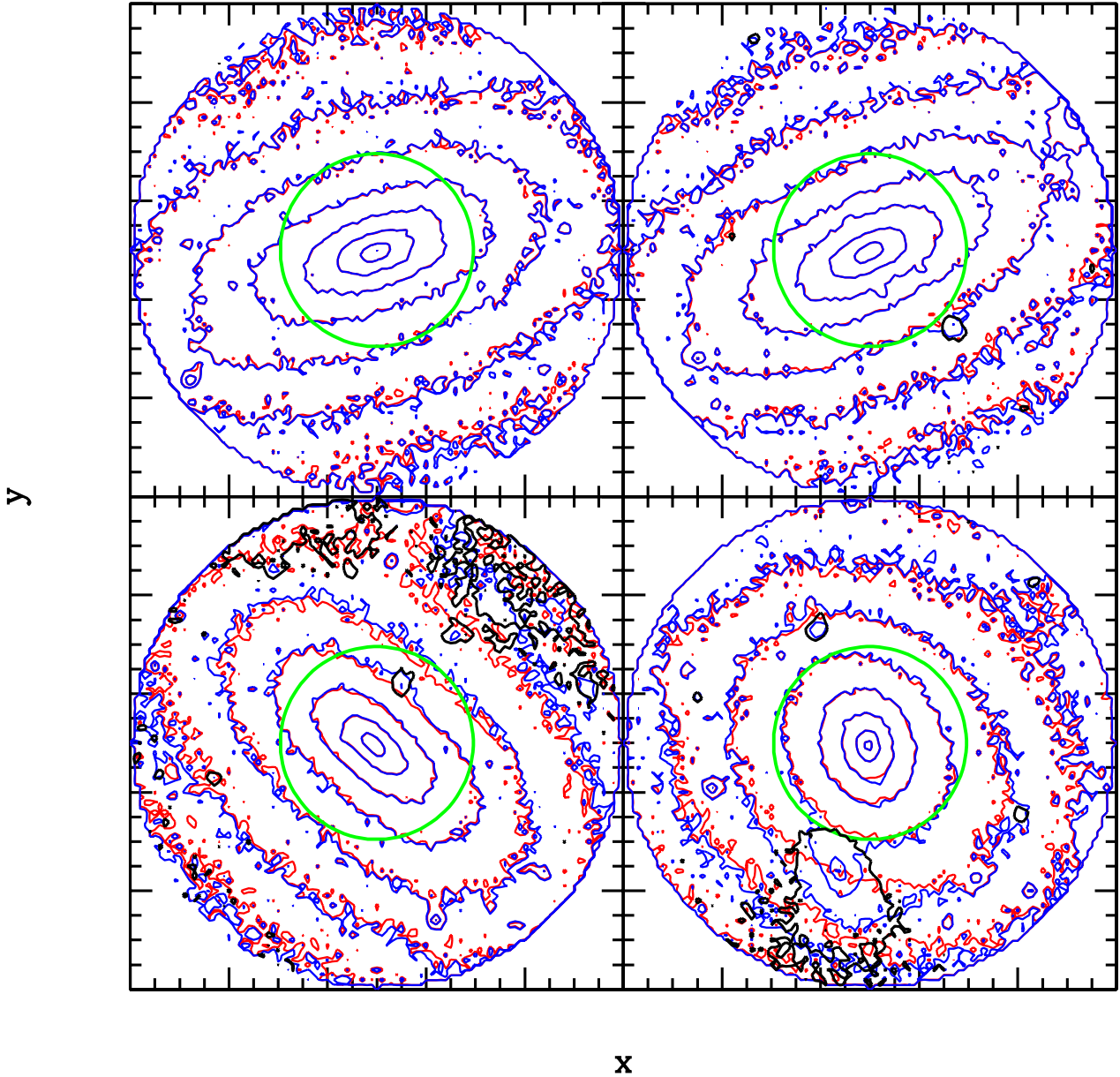}
\caption{\label{fig:simulation} Density contours of projected mass centered on halos taken
from dark matter only simulations \citep{2004MNRAS.352..535D}. The radius of the window
is the virial radius, and the green circle marks the typical radius where lensed images will 
be formed. The red density contours are due to the halo mass interior to the virial sphere, 
while the blue contours are due to all projected mass within a cylinder of roughly 90 Mpc. The
black contours mark regions where the fractional mass excess due to the line of sight structures
(and not the mass within the virial sphere) amount to $25\%$ of total.  The top two panels show
average lines of sight, while the bottom panels the two (out of 100) where los material makes the
most contribution.
}
\end{figure}

Line of sight structures are more likely to make a contribution away from mass concentrations 
within the cluster, where cluster projected densities are lower. This can be illustrated with 
dark-matter N-body simulations. The blue lines in Figure~\ref{fig:simulation} are the isodensity 
contours of the total projected mass in a cylinder centered on a halo whose virial radius is the
radius of the window, while the red lines are the contours of the projected mass inside the 
virial sphere of the cluster. We caution that  these plots were made with a limited line of sight
depth of about comoving 90 Mpc (Simulations courtesy J\"urg Diemand; \cite{2004MNRAS.352..535D}).
The black contours mark regions where the fractional mass excess due to the line of sight 
structures (and not the mass within the virial sphere) amount to $25\%$ of total.  The top two 
panels show examples where the contribution from the los material is typical, while the bottom two
panels present two cases with the most contribution (out of a total of 100 lines of sight). Even though
the length of the cylinder is not large, the plots show that los structures cannot make a significant 
contribution where the cluster density is high. However, such structures can make a significant
contribution at some distance away from the cluster centre.     

In A1689 we might be seeing such a line of sight structure. After subtracting the mass 
reconstruction based on high-$z$ sources (HRS) from that based on low-$z$ sources (LRS) we see a 
mass concentration about 30 arcsec, or 100 kpc from cluster center. It is statistically significant
(it contributes to the tail of the distribution shown in Fig.~\ref{fig:a1689_los} which extend beyond the
right edge of the plot) but is not associated with bright cluster galaxies. We interpret it as arising 
from  a structures between the $z\approx 2$ and $3$.

If not line of sight structure, what else can be responsible for mass-light offsets seen in 
A3827 and A2218?  Offsets could be intrinsic to the cluster, and be due to manifestations 
of known physics, like gravity, and hydrodynamics of the gas, or new physics, such as self-scattering 
of dark matter. Offsets in merging clusters have been observed, but mostly between the dark matter and 
the X-ray emitting gas components \citep{2006ApJ...648L.109C,2013MNRAS.429..833H,2012ApJ...758..128C}. 
In the outskirts of Abell 2744 a separation between dark matter and galaxy components is also seen
\citep{2011MNRAS.417..333M}, and in the merging cluster CL0152-1357 an offset between Sunyaev-Zel'dovich 
effect and X-ray peaks has been detected \citep{2012ApJ...748...45M}.  Most of these offsets are
on larger scales then what we detect in this work. For smaller scale offsets early stage mergers are
probably not the cause, and different set of causes has to be considered.

One of the possibly relevant gravitational effects is the oscillation or wobbling of a galaxy, 
such as a BCG around the bottom of the gravitational potential. This has been observed in 
a sample of galaxy clusters as a displacement of the BCG from the lensing centroid
\citep{2012MNRAS.426.2944Z}. The distribution is displacements is wide, and peaks at
roughly 10~kpc.  Whether this is a likely explanation for the offsets in A3827 and A2218 is yet 
to be determined---the observed offsets are not for central cluster galaxies.

It is less likely, but still possible that the offsets are a consequence of tidal effects. 
These would strip the material from the galaxy symmetrically in the leading and 
trailing directions. Since the offsets in A3827 and A2218 do not show such symmetry, 
tidal effects are probably not the main cause.

Dynamical friction would create an asymmetric structure and would preferentially 
distort the distribution of dark matter and not stars if the former has a more extended 
distribution. A numerical simulation would be required to test this possibility.

The formation of a galaxy cluster is a complex process involving hydrodynamics of gas. 
It is possible that  star formation induced by galaxy mergers within clusters would result 
in stars and dark matter halos offsets.

Finally, if dark-matter has non-negligible self-interaction cross-section, dark-matter 
particles of the galaxy halo would experience a drag force as the galaxy moves within
the halo of the cluster. The nature of the resulting dark-matter features induced by these 
interactions may be consistent with those observed in A3827 and A2218, but detailed 
simulations are required \citep{2014MNRAS.437.2865K}.

\section{Acknowledgement}
LLRW would like to acknowledge the hospitality of ITP, Zurich.

%%%%%%%%%%%%%%%%%%%%%%%%%%%%%%%%%%%%%%%%%%%%%%%%%%%%%%%%%%%%%%%%%%%%%%%%%%

\bibliographystyle{mn2e}

\def\apj{ApJ}
\def\apjl{ApJL}
\def\aj{AJ}
\def\mnras{MNRAS}
\def\aap{A\&A}

\bibliography{ms}

\end{multicols}
\newpage

\newpage

\newpage

\newpage

\end{document}